\documentclass[12pt,a4paper]{revtex4}

\usepackage{amsmath}
\usepackage{amssymb}
\usepackage{graphicx}
\usepackage{hyperref}

\usepackage{amscd,amsthm}

\usepackage{epsfig}
\usepackage{pstricks}

\usepackage{cases}
\usepackage[left=2.5cm,top=2.9cm,right=2.5cm,bottom=2.9cm]{geometry}
\usepackage{comment}
\usepackage{commath}
\usepackage{makecell}

\renewcommand{\text}[1]{\mathrm{#1}}
\newcommand{\Ai}{\mathrm{Ai}}
\newcommand{\Li}{\mathrm{Li}}

\newcommand{\doidoi}[2]{\href{http://dx.doi.org/#1}{#2}}

\newcommand{\rme}{\mathrm{e}}
\newcommand{\rmd}{\mathrm{d}}

\newtheorem{theorem}{Theorem}[section]

\newtheorem{conjecture}[theorem]{Conjecture}
\newtheorem{property}[theorem]{Property}
\newtheorem{appKPZ}[theorem]{Application to KPZ}

\theoremstyle{remark}
\newtheorem{remark}[theorem]{Remark}

\theoremstyle{definition}

\usepackage{tikz}
\usetikzlibrary{patterns,hobby}

\newcommand \be  {\begin{equation}}
\newcommand \ee  {\end{equation}}
\newcommand \bea {\begin{eqnarray} }
\newcommand \eea {\end{eqnarray}}

\begin{document}

\title{Systematic time expansion for the Kardar-Parisi-Zhang equation, 
linear statistics of the GUE at the edge and trapped fermions
}
\author{
Alexandre Krajenbrink$^\spadesuit$, Pierre Le Doussal$^\spadesuit$ and Sylvain Prolhac$^\diamondsuit$
}

\address{
$^\spadesuit$Laboratoire de Physique Th\'eorique de l'\'Ecole Normale Sup\'erieure,
PSL University, CNRS, Sorbonne Universit\'es
24 rue Lhomond, 75231 Paris Cedex 05, France\\
$^\diamondsuit$Laboratoire de Physique Th\'eorique, IRSAMC, UPS,  Universit\'e de Toulouse, France
}
\begin{abstract}
We present a systematic short time expansion for the generating function of the one point height 
probability distribution for the KPZ equation with droplet initial condition, which goes much beyond previous studies. The expansion is checked against a numerical evaluation of the known exact Fredholm determinant expression. We also obtain the next order term for the Brownian initial condition. Although initially devised for short time, a resummation of the series allows to obtain also the {\it long time large deviation function}, found to agree with previous works using completely different techniques.  Unexpected similarities with stationary large deviations of TASEP with periodic and open boundaries are discussed. Two additional applications are given. (i) Our method is generalized to study the linear statistics of the {Airy point process}, i.e. of the GUE edge eigenvalues. We obtain the generating function of the cumulants of the empirical measure to a high order. The second cumulant is found to match the result in the bulk obtained from the Gaussian free field by Borodin and Ferrari \cite{BorodinFerrari,borodinclt}, but we obtain systematic corrections to the Gaussian free field (higher cumulants, expansion towards the edge). This also extends a result of Basor and Widom \cite{basor1999determinants} to a much higher order. We obtain {large deviation functions} for the {Airy point process} for a
variety of linear statistics test functions. (ii) We obtain results for the {\it counting statistics
of trapped fermions} at the edge of the Fermi gas in both the high and the low temperature limits.
\end{abstract}

\maketitle

\tableofcontents

\begin{section}{Introduction and main results}

\subsection{Overview}

Recent exact results on a number of models in the 1D Kardar-Parisi-Zhang (KPZ) universality class
allows insight into non-equilibrium dynamics for a broad class of systems
\cite{HH-TakeuchiReview}. Such models \cite{spohn2000,png,Johansson2000,corwinsmallreview} range from
discrete particle transport models, such as the TASEP \cite{spohnTASEP},
to continuum models of stochastic interface growth, such as the KPZ equation \cite{KPZ,reviewCorwin,directedpoly,HairerSolvingKPZ}. In all of them, an analog of a "height field" can be defined, with common universal
behavior at large scale defining the KPZ class. Although exact formulas valid at all time exist for some observables, such as the probability distribution function (PDF) of the (centered) 
height $H(t)$
at a given space-time point for some initial conditions, extracting from them the physical information about the stochastic process may prove quite difficult. It often amounts to study functional (i.e. Fredholm) determinants,
which can be evaluated numerically \cite{bornemann2009numerical}, but from which explicit formulas are hard to extract. While the long time asymptotics for the typical fluctuations (i.e. the regime where $H(t) \sim t^{1/3}$) has been obtained in several cases, leading to the celebrated Tracy-Widom distribution of random matrix theory \cite{TW}, the short and intermediate time behavior, as well as the regime of large fluctuations has only been studied recently.

The short time expansion of the KPZ equation formally identifies with perturbation theory in the noise.
Naive perturbation theory straightforwardly shows that typical behavior is similar to its linear version,
the Edwards-Wilkinson (EW) equation, with Gaussian fluctuations $H(t) \sim t^{1/4}$. Study of the higher cumulants of the height shows, however, that there exists a non trivial short time large deviation regime,
$H(t) \sim \mathcal{O}(1)$, where the PDF of the height takes the form $P(H,t) \sim \exp(- \Phi(H)/t^{1/2})$,
and which has required to develop new theoretical methods. The first is the weak noise theory (WNT), which has allowed to obtain numerically the large deviation function
$\Phi(H)$, and analytically its tails for large $|H|$ for a
variety of initial conditions \cite{Korshunov,Baruch,MeersonParabola,janas2016dynamical,meerson2017randomic,Meerson_Landau,Meerson_flatST}.
The second method uses the known exact solutions of the KPZ equation
\cite{SS10,CLR10,DOT10,ACQ11,CLDflat,SasamotoStationary,SasamotoStationary2,BCFV},
leading to exact formula for the rate function
$\Phi(H)$ for arbitrary $H$. It has been achieved for the droplet initial condition
\cite{le2016exact} (with an impressive confirmation from high precision numerics \cite{NumericsHartmann}), for the Brownian initial condition \cite{krajenbrink2017exact}, and for the half-space KPZ equation with droplet initial condition
\cite{krajenbrink2018large}.

In this paper we develop a systematic expansion of the KPZ equation at short time,
using the exact solutions. 
We obtain an expansion of various generating functions associated to the height $H(t)$ 
to a much higher order than previous results. We focus on the droplet initial condition, with some
additional results for the Brownian initial condition. We use a direct method starting from the known Fredholm determinant representation in Section \ref{section sw systematic},
and the cumulant method introduced in \cite{KrajLedou2018} in Section \ref{sec:cumulants}. For the
droplet initial condition we find that the additional orders lead to a very good approximation
of the exact Fredholm determinant, up to large time. We also obtain the large deviation function of $H$ and the generating function of the cumulants of $H$ to the next order in $t$.

There are several other outcomes of our study, some unexpected. First, a careful examination
of the structure of the short time expansion allows to obtain some results {\it at long time}. 
This is relevant to the recent works which have studied the large deviations at long time $t \gg1$
\cite{LargeDevUs,sasorov2017large,CorwinGhosalTail,JointLetter,KrajLedou2018}. The 
left tail of the PDF of $H(t)$ was argued quite generally to take
the form $\log P(H,t) \simeq - t^2 \Phi_-(H/t) $ for large negative fluctuations $- H \sim t$ when $t \gg 1$.
\cite{LargeDevUs}. In recent works the explicit expression of $\Phi_-(z)$ for droplet initial condition
in the full space was obtained (i) using a WKB type approximation \cite{sasorov2017large} on
a non local Painlev\'e type equation representation of the exact solution derived in \cite{ACQ11} (ii)
using Coulomb gas methods \cite{JointLetter}. Remarkably, we obtain here a third, and completely independent derivation of this result. 

The second outcome relates to random matrix theory. There has been a lot
of interest in linear statistics and central limit theorems for eigenvalues of large
random matrices in the Gaussian Unitary Ensemble (GUE) \cite{mehta}, see e.g. \cite{johansson2015gaussian,grabsch2017truncated,grabsch2017truncated2} and references therein. Interesting connections to the 2D Gaussian free field have been
demonstrated, see e.g. \cite{BorodinFerrari,borodinclt, BWZ}. There are not so many results concerning the linear
statistics near the edge of the GUE, see however \cite{basor1999determinants}.
Remarkably, there is a connection between the KPZ equation with
droplet initial condition and the Airy point process, which describes the (scaled) eigenvalues of
the GUE at the edge \cite{SS10,CLR10,DOT10,ACQ11,dean2015finite,borodin2016moments,borodinVertexASEPmoments}.
Here we are able to use that connection to make detailed
predictions for (i) the higher cumulants of the linear statistics of the Airy point process
(ii) the large deviations.

The third outcome relates to non-interacting fermions in a trap
at finite temperature. At the edge of the Fermi gas, where the density vanishes, the quantum 
and thermal fluctuations are both important. Interesting questions pertain to the
position of the rightmost fermion and to the counting statistics (the number of
fermions in an interval). At zero temperature they have been well studied
\cite{eisler_prl, marino_prl, castillo, CDM14, marino_pre, mehta}. Recently
it was found that the finite temperature problem can be related to the solution
of the KPZ equation with droplet initial condition \cite{dean2015finite,dean2016noninteracting}.
Our results for the systematic short time expansion for KPZ can thus be transported
and extended to obtain the high temperature expansion for the full counting statistics 
of the fermions, well beyond previous results \cite{le2016exact,KrajLedou2018}.
Similarly our large time solution can be extended to obtain the large deviations
of the number of fermions in an interval at low temperature.

Finally, we observe unexpected similarities in the expressions of large deviations for the KPZ equation on an infinite \cite{le2016exact} or semi-infinite \cite{krajenbrink2018large} line at short time and for TASEP with periodic \cite{DerridaLebowitz,P2016,mallick2018brownian} or open geometry \cite{LazarescuMallick,LazarescuMallick2,Lazarescu,CrampeNepomechie} in the long time limit. In terms of KPZ universality in finite volume, this suggests the existence of a duality between stationary fluctuations at the KPZ fixed point and fluctuations in the vicinity of the Edwards-Wilkinson fixed point. Such a duality is not predicted by the usual scaling theory of the KPZ equation. We stress that we do not have at the moment a precise statement about what quantities should be related by this hypothetical duality for general initial condition.\\

Before we detail the aim of the paper and summarize the main results, let us introduce more explicitly 
the model studied here.

\subsection{The model: KPZ equation and initial conditions} 
	In this paper we consider the Ito solution $Z(x,t)$ of the stochastic heat equation
	\begin{equation}
	\label{SHE}
	\partial_{t}Z=\partial_{x}^{2}Z+\sqrt{2}\,\xi Z
	\end{equation}
	with $\xi$ Gaussian white noise, $\mathbb{E}[\xi(x,t)]=0$, $\mathbb{E}[\xi(x,t)\xi(x',t')]=\delta(x-x')\delta(t-t')$. The corresponding KPZ equation for the height field $h(x,t)$ defined through Cole-Hopf transform $Z(x,t)=\rme^{h(x,t)}$ is
	\begin{equation}
	\label{KPZ}
	\partial_{t}h=\partial_{x}^{2}h+(\partial_{x}h)^{2}+\sqrt{2}\,\xi\;.
	\end{equation}
	Exact solutions have been found for several initial conditions, notably
flat, droplet and stationary
\cite{SS10,CLR10,DOT10,ACQ11,CLDflat,SasamotoStationary,SasamotoStationary2,BCFV},
and, remarkably, can be expressed using Fredholm determinants or Pfaffians. 
The typical behavior of the KPZ height fluctuations has been obtained from them, and the scaled PDF of $h(x,t)$ converges in the long time limit to the so-called Tracy Widom distributions (i.e. the distributions of the largest eigenvalues of standard Gaussian random matrix ensembles). \\

We study in this paper the short time expansion of the one-point statistics for a related height function $H(t)$. We consider two types of initial conditions, for which exact solutions are known:
	
	\begin{enumerate}
	\item The sharp wedge initial condition, corresponding to droplet growth, defined for the stochastic heat equation as $Z(x,0)=\delta(x)$.
	The first moment is normalized as $\langle Z(x,t)\rangle=\frac{\rme^{-\frac{x^{2}}{4t}}}{\sqrt{4\pi t}}$. For this initial condition, we define the shifted height $H$ as
	\begin{equation}
	H(x,t)=\log\frac{Z(x,t)}{\langle Z(x,t)\rangle}+\frac{t}{12}=h(x,t)+\frac{t}{12}+\log\sqrt{4\pi t}+\frac{x^{2}}{4t}\;.
	\end{equation}
	The statistics of $H(x,t)$ is known to be independent from $x$.	We can thus focus on $x=0$ and consider only $H(t)=H(0,t)$. 
	
	\item The Brownian initial condition with drift $w$, defined as $h(x,0) = B(x) - w |x|$, where $B$ is a unit double-sided Brownian motion with $B(0)=0$. The case $w=0^+$ corresponds to the invariant measure of the KPZ equation, and is called the stationary initial condition. We will be interested in the height at point $x=0$, and we define the shifted height $H(t)=h(0,t)+\frac{t}{12}$.

	\end{enumerate}

\subsection{Aim and main results} 
\label{section aims}

Our aim in this paper is twofold. First we construct a systematic short time expansion for
the solutions to the KPZ equation in the cases where they can be obtained exactly in terms of Fredholm determinants, namely for the droplet and Brownian initial condition. Remarkably, it will also allow us to obtain results at long time in the droplet case. Second, we use the 
connection between the droplet initial condition solution to the Airy point process, and generalize our calculation
to obtain exact results for the linear statistics at the edge of the GUE and for the counting statistic of trapped fermions. 

\subsubsection{KPZ equation} 

In this paper we obtain the short time expansion for a doubly exponential generating function $Q_{t}(\sigma)$. For each initial condition, $Q_{t}(\sigma)$ is defined in accordance with the form of the exact solution. For the droplet initial condition, we define $Q_{t}(\sigma)$ as the expectation over the KPZ evolution,
\begin{equation}
	\label{G sw}
	Q_{t}(\sigma) = \mathbb{E}_{\, \mathrm{KPZ}}\left[\exp\left(\frac{\sigma}{\sqrt{4 \pi t}}\,\rme^{H(t)}\right) \right]\;.
\end{equation}
Using known exact results summarized in the next section, we derive the new representation \eqref{G'[L]} for $Q_{t}(\sigma)$, from which the short time expansion can be performed systematically for $\sigma<1$. At leading orders in $t$, we obtain
\begin{equation}
	\label{G short time expansion sw t1/2}
	\log Q_{t}(\sigma)=
	\frac{\Li_{5/2}(\sigma)}{\sqrt{4\pi t}}
	+\frac{1}{4\pi}\int_{0}^{\sigma}\!\frac{\rmd u}{u}\,\Li_{1/2}(u)^{2}
	+\Big(\frac{\Li_{1/2}(\sigma)^{3}}{12\pi^{3/2}}+\frac{\Li_{-1/2}(\sigma)}{24\sqrt{\pi}}\Big)\sqrt{t}+\mathcal{O}(t)\;.
\end{equation}
The expansion up to order $t^{3}$ is given in \eqref{G short time expansion}, and a conjecture for the form of the expansion to all orders in \eqref{conjecture G}. 
The leading order $\mathcal{O}(t^{-1/2})$ recovers the result of \cite{le2016exact}.

Comparison with the exact Fredholm determinant representation \eqref{G[M] sw} below, numerically evaluated using Bornemann's method \cite{bornemann2009numerical}, shows a very good agreement. The plots of Figure \ref{fig Gt[sigma]} (see especially the insets for $t=10$) indicate however that the expansion \eqref{G short time expansion sw t1/2} has only an asymptotic nature: adding more terms $\propto t^{m}$ in the expansion only reduces the difference to the exact value \eqref{G[M] sw} for small enough orders $m\leq n(t,\sigma)$, before the expansion begins to depart from the exact value when adding the orders $\propto t^{m}$ with $m>n(t,\sigma)$ of the expansion. Equivalently, for fixed values $t$, $\sigma$, the short time expansion can only approximate $\log Q_{t}(\sigma)$ within a fixed range $\epsilon(t,\sigma)>0$. As usual for asymptotic series, $n(t,\sigma)\to\infty$ and $\epsilon(t,\sigma)\to0$ when $t\to0$ for fixed $\sigma$. Additionally, the plots in Figure \ref{fig Gt[sigma]} also suggest that $n(t,\sigma)\to\infty$ and $\epsilon(t,\sigma)\to0$ when $\sigma\to-\infty$ with fixed $t$.

The point $t=0$ thus corresponds to an essential singularity of the function $Q_{t}(\sigma)$, and the expansion of $Q_{t}(\sigma)$ around this point has a radius of convergence equal to zero (\textit{i.e.} there is no $t\neq0$ such that the full perturbative series \eqref{G short time expansion sw t1/2} converges to the exact value \eqref{G[M] sw} of $Q_{t}(\sigma)$). However, the expansion \eqref{G short time expansion sw t1/2}, whose summation is equal to the analytic part in the variable $t^{1/2}$ of $Q_{t}(\sigma)$ with essential singularities at $t=0$ removed, does have a non-zero radius of convergence $\tau(\sigma)$ in the variable $t$. A numerical evaluation of the first coefficients of the short time expansion for various values of $\sigma$ suggest that $\tau(\sigma)$ is uniformly bounded from below away from $\sigma=1$, but converges to zero when $\sigma\to1^{-}$. This is supported by conjecture \eqref{conjecture G}, which implies $\tau(\sigma)\simeq\pi^{2}\log(-\sigma)$ when $\sigma\to-\infty$ and $\tau(\sigma)\sim(1-\sigma)^{3}$ when $\sigma\to1^{-}$.\\

\begin{figure}
	\begin{center}
		\begin{tabular}{lll}
			\begin{tabular}{c}\includegraphics[width=75mm]{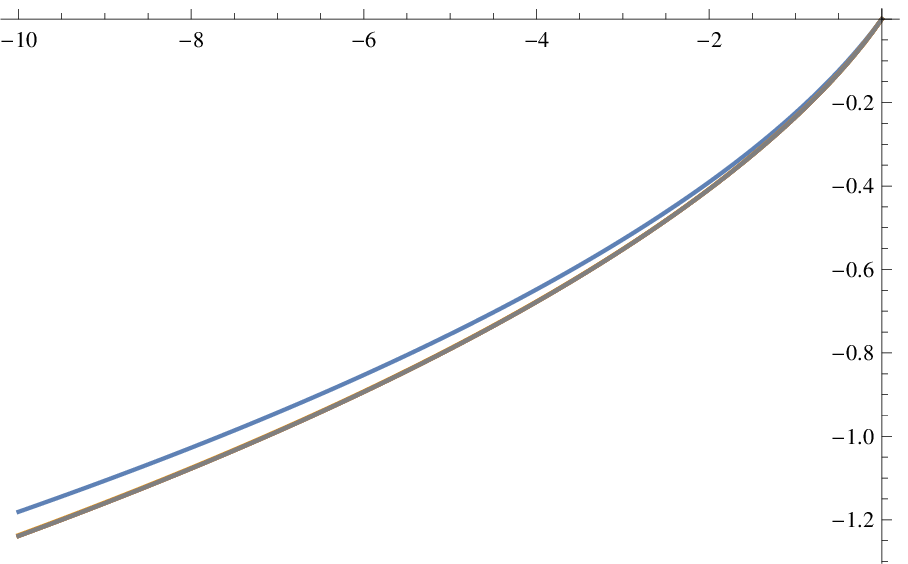}\end{tabular}
			&&
			\begin{tabular}{c}\includegraphics[width=75mm]{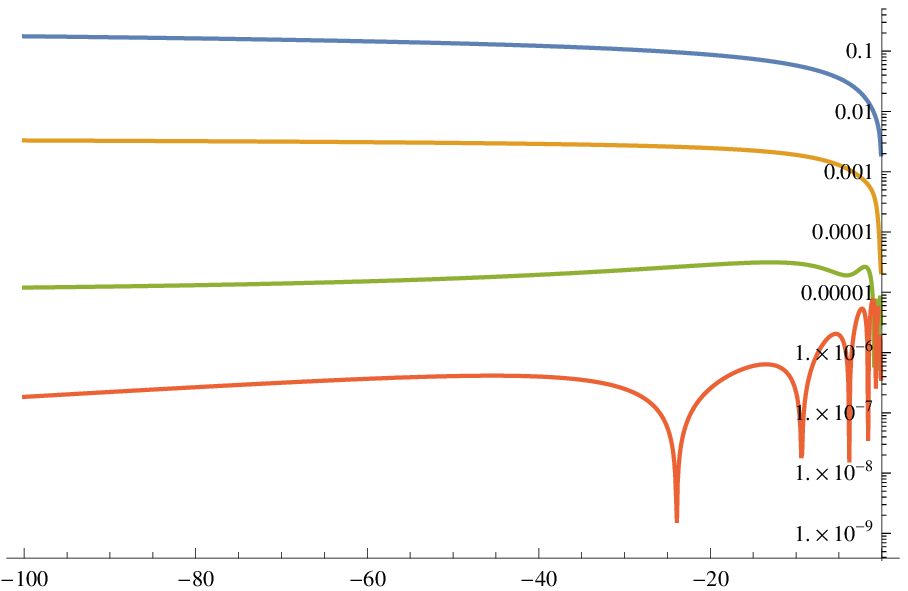}\end{tabular}\\
			\begin{tabular}{c}\includegraphics[width=75mm]{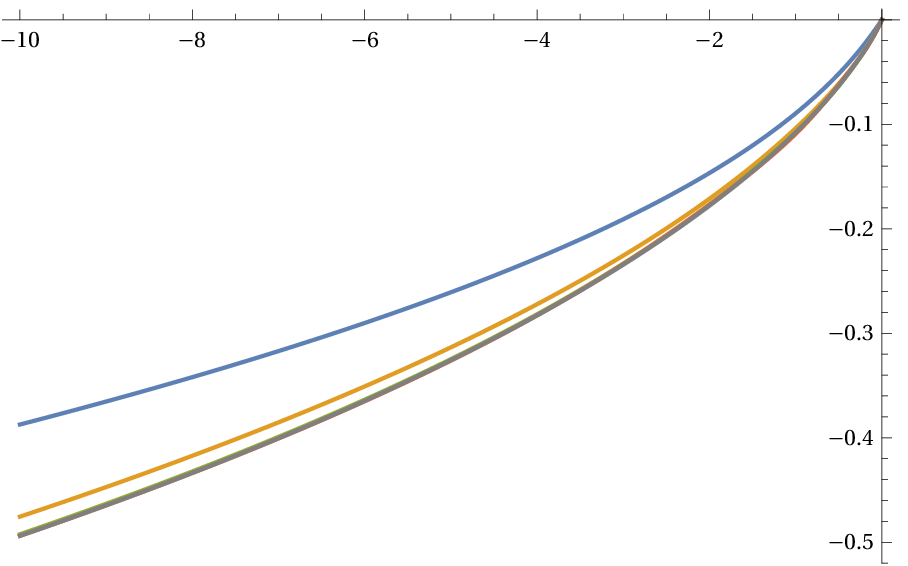}\end{tabular}
			&&
			\begin{tabular}{c}\includegraphics[width=75mm]{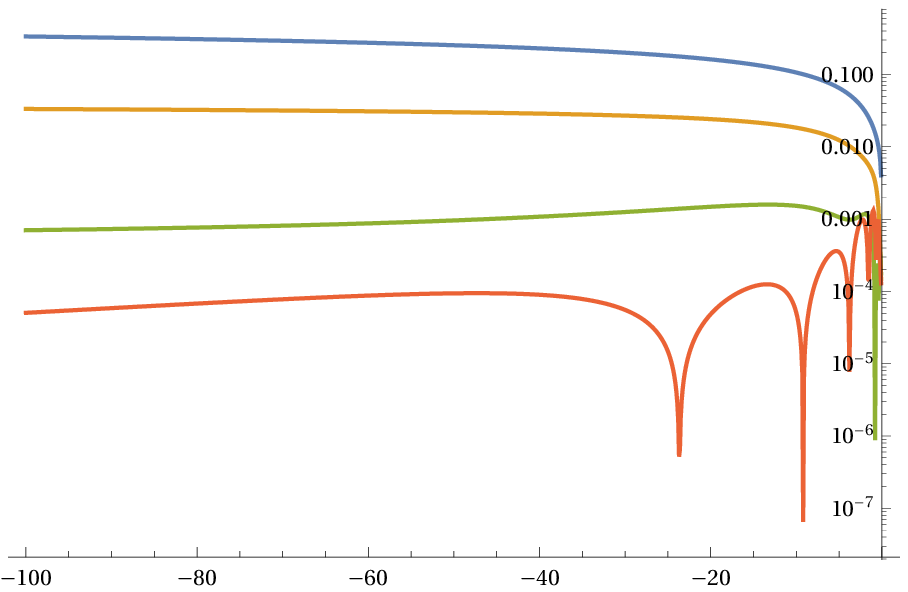}\end{tabular}\\
			\begin{tabular}{c}\includegraphics[width=75mm]{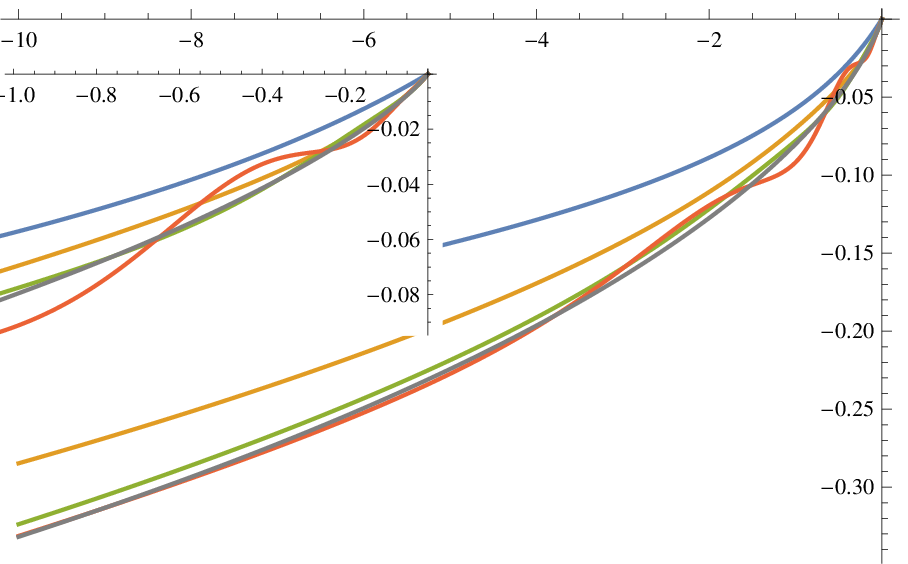}\end{tabular}
			&&
			\begin{tabular}{c}\includegraphics[width=75mm]{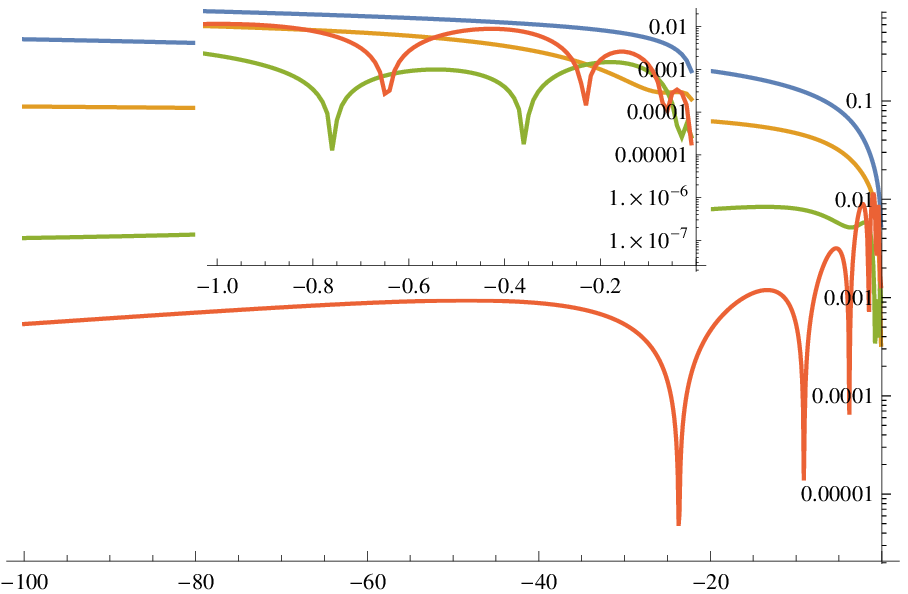}\end{tabular}
		\end{tabular}
	\end{center}
	\caption{Plot of the generating function $\log Q_{t}(\sigma)$, defined in \eqref{G[M] sw} as a function of the parameter $\sigma \leq 0$ for $t=1$ (top), $t=5$ (middle) and $t=10$ (bottom). On the left, the lowest, grey curve corresponds to the numerical evaluation of the Fredholm determinant \eqref{G[M] sw} while the blue, orange, green and red curves (starting from the top) correspond to the short time expansion \eqref{G short time expansion} truncated to respective order $t^{0}$, $t$, $t^{2}$ and $t^{3}$. On the right, the logarithm of the absolute value of the difference between the short time expansion and the numerical evaluation of the Fredholm determinant is plotted for a larger range of $\sigma$, with the same color code as on the left for the various truncations of the expansion. Insets show the situation in more details in the range $-1<\sigma<0$ for $t=10$.}
	\label{fig Gt[sigma]}
\end{figure}

For the Brownian initial condition, $Q_{t}(\sigma)$ is defined as the expectation value over the KPZ evolution, the Brownian motion $B$ characterizing the initial condition, and an additional random variable $\chi$ independent of $H$ with probability density $p(\chi) =  \rme^{-2 w \chi - \rme^{-\chi}} / \Gamma(2 w)$,
\begin{equation}
	\label{G stat}
	Q_{t}(\sigma) = \mathbb{E}_{\, \mathrm{KPZ},\, B, \, \chi}\left[\exp\left(\frac{\sigma}{t}\,\rme^{H(t)+\chi}\right) \right]\;.
\end{equation}
Using the cumulant method introduced in \cite{KrajLedou2018,krajenbrink2018large} we obtain for fixed $\tilde{w}=w\sqrt{t}$ the short time expansion 
\begin{equation}
\begin{split}
	\log Q_{t}(\sigma)  =&  \frac{1}{ \sqrt{t}} \int_{-\infty}^{+\infty}\frac{\rmd k}{2\pi} \, 
{\rm Li}_2\Big(\frac{\sigma \rme^{-k^2}}{k^2 + \tilde w^2}\Big)  -  \int_{-\infty}^{+\infty}\frac{\rmd k}{2\pi} \, \frac{\tilde{w}}{k^2 + \tilde w^2} 
\log\Big(1- \frac{\sigma \rme^{-k^2}}{k^2 + \tilde w^2}\Big) \\
& + \frac{1}{4 \pi^2}  \int_0^{+\infty} \mathrm{d} u \, u \, \bigg|\int_{-\infty}^{\infty} \mathrm{d}k \,  \rme^{i k u} \, \log\Big(1-\frac{\sigma \rme^{-k^2}}{k^2+ \tilde w^2}\Big) \bigg|^2
+ \mathcal{O}(\sqrt{t}) \\
\end{split}
\end{equation}
The leading order agrees with the result obtained in \cite{krajenbrink2017exact}. The limit $\tilde w=0$
relevant for stationary KPZ is discussed in Section \ref{subsubsec:Pstat}.

In addition we also study in Section \ref{sec:Fts} another generating function, the exponential generating function $F_{t}(s)$, whose short time expansion is related to that of $Q_{t}(\sigma)$. For droplet initial condition, we define
\begin{equation}
	\label{F sw}
	F_{t}(s)=\mathbb{E}_{\,\mathrm{KPZ}}\left[\exp\left(-s\,\frac{H(t)}{\sqrt{t}}\right)\right]
\end{equation}
and obtain for $s>0$ the short time expansion
\begin{eqnarray}
\label{F short time expansion sw t0}
\log F_{t}(s) && = \frac{1}{\sqrt{t}} \left[ \dfrac{\mathrm{Li}_{5/2}(\sigma)-\mathrm{Li}_{3/2}(\sigma)}{\sqrt{4\pi }}+\frac{\mathrm{Li}_{3/2}(\sigma)}{\sqrt{4\pi}}\log \left( \frac{\mathrm{Li}_{3/2}(\sigma)}{\sigma} \right) \right] 
\\
&& + \frac{1}{4\pi}\int_{0}^{\sigma}\!\frac{\rmd u}{u}\,\Li_{1/2}(u)^{2} +\frac{1}{2}\log \left( \frac{\mathrm{Li}_{3/2}(\sigma)}{\mathrm{Li}_{1/2}(\sigma)} \right) + \mathcal{O}(\sqrt{t}) \nonumber 
\end{eqnarray}
where $\sigma$ is related to $s$ as (for $s \geq - \zeta(3/2)/\sqrt{4 \pi}$, see Section  \ref{sec:Fts})
\begin{equation}
\label{s[sigma]}
s=-\frac{\mathrm{Li}_{3/2}(\sigma)}{\sqrt{4\pi}}\;.
\end{equation}
From \eqref{F short time expansion sw t0} we can obtain the short time expansion of the cumulants 
$\mathbb{E}_{\mathrm{KPZ}}\left[ H(t)^q\right]^c$ of the height, see Eq. 
\eqref{cumH} and Table \ref{Table1_cum} which agree with the values obtained in \cite{CLR10} (see Eqs. (11) and (12) there). In principle from our result
for $Q_t(\sigma)$ to order $t^3$ one can obtain $F_t(s)$ to the same order. Since it is tedious
we give here explicitly only the first three orders (the order $\mathcal{O}(\sqrt{t})$ is displayed in  \eqref{mu1}).

\medskip


We also obtain the corrections to the short time large deviation expression for $P(H,t)$ in the form
\be
\log P(H,t) = - \frac{1}{\sqrt{t}} \Phi(H) + \Phi_0(H) + \mathcal{O}(\sqrt{t}) 
\ee
both for droplet and Brownian initial condition. For droplet initial condition the result is displayed in a parametric
form in 
\eqref{param_phi_0_1} for $H<H_c=\log \zeta(3/2)$ in and \eqref{param_phi_0_2}
for $H>H_c$. The leading order rate function $\Phi(H)$ is identical to the
one obtained in \cite{le2016exact}. The result for the Brownian initial condition is given in a parametric
form in Eqs \eqref{Phi0stat}, \eqref{parametric} for arbitrary $\tilde w$. The leading
order rate function $\Phi(H)$ is identical to the
one obtained in \cite{krajenbrink2017exact}. The expressions displayed 
here are valid for $H<H_{c}(\tilde w)$ given in \eqref{Hcw}. While
$\Phi(H)$ was also obtained for $H>H_c$ in \cite{krajenbrink2017exact},
we leave to the brave and courageous to extend it to $\Phi_0(H)$.

Finally, upon careful examination of the structure of the cumulant expansion 
in Section \ref{sec:structure} we obtain, quite remarkably, the large deviation form for the left tail of $P(H,t)$ {\it in the large time limit}, for $-H \sim t$ as
\be \label{Plarget} 
\log P(H,t)=  - t^2 \Phi_-(z) +\mathcal{O}(1) \quad , \quad z=\frac{H}{t} <0 
\ee
with the function $\Phi_-(z)$ given in \eqref{eq:Phi}.
This recovers, by a completely different method, exactly the result of  \cite{sasorov2017large} using a non-local Painleve equation method and of \cite{JointLetter} using a Coulomb gas method. 
We have not been able to obtain the equivalent result for the stationary initial condition but
we are able (see Section \ref{subsec:statlargetime}) to obtain the two leading orders 
of the rate function $\Phi^{\rm stat}_-(z)$ in the large $|z|$ expansion, which
is consistent with the conjecture $\Phi^{\rm stat}_-(z) = \Phi^{\rm droplet}_-(z)=\Phi_-(z)$
discussed in \cite{JointLetter}.

\subsubsection{Linear statistics of the Airy point process (edge of GUE)}

Linear statistics of the GUE adresses the evaluation of averages of the type $\mathbb{E}_{\rm GUE}[ \exp (\sum_{i=1}^N F(\lambda_i) )]$ for suitably chosen functions $F(\lambda)$, where the $\lambda_i$ are the eigenvalues of a $N \times N$ GUE random matrix. Here we will be interested in the edge of the GUE in the large $N$ limit, which is described by the so-called
Airy point process \cite{TW}. Normalizing the GUE measure at large $N$ to a support $[-2,2]$
we recall that the Airy point process is the limit point process $\lbrace \mathbf{a}_i \rbrace$ obtained by writing
\be
\lambda_i = 2 + \frac{\mathbf{a}_i}{N^{2/3}} 
\ee
as $N \to \infty$, where the eigenvalues are ordered as $\lambda_{1}<\ldots<\lambda_{N}$. Let us denote equivalently by $\mathbb{E}_{\Ai}[\dots]$ or $\overline{\cdots}$ 
averages over the Airy point process. Introducing $\mu(a)= \sum_i \delta(a-\mathbf{a}_i)$ its empirical measure, we recall
that its mean density is given by $\overline{\mu(a)}=K_{\Ai}(a,a)$ in terms of the
Airy kernel defined in Eq. \eqref{KAi}. It decays (stretched) exponentially for $a \to +\infty$ and 
behaves as $\frac{1}{\pi} \sqrt{-a}$ for $a \to -\infty$, leading to an accumulation of the $a_i$ for large negative values, $\overline{a_i} \simeq - (3 \pi/2)^{2/3}\,i^{2/3}$ for $i\to\infty$.\\

Remarkably, the KPZ problem and the linear statistics of the Airy point process are related. The exact solution for the generating function \eqref{G sw} of the KPZ equation with droplet initial condition \cite{SS10, ACQ11, DOT10, CLR10} can be written as an average over the Airy point process \cite{borodin2016moments} (a type of relations which appear in a more general setting, see \cite{borodinVertexASEPmoments})
\be \label{relKPZ} 
Q_t(\sigma) = \mathbb{E}_{\Ai} \left[\prod_{i=1}^{+\infty} \frac{1}{1 - \sigma e^{t^{1/3} \mathbf{a}_i} } \right]\;,
\ee 
which also appears in the context of fermions \cite{dean2015finite} (see below).\\

We exploit this connection in Section \ref{sec:appGUE} to study the linear statistics for the Airy point process,
namely averages over the Airy point process of the type $\mathbb{E}_{\Ai} [\exp(\sum_{i=1}^{+\infty} f(\mathbf{a}_i))]$.
Since explicit expressions are difficult to obtain for arbitrary function $f(x)$, we focus on the following two problems:

\begin{itemize} 

\item We calculate the following average in an expansion for small value of an arbitrary parameter $t$
\be
\mathbb{E}_{\Ai} \left[\exp\left( \sum_{i=1}^{+\infty} f(t^{1/3} \mathbf{a}_i)\right)\right]
\ee
for a large class of functions $f$, which allows us to obtain the cumulants of the scaled empirical measure $\mu(a t^{-1/3})$ of the Airy point process, in an expansion in small $t$, i.e. in an expansion from $a t^{-1/3} \ll -1$ the matching region with the bulk, and towards the edge. The second cumulant is given, to leading order in small $t$, by the formula 
\eqref{resGFF} where ${\sf H}$ is the integrated empirical measure \eqref{defH}. 
We show that it matches 
the result in the bulk of the GUE spectrum obtained from the Gaussian free field correspondence by Borodin and Ferrari \cite{BorodinFerrari,borodinclt}. Here we obtain systematic corrections to the Gaussian free field (higher cumulants, expansion towards the edge), see results in Section \ref{subsub:deeperGFF}. This also extends a classical result of Basor and Widom to a much higher order \cite{basor1999determinants}. 

\item

We calculate the following average in an expansion {\it for large $t$}
\begin{equation}
\mathbb{E}_{\Ai}\left[ \exp\left(- t \sum_{i=1}^{+\infty} \phi(t^{-2/3} \mathbf{a}_i)\right)\right] \sim \exp\left(- t^2 {\cal F}(\phi)\right)
\end{equation}
This is a large deviation result, with a rate function
${\cal F}(\phi)$ which we obtain explicitly for a class of functions $\phi$. It is an extension
of the large time large deviation formula for KPZ with droplet initial condition \eqref{Plarget}. 
The formula are given in Section \ref{sec:largedevAPP}.

\end{itemize} 

Both results, including the KPZ results for the droplet initial condition mentioned above, are obtained in a single framework by extending the calculation of the generating function \eqref{G sw} to 
\begin{equation}
	\label{G sw2}
	Q_{t,\beta}(\sigma) = \mathbb{E}_{\mathrm{Ai	}}\left[  \exp\left( \sum_{i=1}^\infty \beta g(\sigma e^{t^{1/3} \mathbf{a}_i})\right)\right] = \mathrm{Det}\left[1- (1- \rme^{\beta \hat g_{t,\sigma}}) K_\Ai \right] 
\end{equation}
where from \eqref{relKPZ} one sees that \eqref{G sw} is recovered for the choice $g(x)=g_{\rm KPZ}(x):=- \log(1-x)=\mathrm{Li}_1(x)$ and $\beta=1$. 
We have defined $\hat g_{t,\sigma}(a)=g(\sigma e^{t^{1/3} a})$.
The second identity in \eqref{G sw2} holds from well known properties of the Airy point process and generalises to any determinantal point process with associated kernel $K$ and any function $g$
(see Refs. \cite{johansson2005random,borodin2009determinantal,forrester2010log,borodin2016moments}
for reviews on determinantal point processes.

\subsubsection{Counting statistics of trapped fermions}

In Section \ref{fermion_large_dev} we consider $N$ non-interacting fermions in an harmonic trap.
Near the edge of the Fermi gas, $x_{\mathrm{edge}}$, the fermion density (of Wigner semi-circle mean shape)
vanishes. There are however mesoscopic
quantum fluctuations on distances of order $w_N=N^{-1/6} /\sqrt{2}$ near the edge, and thermal fluctuations are
important in the temperature regime $T= N^{1/3}/b$ where $b$ is the reduced inverse temperature
parameter \cite{dean2015finite,dean2016noninteracting}. Eq. \eqref{relKPZ} allows to
relate this finite $T$ fermion problem to the finite time KPZ solution with droplet initial conditions and to extend the KPZ results to the fermions \cite{dean2015finite,le2016exact}.

Here we study the PDF of $N(s)$, the number of fermions in the interval $\xi_i = b \frac{x_i-x_{\mathrm{edge}}}{w_N} \in [s,+\infty[$ as $s$ is varied. We first obtain its Laplace transform $\langle e^{- \lambda N(s) } \rangle$ in a high temperature expansion $b \ll1$, up to $\mathcal{O}(b^6)$, in Eqs. \eqref{Nexp1}, \eqref{Lifermions}. For $b \ll 1$ the typical position of the rightmost fermion is $\xi_{\rm typ}=\log( \frac{1}{b^{3/2} \sqrt{4 \pi}})$: its fluctuation are
Gumbel distributed to leading order, and here we obtain the systematic corrections to the Gumbel distribution, see Eq. \eqref{pmax}.
Accordingly, $N(s) = \mathcal{O}(1)$ in that region, and $N(s= \xi_{\rm typ} + \hat s)$ is Poisson distributed to leading order, with corrections 
\bea
&& \log \langle e^{- \lambda N(\xi_{\rm typ} + \hat s) } \rangle = - e^{- \hat s}  (1- e^{-\lambda}) 
+ \frac{1}{2} \sqrt{\frac{\pi}{2}}  e^{-2 \hat s}  (1- e^{-2 \lambda})  b^{3/2} + \mathcal{O}(b^3)
\eea
the higher orders are obtained in \eqref{Ntyp}.
Next we obtain explicit the small $b$ expansion formula for the cumulants $\langle  N(s)^p  \rangle^c$
in the region $s = \mathcal{O}(1)$ where $N(s)$ is large, see Eq. \eqref{expcumN}. Our high temperature results
extend those of \cite{le2016exact} to higher orders. \\

Finally, we study the low $T$ region, large $b$. There is an interesting large deviation regime
for the quantity ${\cal N}=N(s=z b^3)$ for fixed $z<0$. The typical value of ${\cal N}$
is given by the semi-circle estimate ${\cal N}_{\rm typ} = \frac{2}{3 \pi} (-z)^{3/2} b^3 \gg 1$, i.e. 
it is large. However ${\cal N}$ can fluctuate on the same scale $b^3$, and 
its PDF takes the large deviation form at large $b$
\bea 
P({\cal N}) \sim \exp \left( - b^6 F(\nu) \right) \quad , \quad \nu = {\cal N}/b^3
\eea 
where $F(0)=\Phi_-(z)$ in \eqref{Plarget} describes the large deviation (left tail) of the PDF of 
the position of the righmost fermion, see \eqref{sb2}. We obtain $F(\nu)$ as the Legendre transform of the rate function $\Phi(z,\tilde \lambda)$ associated to the large deviation form of the Laplace transform of the PDF of ${\cal N}$.
The calculation of $\Phi(z,\tilde \lambda)$ is an extension of the one of 
$\Phi_-(z)$ in \eqref{Plarget}, with $\Phi(z,\tilde \lambda)=\Phi_-(z)$ for $\tilde \lambda>-z$
and leads to formula \eqref{Philambda}. We obtain explicitly the leading terms for $F(\nu)$ 
in an expansion at large negative $z$
\bea \label{expandF0} 
F(\nu) = \frac{4}{15 \pi} (-z)^{5/2} F_0(\tilde \nu) - \frac{z^2}{2 \pi^2} F_1(\tilde \nu) 
+ \dots 
\eea 
with $F_0(\tilde \nu)=1 - \frac{5}{2} \tilde \nu + \frac{3}{2} \tilde \nu^{5/3}$ with 
$\tilde \nu={\cal N}/{\cal N}_{\rm typ}$, and the higher corrections $F_n(\tilde \nu)$
are discussed there. 

\subsection{Known exact solutions for the KPZ equation}
\label{recall}

Let us recall here some details about the solution of the KPZ equation for all times $t$ with
droplet and Brownian initial conditions which will be useful in the rest of the paper.

\begin{enumerate}
\item \textit{Droplet initial condition } \\
The moment generating function \eqref{G sw} of $\rme^H$ is given by the Fredholm determinant \cite{SS10, ACQ11, DOT10, CLR10}
\begin{equation}
\label{G[M] sw}
Q_{t}(\sigma)=\mathrm{Det}[I-M_{t,\sigma}]_{\mathbb{L}^2(\mathbb{R})}\;.
\end{equation}
The kernel of the integral operator $M_{t,\sigma}$, equal to
\begin{equation} \label{M sw} 
M_{t,\sigma}(u,u') = \frac{\sigma}{ \sigma - \rme^{- t^{1/3} u}} K_{\rm Ai}(u,u')\;,
\end{equation}
is the product of a Fermi factor and of the Airy kernel 
\begin{equation} \label{KAi} 
K_{\Ai}(u,u')=\int_{0}^{\infty} \! \rmd r \; \Ai(r+u) \Ai(r+u')\;.
\end{equation}
\item \textit{Brownian initial condition} \\
For the Brownian initial condition, the exact solution is written in terms of an additional random variable $\chi$ independent of $H$, with probability density $p(\chi) =  \rme^{-2 w \chi - \rme^{-\chi}} / \Gamma(2 w)$. The moment generating function of $\rme^{H(t)+\chi}$ is again a Fredholm determinant \cite{SasamotoStationary,SasamotoStationary2,BCFV},
\begin{equation} \label{QGamma} 
Q_{t}(t\sigma)=\mathrm{Det}\left[ I-M_{t,\sigma}^{\Gamma} \right]_{\mathbb{L}^2(\mathbb{R})}\;.
\end{equation}
The kernel of the integral operator $M_{t,\sigma}^{\Gamma}$, equal to
\begin{equation} \label{M stat} 
M_{t,\sigma}^{\Gamma}(u,u') = \frac{\sigma}{ \sigma - \rme^{- t^{1/3} u}} K_{\rm Ai, \Gamma}(u,u')\;,
\end{equation}
is the product of a Fermi factor and of the deformed Airy kernel
\begin{equation} \label{KAiGamma}
  K_{\rm Ai, \Gamma}(u,u') = \int_{0}^{+\infty} dr \, \Ai_\Gamma^\Gamma(r+u,t^{-\frac{1}{3}},w,w) \Ai_\Gamma^\Gamma(r+u',t^{-\frac{1}{3}},w,w)\;,
\end{equation}
where the deformed Airy function is equal to
\begin{equation} \label{aigamma} 
\Ai_\Gamma^\Gamma(a,b,c,d) := \frac{1}{2 \pi}\int\limits_{-\infty+i \epsilon}^{+\infty+i \epsilon}\rmd\eta \; \mathrm{exp}\Big(i \frac{\eta^3}{3}+ i a\eta\Big)\frac{\Gamma(i b\eta+d)}{\Gamma(-i b\eta+c)}\;.
\end{equation}
where $\Gamma$ is the  Gamma function and $\epsilon \in [0, \mathrm{Re}(d/b))$ due to the pole of the $\Gamma$ function.
\end{enumerate}
\begin{remark}
The operators on $\mathbb{L}^2(\mathbb{R})$ with kernels $K_{\Ai}$ and $K_{\Ai,\Gamma}$ can be written as 
\begin{equation}
K_{\Ai}={\sf Ai} \, P_0 \, {\sf Ai}, \qquad K_{\Ai,\Gamma}={\sf Ai}_\Gamma^\Gamma \, P_0 \, {\sf Ai}_\Gamma^{\Gamma}\;.
\end{equation}
The projector $P_0$ onto $\mathbb{R}^+$ has kernel $P_{0}(x,y)=\Theta(x)\delta(x-y)$ with $\Theta$ the Heaviside step function, and ${\sf Ai}$ and ${\sf Ai}_\Gamma^\Gamma$ are operators on $\mathbb{L}^2(\mathbb{R})$ with respective kernels
\begin{equation}
{\sf Ai}(x,y)=\Ai(x+y), \qquad {\sf Ai}_\Gamma^\Gamma(x,y)=\Ai_\Gamma^\Gamma(x+y,t^{-1/3},w,w)\;.
\end{equation}
\end{remark}

\end{section}

\section{Direct short time expansion for KPZ with droplet initial condition}
\label{section sw systematic}

In this section, we consider the doubly exponential generating function of the height $Q_{t}(\sigma)$ defined in \eqref{G sw} for the KPZ equation with sharp wedge initial condition. After some manipulations on the exact solution \eqref{G[M] sw}, we obtain the alternative expression \eqref{G'[L]}, from which the short time expansion of $\log Q_{t}(\sigma)$ can be performed in a systematic way. The expansion up to order $t^{3}$ is given in \eqref{G short time expansion}, \eqref{Likpz}. At the end of the section, we also consider the generating function of the height cumulants $F_{t}(\sigma)$ defined in \eqref{F sw}, and the PDF of the height $P(H,t)$.

\begin{subsection}{Doubly exponential generating function of the height $Q_{t,\beta}(\sigma)$}

As explained in the introduction it is advantageous for the short time limit to consider the slightly more general generating function $\log Q_{t,\beta}(\sigma)=q_{t,\beta}(\sigma)$ defined as
\begin{equation}
q_{t,\beta}(\sigma)=\log \mathbb{E}_{\mathrm{Ai	}}\left[  \exp\left(\beta \sum_{j=1}^\infty
g(\sigma e^{t^{1/3} \mathbf{a}_j}) \right)\right]\;,
\end{equation}
where $\lbrace \mathbf{a}_j \rbrace$ forms the Airy point process, $\gamma=t^{-1/3}$, $\beta$ is a scalar parameter which we will use below for bookkeeping purpose (and is set to $\beta=1$ in all applications). Here we consider general functions $g$ with $g(0)=0$ which is mandatory for convergence, another condition being $g(-\infty)<0$. The results for the short time expansion of the KPZ equation with droplet initial condition, for the generating function $Q_t(\sigma)$ defined in \eqref{G sw}, are recovered for the special choice
\be \label{gKPZ} 
g(x) = g_{\rm KPZ}(x) := - \log(1-x)= {\rm Li}_1(x)
\ee

The expectation value over the Airy point process has the Fredholm determinant expression
\begin{align}
\label{G[M] sw expanded}
q_{t,\beta}(\sigma)&=\log {\rm Det}\left[ I - (1- e^{\beta \hat g_{t,\sigma} }) K_{\Ai} \right]\\
&=-\sum_{m=1}^{\infty}\frac{1}{m}\int_{-\infty}^{\infty}\rmd u_{1}\ldots\rmd u_{m}
\Bigg(\prod_{j=1}^{m}\left[1-\exp\left(\beta g(\sigma e^{ u_j/\gamma})\right)\right]\,K_{\Ai}(u_{j},u_{j+1})\Bigg)\;.\nonumber
\end{align}

We recall that $\hat g_{t,\sigma}(a)=g(\sigma e^{t^{1/3} a})$. Here and below cyclicity on the variables $u_{j}$ is assumed, $u_{j+m}\equiv u_{j}$. It is convenient for the following to consider instead the derivative $q_{t,\beta}'(\sigma)$. This leads to a sum over $k=1,\ldots,m$ of terms where the factor $\exp\big(\beta g(\sigma e^{ u_j /\gamma})\big)$ is replaced by $\partial_{\sigma}\exp\big(\beta g(\sigma e^{ u_j/\gamma})\big)$. By cyclicity of the product over $j$, one can make the derivative act only on the factor with $k=1$ after renaming the variables $u_{j}$, and the factor $1/m$ of the Fredholm expansion cancels. Using then 
\begin{equation}
\partial_{\sigma}\exp\left(\beta g(\sigma e^{ u_1/\gamma})\right)=\frac{\gamma}{\sigma}\partial_{u_{1}}\exp\left(\beta g(\sigma e^{u_1/\gamma })\right)
\end{equation}
one finds after partial integration on $u_{1}$
\begin{equation}
q_{t,\beta}'(\sigma)=\frac{\gamma}{\sigma}\sum_{m=1}^{\infty}\int_{-\infty}^{\infty}\rmd u_{1}\ldots\rmd u_{m}
\prod_{j=1}^{m}\left[1-\exp\left(\beta g(\sigma e^{ u_j/\gamma})\right)\right]
\partial_{u_{1}}\Big(\prod_{j=1}^{m}K_{\Ai}(u_{j},u_{j+1})\Big)\;.
\end{equation}
The derivative over $u_{1}$ gives a sum of two terms, containing respectively a factor $\partial_{u_{1}}K_{\Ai}(u_{1},u_{2})$ and a factor $\partial_{u_{1}}K_{\Ai}(u_{m},u_{1})$. Shifting the indices of the variables $u_{j}$ by $1$ in the former, the integrand can be factorized as $(\ldots)\times(\partial_{u_{m}}K_{\Ai}(u_{m},u_{1})+\partial_{u_{1}}K_{\Ai}(u_{m},u_{1}))$. Using $\left(\partial_{u}+\partial_{v}\right)K_{\Ai}(u,v)=-\Ai(u)\Ai(v)$ then leads to
\begin{align}
q_{t,\beta}'(\sigma)=-\frac{\gamma}{\sigma}\sum_{m=1}^{\infty}\int_{-\infty}^{\infty}\rmd u_{1}\ldots\rmd u_{m}&\prod_{j=1}^{m}\Big[1-\exp\big(\beta g(\sigma e^{u_j/\gamma})\big)\Big]\nonumber\\
&\times\Ai(u_{1})\Bigg(\prod_{j=1}^{m-1}K_{\Ai}(u_{j},u_{j+1})\Bigg)\Ai(u_{m})\;.
\end{align}
Writing the Airy kernels explicitly as \eqref{KAi} and expanding the pre-function as 
\begin{equation}
1-\exp\big(\beta g(\sigma e^{ u/\gamma})\big)=-\sum_{n=1}^{\infty}a_{n,\beta}\,\sigma^{n}\rme^{n  u/\gamma}
\end{equation}
the integration over the variables $u_{j}$ can be performed using the Airy propagator
\begin{equation}
\int_{-\infty}^{\infty}\rmd u\,\rme^{nu}\Ai(u+a)\Ai(u+b)=\frac{\rme^{\frac{n^{3}}{12}-\frac{(a+b)n}{2}-\frac{(a-b)^{2}}{4n}}}{\sqrt{4\pi n}}
\end{equation}
After the change of variables $z_{j}\to z_{j}/\sqrt{\gamma}$, one finds (we recall that $\gamma=t^{-1/3}$)
\begin{align}
\label{g' tmp}
q_{t,\beta}'(\sigma)=-\frac{1}{\sigma\sqrt{t}}\sum_{m=1}^{\infty}&(-1)^{m}\sum_{n_{1},\ldots,n_{m}=1}^{\infty}\Bigg(\prod_{j=1}^{m}\frac{a_{n_j,\beta}\,\sigma^{n_{j}}\,\rme^{\frac{t}{12}\,n_{j}^{3}}}{\sqrt{4\pi n_{j}}}\Bigg)\\
&\int_{0}^{\infty}\rmd z_{1}\,\ldots\,\rmd z_{m-1}\,\Bigg(\prod_{j=1}^{m}\rme^{-\frac{(z_{j-1}+z_{j})n_{j}\sqrt{t}}{2}-\frac{(z_{j}-z_{j-1})^{2}}{4n_{j}}}\Bigg)_{z_{0}=z_{m}=0}\;.\nonumber
\end{align}
We note that for $\beta=1$ and $g(x)=\mathrm{Li}_1(x)$, \eqref{g' tmp} is essentially identical to some intermediate expression written in the derivation \cite{DOT10,CLR10} of \eqref{G[M] sw} by the replica method: between \eqref{G[M] sw expanded} and \eqref{g' tmp}, we have essentially just done part of the replica calculation backwards.\\

The change of variables $z_{j}\to z_{j}-\sqrt{t}(\sum_{\ell=1}^{j}n_{\ell})(\sum_{\ell=j+1}^{m}n_{\ell})$ cancels the linear terms in the $z_{j}$ inside the exponentials. Using the identity
\begin{equation}
\label{sum_integers}
\sum_{j=1}^{m}\left(\frac{n_{j}^{3}}{12}+\frac{\left((\sum_{\ell=1}^{j}n_{\ell})(\sum_{\ell=j+1}^{m}n_{\ell})-(\sum_{\ell=1}^{j-1}n_{\ell})(\sum_{\ell=j}^{m}n_{\ell})\right)^{2}}{4n_{j}}\right)=\frac{1}{12}\Big(\sum_{j=1}^{m}n_{j}\Big)^{3}
\end{equation}
valid for any $m\in \mathbb{N}^*$ and $\lbrace n_j \rbrace _{j\in [1,m]}$ a set of integers,
it leads to
\begin{align}
q_{t,\beta}'(\sigma)=-\frac{1}{\sigma\sqrt{t}}\sum_{m=1}^{\infty}&(-1)^{m}\sum_{n_{1},\ldots,n_{m}=1}^{\infty}\rme^{\frac{t}{12}(\sum_{j=1}^{m}n_{j})^{3}}\Big(\prod_{j=1}^{m}\frac{a_{n_j,\beta}\,\sigma^{n_{j}}}{\sqrt{4\pi n_{j}}}\Big)\\
&\left(\prod_{j=1}^{m-1}\int_{\sqrt{t}(\sum_{\ell=1}^{j}n_{\ell})(\sum_{\ell=j+1}^{m}n_{\ell})}^{\infty}\rmd z_{j}\right)\left(\prod_{j=1}^{m}\rme^{-\frac{(z_{j}-z_{j-1})^{2}}{4n_{j}}}\right)_{z_{0}=z_{m}=0}\;.\nonumber
\end{align}
The factor $\rme^{\frac{t}{12}(\sum_{j=1}^{m}n_{j})^{3}}$ can be conveniently written as the action of the operator $\rme^{\frac{t}{12}(\sigma\partial_{\sigma})^{3}}$ on $\sigma^{\sum_{j=1}^{m}n_{j}}$. One has
\begin{align}
q_{t,\beta}'(\sigma)=-\frac{1}{\sigma\sqrt{t}}\,& e^{\frac{t}{12}(\sigma\partial_{\sigma})^{3}}\sum_{m=1}^{\infty}(-1)^{m}\sum_{n_{1},\ldots,n_{m}=1}^{\infty}\Big(\prod_{j=1}^{m}\frac{a_{n_j,\beta}\,\sigma^{n_{j}}}{\sqrt{4\pi n_{j}}}\Big)\\
&\left(\prod_{j=1}^{m-1}\int_{\sqrt{t}(\sum_{\ell=1}^{j}n_{\ell})(\sum_{\ell=j+1}^{m}n_{\ell})}^{\infty}\rmd z_{j}\right)\left(\prod_{j=1}^{m}\rme^{-\frac{(z_{j}-z_{j-1})^{2}}{4n_{j}}}\right)_{z_{0}=z_{m}=0}\;.\nonumber
\end{align}
For any function $f$ such that $F(\varepsilon)=\int_{\varepsilon}^{\infty}\rmd z\,f(z)$ converges, one can write formally
\begin{equation}
F(\varepsilon)=e^{\varepsilon\partial_{z}}F(z)\mid_{z=0}=F(0)+(\rme^{\varepsilon\partial_{z}}-1)F(z)\mid_{z=0}=\int_{0}^{\infty}\rmd z\,f(z)+\frac{1-e^{\varepsilon\partial_{z}}}{\partial_{z}}\,f(z)\mid_{z=0}\;.
\end{equation}
This identity can be used to perform the short time expansion of $q_{t,\beta}'(\sigma)$. Each integral over $z_{j}$ gives two terms, corresponding respectively to the integral from $0$ to $+\infty$ and to the action of the corresponding operator $\frac{\rme^{\varepsilon\partial_{z_{j}}}-1}{\partial_{z_{j}}}$. The $2^{m-1}$ terms can conveniently be labeled by the set $S\subset[\![1,m-1]\!]$ of $j$'s such that the operator $\frac{\rme^{\varepsilon\partial_{z_{j}}}-1}{\partial_{z_{j}}}$ is chosen for $z_{j}$. This leads to
\begin{align}
q_{t,\beta}'(\sigma)=&-\frac{1}{\sigma\sqrt{t}}\,\rme^{\frac{t}{12}(\sigma\partial_{\sigma})^{3}}\sum_{m=1}^{\infty}(-1)^{m}\sum_{n_{1},\ldots,n_{m}=1}^{\infty}\left(\prod_{j=1}^{m}\frac{a_{n_j,\beta}\,\sigma^{n_{j}}}{\sqrt{4\pi n_{j}}}\right)\sum_{S\subset[\![1,m-1]\!]}\\
&\left(\prod_{j\in S}\frac{1-\rme^{\sqrt{t}(\sum_{\ell=1}^{j}n_{\ell})(\sum_{\ell=j+1}^{m}n_{\ell})\partial_{z_{j}}}}{\partial_{z_{j}}}\right)
\left(\prod_{\substack{j=1\\j\not\in S}}^{m-1}\int_{0}^{\infty}\rmd z_{j}\right)
\left(\prod_{j=1}^{m}\rme^{-\frac{(z_{j}-z_{j-1})^{2}}{4n_{j}}}\right)_{\substack{z_{0}=z_{m}=0\\z_{j}=0,\;j\in S}}\;.\nonumber
\end{align}
The various $n_{j}$ are still coupled. In order to disentangle them, we replace $\sigma^{n_{j}}$ by $\sigma_{j}^{n_{j}}$ and interpret the other $n_{j}$ as $\sigma_{j}\partial_{\sigma_{j}}$. One finds
\begin{align}
q_{t,\beta}'(\sigma)=&-\frac{1}{\sigma\sqrt{t}}\,\rme^{\frac{t}{12}(\sigma\partial_{\sigma})^{3}}\Bigg(\sum_{m=1}^{\infty}(-1)^{m}\sum_{S\subset[\![1,m-1]\!]}\Big(\prod_{j\in S}\frac{1-\rme^{\sqrt{t}(\sum_{\ell=1}^{j}\sigma_{\ell}\partial_{\sigma_{\ell}})(\sum_{\ell=j+1}^{m}\sigma_{\ell}\partial_{\sigma_{\ell}})\partial_{z_{j}}}}{\partial_{z_{j}}}\Big)\nonumber\\
&\times\Big(\prod_{\substack{j=1\\j\not\in S}}^{m-1}\int_{0}^{\infty}\rmd z_{j}\Big)
\Big(\prod_{j=1}^{m}\sum_{n=1}^{\infty}\frac{a_{n,\beta}\,\sigma_{j}^{n}}{\sqrt{4\pi n}}\,\rme^{-\frac{(z_{j}-z_{j-1})^{2}}{4n}}\Big)_{\substack{z_{0}=z_{m}=0\\z_{j}=0,\;j\in S}}\Bigg)_{\sigma_{1}=\ldots=\sigma_{m}=\sigma}\;.
\end{align}
We now introduce the number $p\in[\![0,m-1]\!]$ of elements of $S$ and write $S=\{k_{1},\ldots,k_{p}\}$ with $1\leq k_{1}<\ldots<k_{p}\leq m-1$. Setting $k_{p+1}=m$ and $k_{0}=0$, the summation $\sum_{m=1}^{\infty}\sum_{S\subset[\![1,m-1]\!]}$ is equivalent to $\sum_{p=0}^{\infty}\sum_{0=k_{0}<\ldots<k_{p+1}}$, and one has
\begin{align}
q_{t,\beta}'(\sigma)=&-\frac{1}{\sigma\sqrt{t}}\,\rme^{\frac{t}{12}(\sigma\partial_{\sigma})^{3}}\Bigg(\sum_{p=0}^{\infty}\sum_{0=k_{0}<\ldots<k_{p+1}}(-1)^{k_{p+1}} \Big(\prod_{j=1}^{p}\frac{1-\rme^{\sqrt{t}(\sum_{\ell=1}^{k_{j}}\sigma_{\ell}\partial_{\sigma_{\ell}})(\sum_{\ell=k_{j}+1}^{k_{p+1}}\sigma_{\ell}\partial_{\sigma_{\ell}})\partial_{z_{k_{j}}}}}{\partial_{z_{k_{j}}}}\Big)\nonumber\\
&\times\Big(\prod_{j=0}^{p}\int_{0}^{\infty}\rmd z_{k_{j}+1}\ldots\rmd z_{k_{j+1}-1}\prod_{\ell=k_{j}+1}^{k_{j+1}}\sum_{n=1}^{\infty}\frac{a_{n,\beta}\,\sigma_{\ell}^{n}}{\sqrt{4\pi n}}\,\rme^{-\frac{(z_{\ell}-z_{\ell-1})^{2}}{4n}}\Big)\Bigg)_{\substack{z_{k_{0}}=\ldots=z_{k_{p+1}}=0\\\sigma_{1}=\ldots=\sigma_{k_{p+1}}=\sigma}}
\end{align}
The multiple integral over $z_{k_{j}+1},\ldots,z_{k_{j+1}-1}$ is very similar for each $j=0,\ldots,p$. We introduce
\begin{equation}
\mathcal{L}(\sigma_{1},\ldots,\sigma_{k};a,b)=\int_{0}^{\infty}\rmd z_{1}\ldots\rmd z_{k-1}\prod_{\ell=1}^{k}\Big(\sum_{n=1}^{\infty}\frac{a_{n,\beta}\,\sigma_{\ell}^{n}}{\sqrt{4\pi n}}\,\rme^{-\frac{(z_{\ell}-z_{\ell-1})^{2}}{4n}}\Big)_{\substack{z_{0}=a\\z_{k}=b}}\;.
\end{equation}
After renaming the $z_{k_{j}}$ to $z_{j}$, the generating function becomes
\begin{align}
q_{t,\beta}'(\sigma)=&-\frac{1}{\sigma\sqrt{t}}\,\rme^{\frac{t}{12}(\sigma\partial_{\sigma})^{3}}\Bigg(\sum_{p=0}^{\infty}\sum_{0=k_{0}<\ldots<k_{p+1}}(-1)^{k_{p+1}}\Big(\prod_{j=1}^{p}\frac{1-\rme^{\sqrt{t}(\sum_{\ell=1}^{k_{j}}\sigma_{\ell}\partial_{\sigma_{\ell}})(\sum_{\ell=k_{j}+1}^{k_{p+1}}\sigma_{\ell}\partial_{\sigma_{\ell}})\partial_{z_{j}}}}{\partial_{z_{j}}}\Big)\nonumber\\
&\times\Big(\prod_{j=0}^{p}\mathcal{L}(\sigma_{k_{j}+1},\ldots,\sigma_{k_{j+1}};z_{j},z_{j+1})\Big)\Bigg)_{\substack{z_{0}=\ldots=z_{p+1}=0\\\sigma_{1}=\ldots=\sigma_{k_{p+1}}=\sigma}}\;.\nonumber
\end{align}
Writing $(\sum_{\ell=1}^{k_{j}}\sigma_{\ell}\partial_{\sigma_{\ell}})(\sum_{\ell=k_{j}+1}^{k_{p+1}}\sigma_{\ell}\partial_{\sigma_{\ell}})=(\sum_{i=1}^{j}\sum_{\ell=k_{i-1}+1}^{k_{i}}\sigma_{\ell}\partial_{\sigma_{\ell}})(\sum_{i=j}^{p}\sum_{\ell=k_{i}+1}^{k_{i+1}}\sigma_{\ell}\partial_{\sigma_{\ell}})$, we observe that the action of any power of the operator $\sum_{\ell=k_{i}+1}^{k_{i+1}}\sigma_{\ell}\partial_{\sigma_{\ell}}$ on an arbitrary function $\mathcal{L}(\sigma_{k_{i}+1},\ldots,\sigma_{k_{i+1}})$ is equal, after taking all the $\sigma_{\ell}$ equal to $\sigma$, to the action of the same power of the operator $\sigma\partial_{\sigma}$ on $\mathcal{L}(\sigma,\ldots,\sigma)$. Thus, all the parameters $\sigma$ inside a given function $\mathcal{L}$ in $q_{t,\beta}'(\sigma)$ can be set equal from the beginning. Replacing $\sigma_{k_{j}+1},\ldots,\sigma_{k_{j+1}}$ by $\sigma_{j}$ and introducing the function $\mathcal{L}_{k}(\sigma,a,b)=\mathcal{L}(\sigma_{1}=\sigma,\ldots,\sigma_{k}=\sigma;a,b)$, we obtain
\begin{align}
q_{t,\beta}'&(\sigma)=-\frac{1}{\sigma\sqrt{t}}\,\rme^{\frac{t}{12}(\sigma\partial_{\sigma})^{3}}\Bigg(\sum_{p=0}^{\infty}\sum_{0=k_{0}<\ldots<k_{p+1}}(-1)^{k_{p+1}}\\
&\Big(\prod_{j=1}^{p}\frac{1-\rme^{\sqrt{t}(\sum_{i=1}^{j}\sigma_{i-1}\partial_{\sigma_{i-1}})(\sum_{i=j}^{p}\sigma_{i}\partial_{\sigma_{i}})\partial_{z_{j}}}}{\partial_{z_{j}}}\Big) \Big(\prod_{j=0}^{p}\mathcal{L}_{k_{j+1}-k_{j}}(\sigma_{j},z_{j},z_{j+1})\Big)\Bigg)_{\substack{z_{0}=\ldots=z_{p+1}=0\\\sigma_{0}=\ldots=\sigma_{p}=\sigma}}\;.\nonumber
\end{align}
Performing the change of variables $k_{1}\to k_{1}+k_{0}$ then $k_{2}\to k_{2}+k_{1}+k_{0}$, and so on until $k_{p+1}\to k_{p+1}+\ldots+k_{0}$ decouples the integers $k_{j}$. Introducing $L_{\beta}(\sigma,a,b)=\sum_{k=1}^{\infty}(-1)^{k-1}\mathcal{L}_{k}(\sigma,a,b)$, or more explicitly
\begin{equation}
\label{L[int]}
L_{\beta}(\sigma,a,b)=\sum_{k=1}^{\infty}(-1)^{k-1}\int_{0}^{\infty}\rmd z_{1}\ldots\rmd z_{k-1}\,\prod_{\ell=1}^{k}\bigg(\sum_{n=1}^{\infty}\frac{a_{n,\beta}\,\sigma^{n}}{\sqrt{4\pi n}}\,\rme^{-\frac{(z_{\ell}-z_{\ell-1})^{2}}{4n}}\bigg)_{\substack{z_{0}=a\\z_{k}=b}}\;,
\end{equation}
we finally arrive at
\begin{align}
\label{G'[L]}
q_{t,\beta}'(\sigma)=\frac{1}{\sigma\sqrt{t}}\,\rme^{\frac{t}{12}(\sigma\partial_{\sigma})^{3}}\Bigg(\sum_{p=0}^{\infty}\Big(\prod_{j=1}^{p}&\frac{\rme^{\sqrt{t}\,(\sum_{i=0}^{j-1}\sigma_{i}\partial_{\sigma_{i}})(\sum_{i=j}^{p}\sigma_{i}\partial_{\sigma_{i}})\partial_{z_{j}}}-1}{\partial_{z_{j}}}\Big)\nonumber\\
&\times\prod_{j=0}^{p}L_{\beta}(\sigma_{j},z_{j},z_{j+1})\Bigg)_{\substack{z_{0}=\ldots=z_{p+1}=0\\\sigma_{0}=\ldots=\sigma_{p}=\sigma}}\;.
\end{align}

\end{subsection}

\begin{subsection}{Algorithm for the short time expansion of $Q_{t,\beta}(\sigma)$}
In this section, we explain how the short time expansion of \eqref{G'[L]} can be performed systematically to arbitrary order in $t$. This leads to the short time expansion \eqref{G short time expansion} for $\log Q_{t}(\sigma)$. The algorithm giving the expansion is based on the following identities verified by the functions $L_{\beta}(\sigma,a,b)$ defined in \eqref{L[int]}:
	\begin{align}
		\label{L[Li]}
		& L_{\beta}(\sigma,0,0)=\frac{\beta }{\pi }\sigma \partial_\sigma \int_{0}^{\infty}\mathrm{d}x\,\sqrt{x}\,g(\sigma e^{-x})\\
		\label{L sym}
		& L_{\beta}(\sigma,a,b)=L_{\beta}(\sigma,b,a)\\
		\label{dadbL}
		& (\partial_{a}+\partial_{b})L_{\beta}(\sigma,a,b)=-L_{\beta}(\sigma,a,0)L_{\beta}(\sigma,0,b)\\
		\label{sdsdaL}
		& \sigma\partial_{\sigma}\partial_{a}L_{\beta}(\sigma,a,b)=\frac{b-a}{2}\,L_{\beta}(\sigma,a,b)-L_{\beta}(\sigma,a,0)\,\sigma\partial_{\sigma}L_{\beta}(\sigma,0,b)\;.
	\end{align}
	These identities are proved in Appendix \ref{Appendix identities L}. From \eqref{G'[L]}, each term in the short time expansion of $\sigma g_{t,\beta}'(\sigma)$ is a polynomial in the variables $\mathfrak{L}_{i,j,k}=(\sigma\partial_{\sigma})^{i}L_{\beta}^{(0,j,k)}(\sigma,0,0)$, $(i,j,k)\geq0$. In terms of the $\mathfrak{L}_{i,j,k}$, the identities \eqref{L[Li]} - \eqref{sdsdaL} imply
		\begin{align}
			\label{L[Li] 2}
			& \mathfrak{L}_{i,0,0}=\frac{\beta }{\pi }(\sigma \partial_\sigma)^{i+1} \int_{0}^{\infty}\mathrm{d}x\,\sqrt{x}\,g(\sigma e^{-x})=\frac{\beta}{2\pi }(\sigma \partial_\sigma)^{i}\int_{-\infty}^{\infty}\mathrm{d}p\, g(\sigma e^{-p^2})\\
			\label{L sym 2}
			& \mathfrak{L}_{i,j,k}=\mathfrak{L}_{i,k,j}\\
			\label{dadbL 2}
			& \mathfrak{L}_{i,j+1,k}+\mathfrak{L}_{i,j,k+1}=-\sum_{\ell=0}^{i}{i\choose\ell}\mathfrak{L}_{\ell,j,0}\mathfrak{L}_{i-\ell,0,k}\\
			\label{sdsdaL 2}
			& \mathfrak{L}_{i+1,j+1,0}=-\frac{j}{2}\,\mathfrak{L}_{i,j-1,0}-\sum_{\ell=0}^{i}{i\choose\ell}\mathfrak{L}_{\ell,j,0}\mathfrak{L}_{i+1-\ell,0,0}\;.
		\end{align}
		At each order in $t$ in \eqref{G'[L]}, the index $k$ of the variables $\mathfrak{L}_{i,j,k}$ can be set to zero everywhere after applying recursively \eqref{dadbL 2}. Then, using \eqref{sdsdaL 2}, the index $j$ can be reduced as well, as long as $i\geq1$. Because of this constraint on $i$, it is not guaranteed that \eqref{sdsdaL 2} will be sufficient in order to make $j=0$ everywhere. In practice, we observed that at least up to order $t^{3}$, using repeatedly \eqref{sdsdaL 2} does eliminate all variables $\mathfrak{L}_{i,j,k}$ with either $j>0$ or $k>0$: only variables of the form $\mathfrak{L}_{i,0,0}$ remain. For the first orders in $t$, we find
		\begin{align}
			q_{t,\beta}'&(\sigma)
			=\frac{1}{\sigma\sqrt{t}}\,\rme^{\frac{t}{12}(\sigma\partial_{\sigma})^{3}}\bigg(
			\mathfrak{L}_{0}
			+\mathfrak{L}_{1}^2\,\sqrt{t}
			+2\,\mathfrak{L}_{1}^{2}\mathfrak{L}_{2}\,t
			+\Big(\frac{4}{3}\,\mathfrak{L}_{1}^{3}\mathfrak{L}_{3}+4\,\mathfrak{L}_{2}^2 \mathfrak{L}_{1}^2-\frac{1}{6}\,\mathfrak{L}_{2}\mathfrak{L}_{3}\Big)\,t^{3/2}\nonumber\\
			&+\Big(\frac{2}{3}\,\mathfrak{L}_{1}^{4}\mathfrak{L}_{4}+8\,\mathfrak{L}_{1}^{3}\mathfrak{L}_{2}\mathfrak{L}_{3}+8\,\mathfrak{L}_{2}^{3}\mathfrak{L}_{1}^2-\frac{1}{3}\,\mathfrak{L}_{1}\mathfrak{L}_{3}^{2}-\frac{1}{3}\,\mathfrak{L}_{1}\mathfrak{L}_{2}\mathfrak{L}_{4}-\mathfrak{L}_{2}^{2}\mathfrak{L}_{3}\Big)\,t^{2}
			+\ldots\bigg)
		\end{align}
		where $\mathfrak{L}_{i}=\mathfrak{L}_{i,0,0}$ has from \eqref{L[Li] 2} the explicit expression
		\begin{equation} \label{Liexplicit} 
\mathfrak{L}_{i}\equiv\mathfrak{L}_{i}(\sigma)=\frac{\beta }{\pi }(\sigma \partial_\sigma)^{i+1} \int_{0}^{+\infty}\mathrm{d}x\sqrt{x} g(\sigma e^{-x})=\beta (\sigma \partial_\sigma)^{i}\int_{-\infty}^{+\infty}\frac{\mathrm{d}p}{2\pi}\, g(\sigma e^{-p^2})
		\end{equation}
Computing explicitly the action of the operator $\rme^{\frac{t}{12}(\sigma\partial_{\sigma})^{3}}$ and integrating with respect to $\sigma$, we observe empirically that the integral at each order can be performed explicitly, except the one at order $t^{0}$. The term at order $t^{0}$ is given by
\begin{equation}
\label{g_exp_lhs1}
 \int_0^\sigma \frac{\mathrm{d}u}{u}\,\mathfrak{L}_{1}(u)^{2}=\frac{\beta^2}{4\pi^2} \int_0^\sigma \mathrm{d}u \, u \left[ \partial_u \int_{-\infty}^{+\infty}\mathrm{d}p\, g(u e^{-p^2})\right]^2 
\end{equation}
We finally obtain our main result for the expansion of $q_{t,\beta}(\sigma)$ defined in \eqref{G[M] sw expanded} for an arbitrary function $g(x)$
		\begin{equation}
		\begin{split}
			\label{G short time expansion}
			&q_{t,\beta}(\sigma)=\frac{\beta}{\pi \sqrt{t}}\int_{0}^{\infty}\mathrm{d}x\,\sqrt{x}\,g(\sigma e^{-x})
			+ \frac{\beta^2}{4\pi^2} \int_0^\sigma \mathrm{d}u \, u \left[ \partial_u \int_{-\infty}^{+\infty}\mathrm{d}p\, g(u e^{-p^2})\right]^2 \\
			&+\Big(\frac{2\mathfrak{L}_{1}^{3}}{3}+\frac{\mathfrak{L}_{2}}{12}\Big)\sqrt{t}
			+\Big(\frac{4}{3}\,\mathfrak{L}_{1}^{3}\mathfrak{L}_{2}+\frac{\mathfrak{L}_{2}^{2}}{12}+\frac{\mathfrak{L}_{1}\mathfrak{L}_{3}}{6}\Big)t\\
			&+\Big(\frac{8}{3}\,\mathfrak{L}_{1}^{3}\mathfrak{L}_{2}^{2}+\frac{\mathfrak{L}_{2}^{3}}{9}+\frac{2}{3}\,\mathfrak{L}_{1}^{4}\mathfrak{L}_{3}+\frac{2}{3}\,\mathfrak{L}_{1}\mathfrak{L}_{2}\mathfrak{L}_{3}+\frac{\mathfrak{L}_{1}^{2}\mathfrak{L}_{4}}{6}+\frac{\mathfrak{L}_{5}}{288}\Big)t^{3/2}\\
			&+\Big(\frac{16}{3}\,\mathfrak{L}_{1}^{3}\mathfrak{L}_{2}^{3}+\frac{\mathfrak{L}_{2}^{4}}{6}+4\mathfrak{L}_{1}^{4}\mathfrak{L}_{2}\mathfrak{L}_{3}+2\mathfrak{L}_{1}\mathfrak{L}_{2}^{2}\mathfrak{L}_{3}+\frac{2}{3}\,\mathfrak{L}_{1}^{2}\mathfrak{L}_{3}^{2}+\frac{4}{15}\,\mathfrak{L}_{1}^{5}\mathfrak{L}_{4}\\
			&\hspace{30mm}+\mathfrak{L}_{1}^{2}\mathfrak{L}_{2}\mathfrak{L}_{4}+\frac{29\mathfrak{L}_{3}\mathfrak{L}_{4}}{720}+\frac{\mathfrak{L}_{1}^{3}\mathfrak{L}_{5}}{9}+\frac{\mathfrak{L}_{2}\mathfrak{L}_{5}}{48}+\frac{\mathfrak{L}_{1}\mathfrak{L}_{6}}{144}\Big)t^{2}\\
			&+\Big(\frac{32}{3}\,\mathfrak{L}_{1}^{3}\mathfrak{L}_{2}^{4}+\frac{4\mathfrak{L}_{2}^{5}}{15}+16\mathfrak{L}_{1}^{4}\mathfrak{L}_{2}^{2}\mathfrak{L}_{3}+\frac{16}{3}\,\mathfrak{L}_{1}\mathfrak{L}_{2}^{3}\mathfrak{L}_{3}+\frac{8}{5}\,\mathfrak{L}_{1}^{5}\mathfrak{L}_{3}^{2}+\frac{16}{3}\,\mathfrak{L}_{1}^{2}\mathfrak{L}_{2}\mathfrak{L}_{3}^{2}+\frac{7\mathfrak{L}_{3}^{3}}{90}\\
			&+\frac{32}{15}\,\mathfrak{L}_{1}^{5}\mathfrak{L}_{2}\mathfrak{L}_{4}+4\mathfrak{L}_{1}^{2}\mathfrak{L}_{2}^{2}\mathfrak{L}_{4}+\frac{14}{9}\,\mathfrak{L}_{1}^{3}\mathfrak{L}_{3}\mathfrak{L}_{4}+\frac{29}{90}\,\mathfrak{L}_{2}\mathfrak{L}_{3}\mathfrak{L}_{4}+\frac{29}{360}\,\mathfrak{L}_{1}\mathfrak{L}_{4}^{2}+\frac{4}{45}\,\mathfrak{L}_{1}^{6}\mathfrak{L}_{5}\\
			&+\frac{8}{9}\,\mathfrak{L}_{1}^{3}\mathfrak{L}_{2}\mathfrak{L}_{5}+\frac{\mathfrak{L}_{2}^{2}\mathfrak{L}_{5}}{12}+\frac{11}{90}\,\mathfrak{L}_{1}\mathfrak{L}_{3}\mathfrak{L}_{5}+\frac{\mathfrak{L}_{1}^{4}\mathfrak{L}_{6}}{18}+\frac{\mathfrak{L}_{1}\mathfrak{L}_{2}\mathfrak{L}_{6}}{18}+\frac{\mathfrak{L}_{1}^{2}\mathfrak{L}_{7}}{144}+\frac{\mathfrak{L}_{8}}{10368}\Big)t^{5/2}\\
			&+\Big(\frac{64}{3}\,\mathfrak{L}_{1}^{3}\mathfrak{L}_{2}^{5}+\frac{4\mathfrak{L}_{2}^{6}}{9}+\frac{160}{3}\,\mathfrak{L}_{1}^{4}\mathfrak{L}_{2}^{3}\mathfrak{L}_{3}+\frac{40}{3}\,\mathfrak{L}_{1}\mathfrak{L}_{2}^{4}\mathfrak{L}_{3}+16\mathfrak{L}_{1}^{5}\mathfrak{L}_{2}\mathfrak{L}_{3}^{2}+\frac{80}{3}\,\mathfrak{L}_{1}^{2}\mathfrak{L}_{2}^{2}\mathfrak{L}_{3}^{2}\\
			&+\frac{32}{9}\,\mathfrak{L}_{1}^{3}\mathfrak{L}_{3}^{3}+\frac{7}{9}\,\mathfrak{L}_{2}\mathfrak{L}_{3}^{3}+\frac{32}{3}\,\mathfrak{L}_{1}^{5}\mathfrak{L}_{2}^{2}\mathfrak{L}_{4}+\frac{40}{3}\,\mathfrak{L}_{1}^{2}\mathfrak{L}_{2}^{3}\mathfrak{L}_{4}+\frac{16}{9}\,\mathfrak{L}_{1}^{6}\mathfrak{L}_{3}\mathfrak{L}_{4}+\frac{140}{9}\,\mathfrak{L}_{1}^{3}\mathfrak{L}_{2}\mathfrak{L}_{3}\mathfrak{L}_{4}\\
			&+\frac{29}{18}\,\mathfrak{L}_{2}^{2}\mathfrak{L}_{3}\mathfrak{L}_{4}+\frac{10}{9}\,\mathfrak{L}_{1}\mathfrak{L}_{3}^{2}\mathfrak{L}_{4}+\frac{7}{9}\,\mathfrak{L}_{1}^{4}\mathfrak{L}_{4}^{2}+\frac{29}{36}\,\mathfrak{L}_{1}\mathfrak{L}_{2}\mathfrak{L}_{4}^{2}+\frac{8}{9}\,\mathfrak{L}_{1}^{6}\mathfrak{L}_{2}\mathfrak{L}_{5}+\frac{40}{9}\,\mathfrak{L}_{1}^{3}\mathfrak{L}_{2}^{2}\mathfrak{L}_{5}\\
			&+\frac{5}{18}\,\mathfrak{L}_{2}^{3}\mathfrak{L}_{5}+\frac{11}{9}\,\mathfrak{L}_{1}^{4}\mathfrak{L}_{3}\mathfrak{L}_{5}+\frac{11}{9}\,\mathfrak{L}_{1}\mathfrak{L}_{2}\mathfrak{L}_{3}\mathfrak{L}_{5}+\frac{17}{60}\,\mathfrak{L}_{1}^{2}\mathfrak{L}_{4}\mathfrak{L}_{5}+\frac{607\mathfrak{L}_{5}^{2}}{181440}+\frac{8}{315}\,\mathfrak{L}_{1}^{7}\mathfrak{L}_{6}\\
			&+\frac{5}{9}\,\mathfrak{L}_{1}^{4}\mathfrak{L}_{2}\mathfrak{L}_{6}+\frac{5}{18}\,\mathfrak{L}_{1}\mathfrak{L}_{2}^{2}\mathfrak{L}_{6}+\frac{8}{45}\,\mathfrak{L}_{1}^{2}\mathfrak{L}_{3}\mathfrak{L}_{6}+\frac{503\mathfrak{L}_{4}\mathfrak{L}_{6}}{90720}+\frac{\mathfrak{L}_{1}^{5}\mathfrak{L}_{7}}{45}+\frac{5}{72}\,\mathfrak{L}_{1}^{2}\mathfrak{L}_{2}\mathfrak{L}_{7}\\
			&+\frac{77\mathfrak{L}_{3}\mathfrak{L}_{7}}{25920}+\frac{\mathfrak{L}_{1}^{3}\mathfrak{L}_{8}}{216}+\frac{5\mathfrak{L}_{2}\mathfrak{L}_{8}}{5184}+\frac{\mathfrak{L}_{1}\mathfrak{L}_{9}}{5184}\Big)t^{3}+\mathcal{O}(t^{7/2})\;,
			\end{split}
		\end{equation} 
where the $\mathfrak{L}_{i}$ are defined in \eqref{Liexplicit}.
\bigskip

\noindent
{\bf Application to KPZ}. We specify the function $g(x) = g_{\rm KPZ}(x) := - \log(1-x) = {\rm Li}_1(x)$. Then, the short time expansion of the generating function of the height $Q_{t}(\sigma)$ for sharp wedge initial condition defined in \eqref{G sw} is given in terms of (\ref{G short time expansion}) by $\log Q_{t}(\sigma) = q_{t,1}(\sigma)$ with
\begin{equation} \label{Likpz} 
	\mathfrak{L}_{i}=\frac{1}{\sqrt{4\pi}}\,\Li_{\frac{3}{2}-i}(\sigma)\;.
\end{equation}
The first terms of the expansion are written more explicitly in \eqref{G short time expansion sw t1/2}.

%
%
\end{subsection}

\begin{subsection}{Conjecture for the form of the general term} \label{sec:conjecture} 

We have not been able to obtain precise analytical expressions for arbitrary high orders in $t$ of $\log Q_{t,\beta}(\sigma)$. The expansion \eqref{G short time expansion} up to order $t^{3}$ however leads us to the following conjecture for the form of the general term of the expansion.
		\begin{conjecture}
			For $\sigma<1$, the short time expansion of $\log Q_{t,\beta}(\sigma)$ has the form
			\begin{align}
				\label{conjecture G}
			\log 	Q_{t,\beta}(\sigma)=&\frac{1}{\pi \sqrt{t}}\int_{0}^{+\infty}\mathrm{d}x\sqrt{x} g(\sigma e^{-x})
			+ \frac{1}{4\pi^2} \int_0^\sigma \mathrm{d}u \, u \left[ \partial_u \int_{-\infty}^{+\infty}\mathrm{d}p\, g(u e^{-p^2})\right]^2
			\\&+\sum_{r=1}^{+\infty}\sum_{q=0}^{\frac{r}{2}+1}\frac{2^{r+1-3q}}{(2q+1)!}\,t^{r/2}
				\times\sum_{\substack{r+2\geq n_{1}\geq\ldots\geq n_{r+2-2q}\geq0\\n_{1}+\ldots+n_{r+2-2q}=r-1+q}}\frac{c_{r,q}(\mathbf{n})}{S(\mathbf{n})}\,\prod_{j=1}^{r+2-2q}\frac{\mathfrak{L}_{n_{j}+1}}{n_{j}!}\;.\nonumber
			\end{align}
			The notation $\mathbf{n}$ is a shorthand for $(n_{1},\ldots,n_{r+2-2q})$. The coefficients $c_{r,q}(\mathbf{n})$ are positive integers, and $S(\mathbf{n})$ is a symmetry factor equal to $S(\mathbf{n})=m_{1}!\ldots m_{k}!$ when the $n_{j}$'s take $k$ distinct values with multiplicities $m_{1},\ldots,m_{k}$.
		\end{conjecture}

The coefficients $c_{r,q}(\mathbf{n})$ extracted from \eqref{G short time expansion} are listed in the Appendix \ref{Appendix coeffs crq}. The fact that all the coefficients $c_{r,q}(\mathbf{n})$ appear to be positive integers suggests the existence of a combinatorial interpretation. We have not managed to guess exact expressions for the coefficients $c_{r,q}(\mathbf{n})$, except for $c_{r,0}(\mathbf{n})=(r-1)!$, which is easily spotted in Table \ref{table crq}, and is crucial for the considerations in Section \ref{sec:structure}.
		 
In the special case of the KPZ equation with sharp wedge initial condition, where $\mathfrak{L}_{i}$ is given by \eqref{Likpz}, the radius of convergence of the short time expansion can be extracted from \eqref{conjecture G} in the limits $\sigma\to1$ and $\sigma\to-\infty$. When $\sigma\to1$, $\mathfrak{L}_{i}\simeq\frac{\Gamma(i-1/2)}{\sqrt{4\pi}\,(1-\sigma)^{i-1/2}}$, and each term contributing to the coefficient of $t^{r/2}$ in \eqref{conjecture G} is at leading order proportional to $(1-\sigma)^{-3r/2}$. The radius of convergence is thus proportional to $(1-\sigma)^{3}$ when $\sigma\to1^{-}$. When $\sigma\to-\infty$, $\mathfrak{L}_{i}\simeq-\frac{(\log(-\sigma))^{3/2-i}}{\sqrt{4\pi}\,\Gamma(5/2-i)}$, and only $q=0$ contributes to \eqref{conjecture G} at leading order. Each term contributing to the coefficient of $t^{r/2}$ is at leading order proportional to $(\log(-\sigma))^{2-r/2}$, and the radius of convergence is thus proportional to $\log(-\sigma)$ when $\sigma\to-\infty$. The resummation \eqref{eq:Phi} below gives more precisely a radius of convergence in the variable $t$ of order $\pi^{2}\log(-\sigma)$.
			\end{subsection}

\subsection{Height distribution $P(H,t)$ and generating function $F_{t}(s)$} 

In this section, we consider the probability distribution function $P(H,t)$ of the KPZ height field with droplet initial condition, and the associated moment generating function $F_t(s)$ defined in   \eqref{F sw}. By consistency with our results, as sketched below, they admit the following short time expansion
\begin{align} \label{expansionP} 
&P(H,t)=t^{-1/4} \exp \left(  -\frac{\Phi(H)}{\sqrt{ t}} +\Phi_0(H)+\sum_{j=1}^{+\infty} \Phi_{j}(H) t^{\frac{j}{2}}  \right)\\
&F_t(s)=\mathbb{E}_{\rm KPZ}\left[ \exp \left( -\frac{s H}{\sqrt{t}}\right)  \right]=\exp \left( \frac{\lambda(s)}{\sqrt{t}}+\mu_0(s)+\sum_{j=1}^\infty \mu_j(s)t^{j/2} \right) \label{large_dev_gen_function}
\end{align}
The leading order $\Phi(H)$ (resp. $\lambda(s)$), already obtained in \cite{le2016exact}, agrees with our present results. Here we explicitly determine the next order corrections $\Phi_0$ (resp. $\mu_0$), but the
method can be extended to obtain iteratively the higher orders. 

Our starting point is the short time expansion of the generating function of $e^H$ 
 \begin{equation}
\label{large_dev_distribution}
Q_t(\sigma)= \mathbb{E}_{\, \mathrm{KPZ}}\left[\exp\left(\frac{\sigma}{\sqrt{4 \pi t}}\,\rme^{H(t)}\right) \right]=\exp \left( -\frac{\Psi(\sigma)}{\sqrt{ t}} +\Psi_0(\sigma)+\sum_{j=1}^{+\infty} \Psi_{j}(\sigma) t^{\frac{j}{2}} \right) 
\end{equation}
where we have defined the notations
\begin{equation}
\label{notation_KPZ}
\begin{split}
 &\Psi(\sigma)=-\frac{\mathrm{Li}_{5/2}(\sigma)}{\sqrt{4\pi }}\qquad , \qquad \Psi_0(\sigma)=\frac{1}{4\pi}\int_{0}^{\sigma}\!\frac{\rmd u}{u}\,\Li_{1/2}(u)^{2}
\end{split}
\end{equation}
and similar notations for the higher order functions which can be read from
Eqs. \eqref{G short time expansion} and \eqref{Likpz} . Since here we only consider the leading order correction, we drop all terms above $\Phi_0$, $\mu_0$
and $\Psi_0$.  All equalities will then have to be understood with an $\mathcal{O}(\sqrt{t})$ error term. In the rest of the section, we first evaluate the height distribution and then the generating function of $H$.

\subsubsection{Short time expansion of $P(H,t)$}

The l.h.s of \eqref{large_dev_distribution} can be written as an integral involving the height distribution of $H$,
\begin{equation}
 \mathbb{E}_{\, \mathrm{KPZ}}\left[\exp\left(\frac{\sigma}{\sqrt{4 \pi t}}\,\rme^{H(t)}\right) \right]=\int_{-\infty}^{\infty}\mathrm{d}H\, P(H,t)\exp\left(\frac{\sigma\,\rme^{H}}{\sqrt{4 \pi t}}\right) 
\end{equation}
Inserting the expansion \eqref{expansionP}, this leads to
\begin{equation}
\begin{split}
t^{-1/4} \int_\mathbb{R}\mathrm{d}H\, &\exp\left(\frac{1}{\sqrt{t}}\left[ \frac{\sigma\,\rme^{H}}{\sqrt{4 \pi }} -\Phi(H)\right]  +\Phi_0(H) \right) =\exp \left( -\frac{\Psi(\sigma)}{\sqrt{ t}} +\Psi_0(\sigma)\right) 
\end{split}
\end{equation}
At leading order in $t$, doing a saddle point evaluation, one recovers the result of \cite{le2016exact}, extended in \cite{krajenbrink2018large}, which we express in a parametric form ($H^*$ indicates the saddle point)
\begin{equation}
\begin{split} \label{parametric1} 
 e^{H^*}=-\sqrt{4\pi}\,\Psi'(\sigma) \quad , \quad 
\Phi(H^*)=\Psi(\sigma)-\sigma \Psi'(\sigma)
\end{split}
\end{equation}
The next term in the short time expansion results from the Gaussian integration around the saddle point (once again the overall error is of order $\mathcal{O}(\sqrt{t})$ and does not contribute to this order)
\begin{equation}
\begin{split}
\int_{-\infty}^{\infty}\mathrm{d}H\,& \exp\left(\frac{1}{\sqrt{t}}\left[ \frac{\sigma \rme^{H}}{\sqrt{4 \pi }} -\Phi(H)\right] +\Phi_0(H)\right) =\\
& \sqrt{\frac{2\pi \sqrt{t}}{\abs{\frac{\sigma e^{H^*}}{\sqrt{4\pi}}-\Phi''(H^*)}}} \exp\left(\frac{1}{\sqrt{t}}\left[\frac{\sigma \rme^{H^*}}{\sqrt{4 \pi }} -\Phi(H^*)\right] +\Phi_0(H^*) +\mathcal{O}(\sqrt{t}) \right)
\end{split}
\end{equation}
Using the following identity (obtained by differentiation of \eqref{parametric1}) for $k=2$
\begin{equation} \label{k}
\forall k\geq 1, \, \Phi^{(k)}(H^*)=\left[ \frac{\Psi'(\sigma)}{\Psi''(\sigma)}\frac{\mathrm{d}}{\mathrm{d}\sigma}\right]^{k-1} \left( -\sigma \Psi'(\sigma) \right)\;,
\end{equation}
one simplifies the prefactor of the exponential as (the minus sign is due to the fact that $\Psi$ is strictly concave \cite{krajenbrink2018large} and ensures that the saddle point is a maximum of the exponential)
\begin{equation}
\sqrt{\frac{2\pi t^{1/2}}{\abs{\frac{\sigma e^{H^*}}{\sqrt{4\pi}}-\Phi''(H^*)}}}  = \sqrt{-\frac{2\pi t^{1/2}\Psi''(\sigma)}{\Psi'(\sigma)^2}}
\end{equation}
We find by identification
\begin{equation}
\Phi_0(H^*)=\Psi_0(\sigma)+\frac{1}{2}\log \left( -\frac{\Psi'(\sigma)^2}{ \Psi''(\sigma)} \right)
- \frac{1}{2} \log 2 \pi
\end{equation}
We now summarize the above results for the first orders of the distribution of $H$ (dropping the $^*$).

\medskip

{\bf Explicit formula} 

\medskip

\begin{figure}
	\begin{center}
		\begin{tabular}{lll}
			\begin{tabular}{c}\includegraphics[width=78mm]{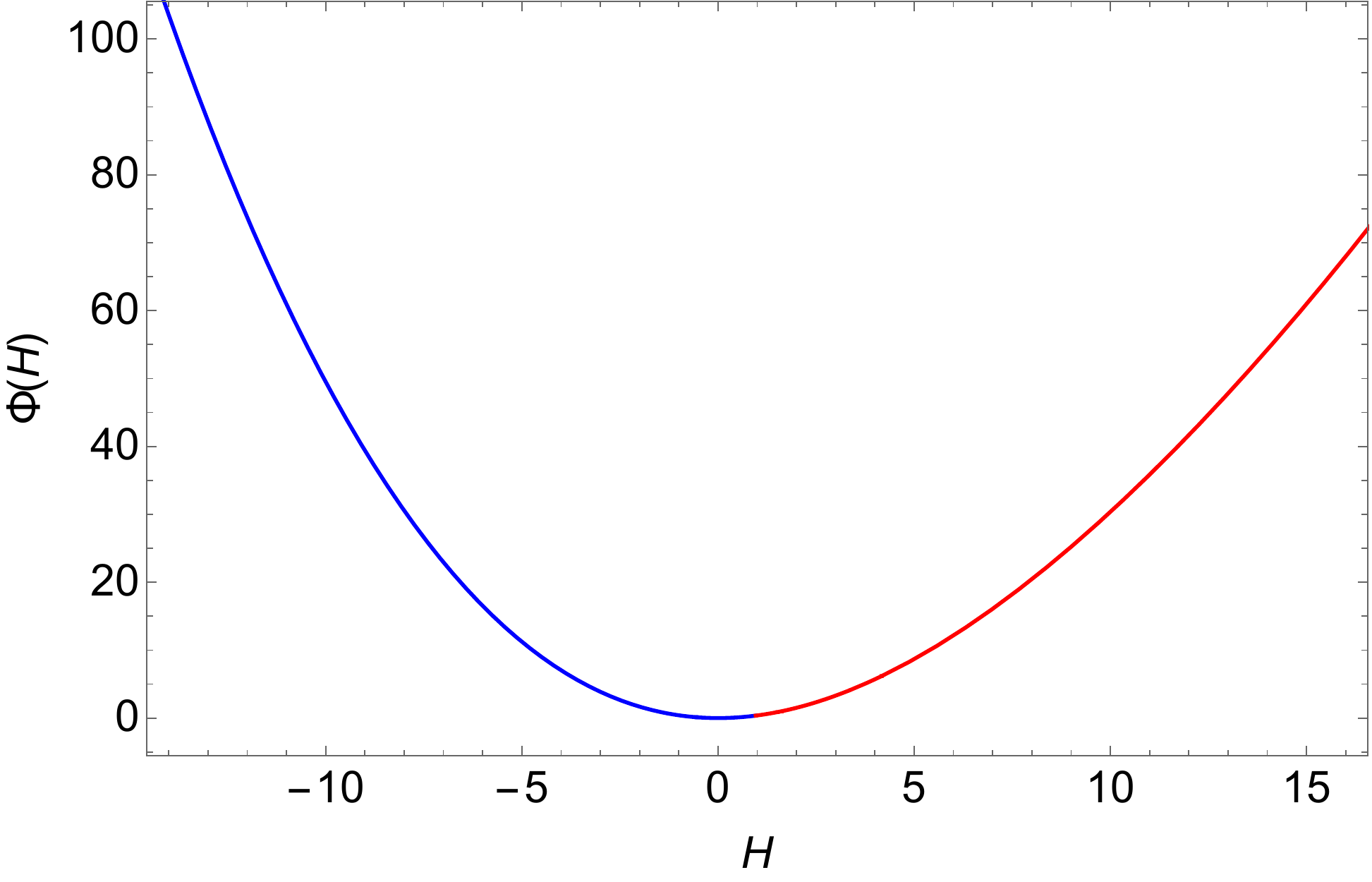}\end{tabular}
			&&
			\begin{tabular}{c}\includegraphics[width=78mm]{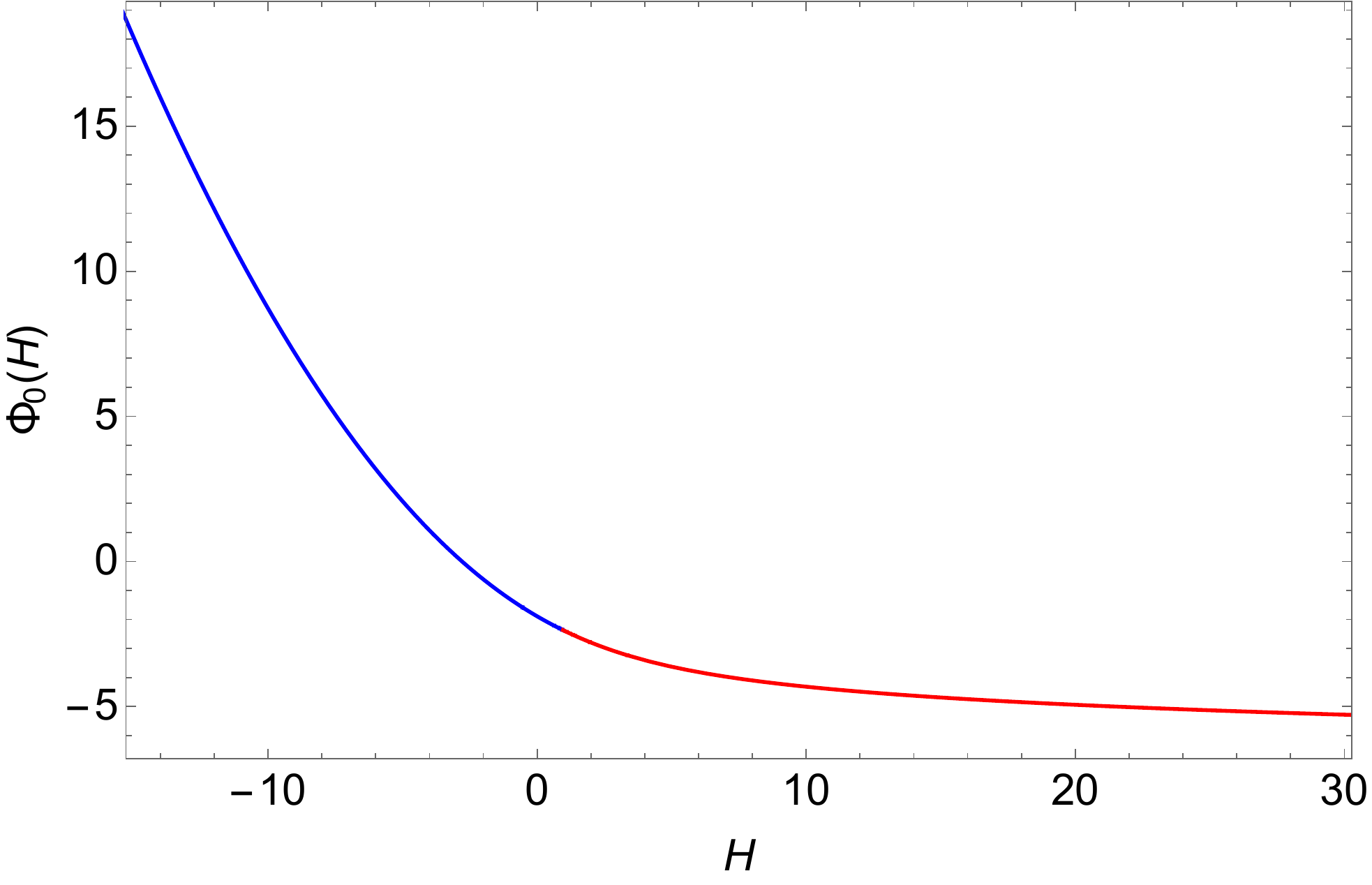}\end{tabular}\\
		\end{tabular}
	\end{center}
	\caption{Plot of the functions $\Phi(H)$ and $\Phi_0(H)$ defined in Eqs. \eqref{param_phi_0_1} and \eqref{param_phi_0_2}  for $H<H_c$ in blue, and $H>H_c$ in red. Both functions are convex and the branching at $H_c=\log \zeta(3/2)$ is smooth.}
	\label{fig Phi}
\end{figure}

Using the exact expressions of $\Psi$ and $\Psi_0$ of \eqref{notation_KPZ}, and the analysis
performed in \cite{le2016exact,krajenbrink2018large}, we obtain the parametric representation for the two leading terms of the distribution of $H$ for the droplet initial condition 
\begin{equation}
\label{param_phi_0_1}
\begin{split}
& e^H=\frac{\mathrm{Li}_{3/2}(\sigma)}{\sigma}\\
&\Phi(H)=\dfrac{\mathrm{Li}_{3/2}(\sigma)-\mathrm{Li}_{5/2}(\sigma)}{\sqrt{4\pi }}\\
&\Phi_0(H)=\frac{1}{4\pi}\int_{0}^{\sigma}\!\frac{\rmd u}{u}\,\Li_{1/2}(u)^{2} +\frac{1}{2}\log \left( \frac{\mathrm{Li}_{3/2}(\sigma)^2}{\mathrm{Li}_{1/2}(\sigma)-\mathrm{Li}_{3/2}(\sigma)} \right)
- \log( 2 \pi^{3/2}) \\
\end{split}
\end{equation}
Note that
the range of variation of $\sigma$ is $]-\infty,1]$. The above result is valid for
$H< H_c = \log \zeta(3/2)$. For $H>H_c$, one needs to introduce the analytic continuations of the polylogarithm, see e.g. \cite{vepvstas2008efficient} Eq. (11.3) or \cite{krajenbrink2018large} Eq. (33), that are
\begin{equation} \label{change}
\begin{split}
&\mathrm{Li}_{1/2}(\sigma) \to \mathrm{Li}_{1/2}(\sigma)-\sqrt{-\frac{4\pi}{\log \sigma} }\\
&\mathrm{Li}_{3/2}(\sigma) \to \mathrm{Li}_{3/2}(\sigma)+\sqrt{-16 \pi \log \sigma}\\
&\mathrm{Li}_{5/2}(\sigma) \to \mathrm{Li}_{5/2}(\sigma)- \frac{8 \sqrt{\pi}}{3} (- \log \sigma)^{3/2}\\
\end{split}
\end{equation}
We symbolically write these continuations as $\mathrm{Li}\to \mathrm{Li}+\Delta$. The interpretation of this shift $\Delta$ comes from the fact that the polylogarithms have a branch cut at $\sigma=1$ and one can extend them in an upper Riemann sheet by adding the contribution of the residue at the branching point which is $\Delta$. Taking into account the shift $\Delta$, the solution for $H>H_c$ is given by the following parametric system (where the range of $\sigma$ is now $]0,1]$)
\begin{equation}
\label{param_phi_0_2}
\begin{split}
& e^H=\frac{\mathrm{Li}_{3/2}(\sigma)+\sqrt{-16 \pi \log \sigma}}{\sigma} \\
&\Phi(H)=\dfrac{\mathrm{Li}_{3/2}(\sigma)-\mathrm{Li}_{5/2}(\sigma)+\sqrt{-16 \pi \log \sigma} + \frac{8 \sqrt{\pi}}{3} (- \log \sigma)^{3/2}}{\sqrt{4 \pi}} \\
 &  \Phi_0(H)= K + \frac{1}{4\pi}\int_{1}^{\sigma}\!\frac{\rmd u}{u}\, \Big[\Big(\Li_{1/2}(u)-\sqrt{-\frac{4\pi}{\log u}}\Big)^{2} - \frac{\pi}{-\log u}\Big] \\
& + \frac{1}{2}\log \left( \frac{(\mathrm{Li}_{3/2}(\sigma)+\sqrt{-16 \pi \log \sigma} )^2}{(-\mathrm{Li}_{1/2}(\sigma)+\mathrm{Li}_{3/2}(\sigma)+\sqrt{-16 \pi \log \sigma}+\sqrt{-\frac{4\pi}{\log \sigma}})\sqrt{- \log \sigma} } \right)\;,\\
\end{split}
\end{equation}
where $K$ is a constant given by
\bea \label{constanteK} 
K = \frac{1}{4\pi}\int_{0}^{e^{-1}}\!\frac{\rmd u}{u}\,\Li_{1/2}(u)^{2} + 
\frac{1}{4\pi}\int_{e^{-1}}^{1}\!\frac{\rmd u}{u}\, \Big[\Li_{1/2}(u)^{2} - \frac{\pi}{-\log u}\Big] - \log( 2 \pi^{3/2})\;.
\eea 
It is shown in the Appendix \ref{sec:analytical} that this is the proper continuation to
$H>H_c$, and that $\Phi(H)$ and $\Phi_0(H)$ are analytic at $H=H_c$. The plots of both functions $\Phi(H)$ and $\Phi_0(H)$ are presented in Fig. \ref{fig Phi}.\\ ~\\

\medskip
{\bf Asymptotic behaviour of the rate functions for $H \to \pm \infty$} 

\medskip

\begin{itemize}
\item The asymptotics near $H=-\infty$ (i.e. $\sigma=-\infty$) are as follows.
Let us use \eqref{largezLi} and denote $\sigma=-e^s$. Then one has from 
the first line of \eqref{param_phi_0_1}
\bea
s \simeq - H + \frac{3}{2} \log(-H) + \log \frac{4}{3 \sqrt{\pi}} 
\eea 
Inserting in the second line we find
\bea
\Phi(H)=\frac{4 (-H)^{5/2}}{15 \pi }-\frac{(-H)^{3/2} \left(-3 \log (-H)  +2+\log
   \left(\frac{9 \pi }{16}\right)\right)}{3 \pi}+\mathcal{O}((-H)^{1/2})
\eea 
Insering in the third line we find
\be
\Phi_0(H)=  \frac{H^2}{2 \pi ^2} + \frac{H
   \left(\log \left(\frac{16}{9 \pi }\right)+3 \log(-H) \right)}{2 \pi^2}+\mathcal{O}( \log (-H) )
\ee

\item The asymptotics near $H=+\infty$ (i.e. $\sigma=0$) are as follows. Let us denote $\sigma=e^{-s}$ with $s \to +\infty$. Here we must use the
analytical continuation for $H>H_c$. The first equation in \eqref{param_phi_0_2}
gives
\bea
s \simeq H - \frac{1}{2} \log(16 \pi H) 
\eea 
Inserting in the second line of \eqref{param_phi_0_2} we find
\be
\Phi(H) \simeq \frac{4 H^{3/2}}{3}+\sqrt{H} \left(\log
   \left(\frac{1}{H}\right)+2-\log (16 \pi   )\right)
\ee
Insering in the third line we find
\be
\Phi_0(H) \simeq - \frac{3}{4} \log \log H 
\ee

\end{itemize}

\begin{figure}[h!]
	\begin{center}
\includegraphics[scale=0.55, trim={3cm 7cm 3cm 4cm}]{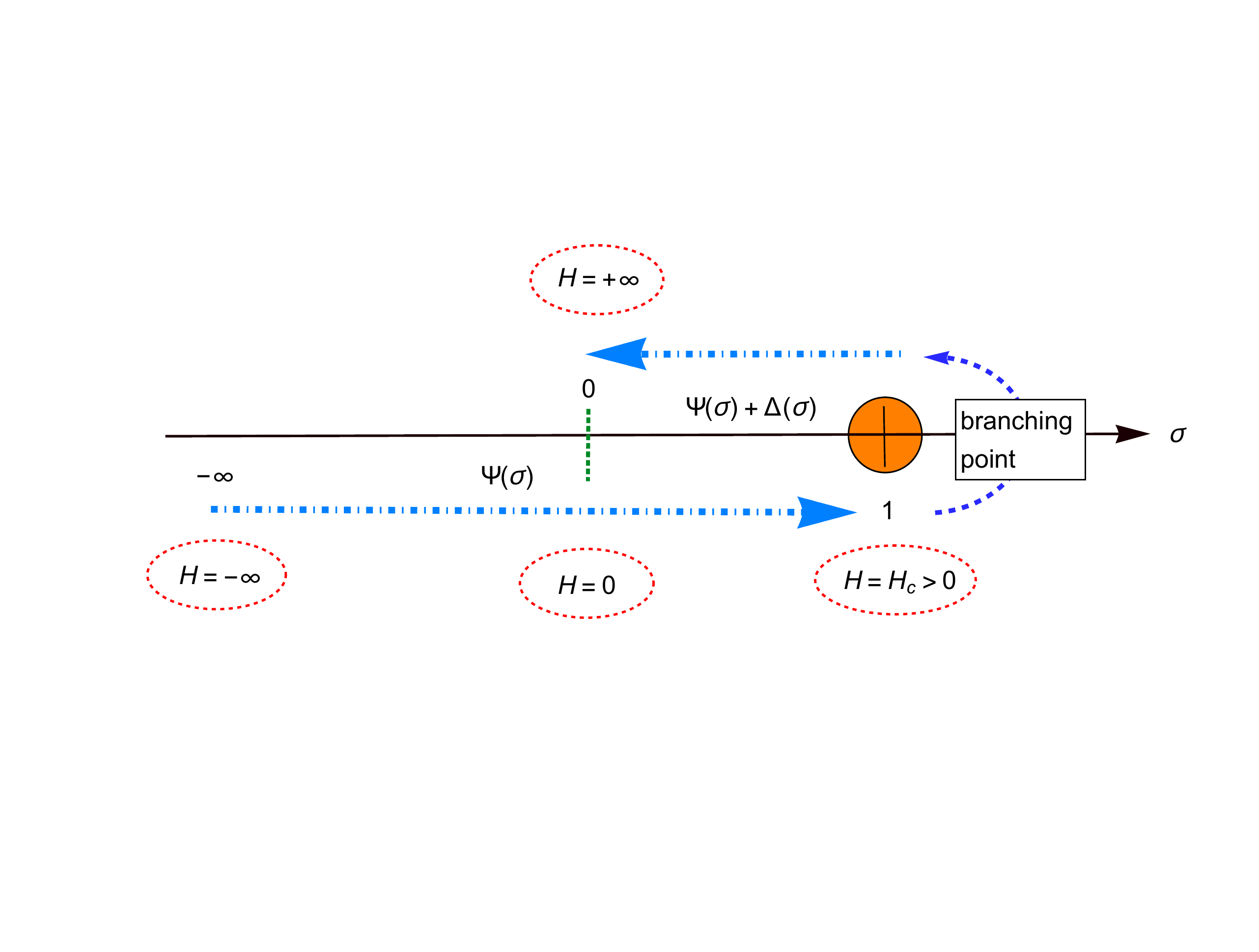}
	\end{center}
	\caption{Schematic representation of the determination of $H$ as a function of $\sigma$. The function $\Psi$ is defined up to $\sigma=1$ where it exhibits a branching point. One can turn around this branching point up to the redefinition of $\Psi$ by its continuation $\Psi+\Delta$.}
	\label{schematic_repn}
\end{figure}

\subsubsection{Short time expansion of $F_{t}(s)$}
\label{sec:Fts} 
To obtain the different contributions to $F_t(s)$, we express the middle term of \eqref{large_dev_gen_function} as an integral with the height distribution of $H$.
\begin{equation}
\begin{split}
 \mathbb{E}_{\, \mathrm{KPZ}}\left[ \exp \left( -\frac{s H}{\sqrt{t}}\right)  \right]&=\int_\mathbb{R}\mathrm{d}H\, P(H,t)\exp \left( -\frac{s H}{\sqrt{t}}\right) \\
 \end{split}
\end{equation}
Inserting both large deviation ansatz, we seek the following identification
\begin{equation}
\exp \left( \frac{\lambda(s)}{\sqrt{t}}+\mu_0(s)+\sum_{j=1}^\infty \mu_j(s)t^{j/2} \right) =t^{-1/4} \int_\mathbb{R}\mathrm{d}H\, \exp \left( -\frac{1}{\sqrt{t}}\left[sH+\Phi(H)\right] +\Phi_0(H) \right) 
\end{equation}
The leading order in $t$ is obtained by a saddle point evaluation and the first correction is obtained by the Gaussian integration close to the saddle point. After identification, the final result is expressed parametrically in terms of $H$ or equivalently in terms of $\sigma$ (using Eqs. \eqref{parametric1} and \eqref{k})
\begin{align}
s&=-\Phi'(H^*)=\sigma \Psi'(\sigma)\\
\lambda(s)&= H^* \Phi'(H^*)  - \Phi(H^*)
=-\left[ \sigma \Psi'(\sigma) \log\left( -\sqrt{4\pi} \Psi'(\sigma) \right)+\Psi(\sigma)-\sigma \Psi'(\sigma) \right]\\
\mu_0(s)&=\Phi_0(H^*)-\frac{1}{2}\log \left(\Phi''(H^*) \right)
= \Psi_0(\sigma)+\frac{1}{2}\log \left( \frac{\Psi'(\sigma)}{ \Psi'(\sigma)+\sigma \Psi''(\sigma)} \right)
\end{align}
where we recall that $\Phi''(H)>0$ as $\Phi$ is convex. Higher orders can be obtained by expanding around the saddle point. We find
\begin{eqnarray}
\label{mu1}
&& \mu_{1}(s)=
-\frac{1}{12\,\sigma\Psi'(\sigma)}
+\frac{5}{12}\,(\sigma\partial_{\sigma})^3\Psi(\sigma)
-\frac{1}{6}\,((\sigma\partial_{\sigma})^2\Psi(\sigma))^3\\
&&\hspace{37mm} -\frac{(\sigma\partial_{\sigma})^4\Psi (\sigma )}{8\,((\sigma\partial_{\sigma})^2\Psi(\sigma))^2}
+\frac{5\,((\sigma\partial_{\sigma})^3\Psi(\sigma))^2}{24\,((\sigma\partial_{\sigma})^2\Psi(\sigma))^3}
\;.\nonumber
\end{eqnarray}
Using the exact expressions of $\Psi$ and $\Psi_0$ of \eqref{notation_KPZ} one obtains the final result. The parametric representation of the two leading terms of the moment generating function of $H$ is 
\begin{equation}
\label{moment_gen_H_cum}
\begin{split}
& s=-\frac{\mathrm{Li}_{3/2}(\sigma)}{\sqrt{4\pi}}\\
&\lambda(s)=\dfrac{\mathrm{Li}_{5/2}(\sigma)-\mathrm{Li}_{3/2}(\sigma)}{\sqrt{4\pi }}+\frac{\mathrm{Li}_{3/2}(\sigma)}{\sqrt{4\pi}}\log \left( \frac{\mathrm{Li}_{3/2}(\sigma)}{\sigma} \right) \\
&\mu_0(s)=\frac{1}{4\pi}\int_{0}^{\sigma}\!\frac{\rmd u}{u}\,\Li_{1/2}(u)^{2} +\frac{1}{2}\log \left( \frac{\mathrm{Li}_{3/2}(\sigma)}{\mathrm{Li}_{1/2}(\sigma)} \right)\\
\end{split}
\end{equation}
The representation is valid for $s\geq -\frac{\zeta(3/2)}{\sqrt{4\pi}}$ and for  $\sigma$ in  $ ]-\infty,1]$. To obtain the representation for values of $s \leq -\frac{\zeta(3/2)}{\sqrt{4\pi}}$, one extends the parametric system as (the range of $\sigma$ is now $]0,1]$)
\begin{equation}
\begin{split}
& s=-\frac{\mathrm{Li}_{3/2}(\sigma)+\sqrt{-16\pi \log \sigma}}{\sqrt{4\pi}}\\
&\lambda(s)=\dfrac{\mathrm{Li}_{5/2}(\sigma)-\mathrm{Li}_{3/2}(\sigma)-\frac{8\sqrt{\pi}}{3}(-\log \sigma)^{3/2}-\sqrt{-16\pi \log \sigma}}{\sqrt{4\pi }}\\
&\hspace*{3cm}+\frac{\mathrm{Li}_{3/2}(\sigma)+\sqrt{-16\pi \log \sigma}}{\sqrt{4\pi}}\log \left( \frac{\mathrm{Li}_{3/2}(\sigma)+\sqrt{-16\pi \log \sigma}}{\sigma} \right) \\
\end{split}
\end{equation}
The continuation of $\mu_0(s)$ can be obtained using similar methods as in the
previous Section (see also Appendix \ref{sec:analytical}).

From this expansion, one can obtain the first few cumulants of $H$ by expanding the system \eqref{moment_gen_H_cum} around $\sigma=0$ and compare them with the ones obtained in  \cite{CLR10} Eqs (11) and (12). Up to $\mathcal{O}(\sigma^4)$, the system  \eqref{moment_gen_H_cum} yields
\begin{equation}
\begin{split}
& s=-\frac{\sigma}{\sqrt{4\pi}}-\frac{\sigma^2}{\sqrt{32\pi} }-\frac{\sigma^3}{\sqrt{108\pi}}\\
&\lambda(s)=\frac{\sigma^2}{8 \sqrt{2 \pi}} + \frac{(27 + 16 \sqrt{3}) \sigma^3}{864 \sqrt{\pi}}  \\
&\mu_0(s)=-\frac{\sigma }{4 \sqrt{2}}+\left(\frac{3}{32}-\frac{1}{3 \sqrt{3}}+\frac{1}{8 \pi }\right) \sigma
   ^2+\left(-\frac{3}{16}-\frac{7}{96 \sqrt{2}}+\frac{5}{12 \sqrt{6}}+\frac{1}{6 \sqrt{2} \pi }\right) \sigma ^3\\
\end{split}
\end{equation}
The inversion of this system up to $\mathcal{O}(s^4)$ reads 
\begin{equation}
\begin{split}
&\sigma=- \sqrt{4\pi } s-\sqrt{2} \pi  s^2+\frac{2}{9} \left(4 \sqrt{3}-9\right) \pi ^{3/2} s^3\\
&\lambda(s)=\frac{1}{2} \sqrt{\frac{\pi }{2}} s^2+\left(\frac{1}{4}-\frac{4}{9 \sqrt{3}}\right) \pi  s^3\\
&\mu_0(s)=\frac{1}{2} \sqrt{\frac{\pi }{2}} s+\left(\frac{1}{2}+\left(\frac{5}{8}-\frac{4}{3 \sqrt{3}}\right) \pi \right) s^2+\frac{\sqrt{\pi}}{36}  \left(\left(54+33  \sqrt{2}-40 \sqrt{6}\right) \pi -6 \sqrt{2}\right) s^3
\end{split}
\end{equation}
By definition, the cumulants of H are given by
\begin{equation} \label{cumH} 
\mathbb{E}_{\mathrm{KPZ}}\left[ H^q\right]^c=(-1)^qt^{\frac{q-1}{2}}\lambda^{(q)}(0)+(-1)^q t^{q/2}\mu_0^{(q)}(0)+\mathcal{O}(t^{\frac{q+1}{2}})
\end{equation}
For $q=1,2,3$ the leading orders are given in Table \ref{Table1_cum} which confirms the values obtained in \cite{CLR10} Eqs. (11) and (12) for the cumulants of $H$. 
\begin{table}[h!]
\begin{center}
\begin{tabular}{|c|c|}
\hline 
\hspace{0.1cm} Cumulant \hspace{0.1cm} & \hspace{0.1cm} Leading orders \hspace{0.1cm}  \\
\hline 
\hline
&\\[-3ex]
$\mathbb{E}_{\mathrm{KPZ}}\left[ H\right]$ & $- \sqrt{\dfrac{\pi t}{8}}  +\mathcal{O}(t)$ \\ [2ex]
\hline 
&\\[-3ex]
$\mathbb{E}_{\mathrm{KPZ}}\left[ H^2\right]^c$ & $ \sqrt{\dfrac{\pi t}{2}}  +\left(1+\left(\dfrac{5}{4}-\dfrac{8}{3 \sqrt{3}}\right) \pi \right) t +\mathcal{O}(t^{3/2})$ \\ [2ex]
\hline 
&\\[-3ex]
$\mathbb{E}_{\mathrm{KPZ}}\left[ H^3\right]^c$ &$ \left(\dfrac{8}{3 \sqrt{3}}-\dfrac{3}{2}\right) \pi t -\dfrac{\sqrt{\pi}}{6}  \left(\left(54+33  \sqrt{2}-40 \sqrt{6}\right) \pi -6 \sqrt{2}\right)  t^{3/2} +\mathcal{O}(t^2)$ \\ [2ex]
\hline 
\end{tabular} 
\end{center}
\caption{Three lowest order cumulants of the KPZ height field $H$ at short time.} \label{Table1_cum} 
\end{table}

\section{Cumulant expansion for KPZ with droplet and Brownian initial conditions:
small and long time limits}
\label{sec:cumulants} 

In this section we study the cumulants (defined below). First we
extend the cumulant expansion method introduced in \cite{KrajLedou2018,krajenbrink2018large}
to obtain a few terms in a systematic short time expansion of the moment generating function.
This method is more versatile that the one of the previous section, and allows to treat also the Brownian
initial condition, in addition to the droplet initial condition. For the latter, it allows to obtain the few lowest orders
quite easily, and agree with the results of the previous section. However, since it becomes quickly quite challenging we have not attempted the systematic study of the higher orders. In a second part
we list the results for the cumulants from the direct method. We study their (conjectured) general structure
which then allows us to also obtain {\it large time} results. 

\subsection{Presentation of the cumulant method}
\label{sec:cumpres} 

Let us describe the method. As recalled in Section \ref{recall}, the generating function for the KPZ equation for both initial conditions can be 
expressed using a Fredohlm determinant which involves a generic kernel $K$, either \eqref{G[M] sw}
with kernel $K=K_{\Ai}$ in \eqref{M sw} for the droplet initial condition, or \eqref{QGamma} with kernel
$K=K_{\Ai,\Gamma}$ in \eqref{M stat} for the Brownian initial condition, multiplied by a weight function
which here we will take quite general, see \eqref{M sw} and \eqref{M stat} for KPZ. 

We use the fact that, for any kernel $K$ and function $g$, we can expand the following Fredholm determinant in powers of $g$ as 
 \begin{equation}
\label{deter_process}
\log Q_t(\sigma) = \log \mathrm{Det}\left[I- (1- \rme^{\beta \hat g}) K\right] = \sum_{n=1}^\infty \frac{\beta^n \kappa_n(g)}{n!} 
\end{equation}
where for applications we will set $\hat g(a)=\hat g_{t,\sigma}(a)=g(\sigma e^{t^{1/3} a})$. This provides 
an expansion in cumulants for the more general generating function of $Q_t(\sigma)$ defined in 
\eqref{G sw2} in the case where $K$ is the Airy kernel. 
The first three cumulants are given by 
 \begin{equation}
  \label{third_order_expansion}
 \begin{split}
& \kappa_1(g)=\mathrm{Tr}(\hat g K), \qquad \kappa_2(g)=\mathrm{Tr}(\hat g^2 K)-\mathrm{Tr}(\hat g K \hat g K), \\
& \kappa_3(g)=\mathrm{Tr}(\hat g^3 K) -3 \mathrm{Tr}(\hat g K \hat g^2 K)+2 \mathrm{Tr}(\hat g K \hat g K \hat g	K)
 \end{split}
 \end{equation}
and the  formula for the general cumulant is 
\begin{equation}
\label{def_cum_general}
\kappa_n(g)=\sum_{\ell=1}^n \frac{(-1)^{\ell+1}}{\ell}\sum_{\substack{m_1,\dots,m_\ell \geq 1 \\ m_1+\dots+m_\ell=n}}\frac{n!}{m_1!\dots m_\ell !}\mathrm{Tr}(\hat g^{m_1}K \hat g^{m_2}K\dots \hat g^{m_\ell}K)
\end{equation}

This expansion is known as the cumulant expansion
see \cite{KrajLedou2018,krajenbrink2018large} for details. 
It was shown there that the first cumulant $\kappa_1(g)$ allows to
obtain the leading order in the short time expansion. Here we show that this expansion can be pushed in a systematic way to obtain 
the higher order contributions at short time. Each cumulant can be expanded to a given power of time. 
The way the various contributions are organized is summarized, in the case of the droplet initial condition, in the Table \ref{Table1}. 
\begin{table}[h!]
\begin{center}
\begin{tabular}{|c|c|c|c|}
\hline 
\hspace{0.1cm} Cumulant \hspace{0.1cm} & \hspace{0.1cm} First order \hspace{0.1cm} & \hspace{0.1cm} Second order \hspace{0.1cm} & \hspace{0.1cm} Third order \hspace{0.1cm} \\ 
\hline 
\hline
$\kappa_1$ & ${ t^{-1/2}}$ & ${ t^{1/2}} $& $t^{3/2} $\\ 
\hline 
$\kappa_2$ & ${ 1}$ &$ t $& $t^2 $\\ 
\hline 
$\kappa_3$ &$ { t^{1/2}}$ & $t^{3/2} $& $t^{5/2}$ \\ 
\hline 
$\kappa_4$ & $t$ & $t^2$ & $t^3$ \\ 
\hline 
$\kappa_5$ & $t^{3/2}$ & $t^{5/2}$ & $t^{7/2}$ \\ 
\hline 
\end{tabular} 
\end{center}
\caption{Leading orders in times for the contributions to a given order of each cumulant for the droplet (i.e. narrow wedge) initial condition,
$\kappa_n$, up to $n=5$.} \label{Table1} 
\end{table}

\subsubsection{Introduction of the propagators} 

We have seen in Section \ref{recall} that for both initial conditions, the kernels $K_\Ai$ and $K_{\Ai,\Gamma}$ can be factorized into a product of two (deformed) Airy operators. Since the cumulants are expressed in terms of traces involving these operators, it is useful to use the cyclicity of the trace and study the operators ${\sf Ai} \,\hat  g \, {\sf Ai}  : \mathbb{L}^2(\mathbb{R})\to \mathbb{L}^2(\mathbb{R}) $ defined as
\begin{equation}
[ {\sf Ai} \, \hat g \, {\sf Ai}](r,r')=\int_\mathbb{R} \rmd v \, \hat g(v) \Ai(r+v)\Ai(r'+v)
\end{equation}
and similarly for ${\sf Ai}^\Gamma_\Gamma$. Upon Taylor expanding $\hat g_{t,\sigma}$ as  
$\hat g_{t,\sigma}(v) = \sum_{p=1}^{+\infty} \frac{\sigma^{p}}{p!}g^{(p)}(0) \rme^{p v/\gamma } $, we introduce the Airy and deformed Airy propagators as
\begin{equation} \label{prop1} 
\begin{split}
G_{p/\gamma}(r_1,r_2) &=\int_{-\infty}^{+\infty} \mathrm{d}v  \, \rme^{p u/\gamma} \Ai(u+r_1) \Ai(u+r_2) \\
&= \frac{\gamma^{1/2}}{\sqrt{4 \pi p}} \rme^{\frac{p^3}{12\gamma^3} - \frac{p}{\gamma}\frac{r_1+r_2}{2}  - \gamma\frac{(r_1-r_2)^2}{4 p}} \\
& = e^{\frac{p^3}{12\gamma^3} - \frac{p}{\gamma}\frac{r_1+r_2}{2} } 
\int_{-\infty}^{+\infty} \frac{\mathrm{d}k}{2 \pi} \rme^{ - \frac{p}{\gamma} k^2
- i k (r_1-r_2) } 
\end{split}
\end{equation}
which represents the propagator of a Brownian particle in a linear potential,
and
\begin{equation} \label{prop2}
\begin{split}
G_{p/\gamma}(r_1,r_2) &=\int_{-\infty}^{+\infty} \rmd v \, \rme^{p v/\gamma} \Ai_\Gamma^\Gamma(v+r_1,\gamma,w,w) \Ai_\Gamma^\Gamma(v+r_2,\gamma,w,w) \\
&=\rme^{\frac{p^3}{12 \gamma^3}-\frac{p}{\gamma}\frac{r_1+r_2}{2}}\int_{-\infty}^{+\infty}\frac{\rmd k}{2\pi}\frac{\Gamma(-i\gamma k+w-\frac{p}{2})\Gamma(i\gamma k+w-\frac{p}{2})}{\Gamma(-i\gamma k+w+\frac{p}{2})\Gamma(i\gamma k+w+\frac{p}{2})}\rme^{-\frac{p}{\gamma}k^2-ik(r_1-r_2)}
\end{split}
\end{equation}
which is proved in Appendix. Hence we have 
\be \label{AiP0Ai} 
[ {\sf Ai} \, \hat g_{t,\sigma} \, {\sf Ai}](r,r')=  \sum_{p=1}^{+\infty} \frac{\sigma^{p}}{p!}g^{(p)}(0)
G_{p/\gamma}(r,r')
\ee
and the same for ${\sf Ai}^\Gamma_\Gamma$. We can also express the 
operators involving powers of $\hat g$
\be \label{AiP02Ai} 
[ {\sf Ai} \, \hat g_{t,\sigma}^2 \, {\sf Ai}](r,r')= \sum_{p_1,p_2=1}^{+\infty}  \frac{\sigma^{p_1+p_2}}{p_1! p_2!}g^{(p_1)}(0)g^{(p_2)}(0)
G_{(p_1+p_2)/\gamma}(r,r')
\ee
and similarly for higher powers. Note that these propagators satisfy the reproducing property
\begin{property}[Reproducing property]
\begin{equation} \label{repro} 
\begin{split}
G_{(p_1+p_2)/\gamma}(r,r')&=[G_{p_1/\gamma}G_{p_2/\gamma}](r,r')\\
&=[G_{p_1/\gamma}P_0G_{p_2/\gamma}](r,r')+[G_{p_1/\gamma}(1-P_0)G_{p_2/\gamma}](r,r')
\end{split}
\end{equation}
\end{property}

\subsubsection{Expression of the cumulants and diagrammatic representation}
\label{sec:expression} 

Using the reproducing property and the propagators we now obtain explicit formula for the
first three cumulants in terms of the propagators and the projector $P_0$ on $\mathbb{R}^+$. These formula have also a diagrammatic interpretation
which we provide here. In each case, we use some algebra to reduce the
number of terms. Since each term is superficially of a lower order in $1/\gamma=t^{1/3}$
than the total, this also allows to automatically perform the cancellations. 

\begin{itemize}

\item {\it First cumulant} 

From \eqref{third_order_expansion} and \eqref{AiP0Ai}, using the cyclicity of the trace we obtain
\be
\kappa_1(g) =  \sum_{p=1}^{+\infty} \frac{\sigma^{p}}{p!}g^{(p)}(0) \,  \mathrm{Tr}(P_0 G_{p/\gamma}) 
\ee
which has the following diagrammatic representation. Here we integrate the variables over the vertical thick lines. The interpretation of this graph is that we sum all paths starting and ending at the same point on the 
positive axis (i.e. bridges) in time $p/\gamma$. 
\begin{center}
\begin{tikzpicture}
\draw[thick] (0,0)   -- (5,0) ; 

\draw[thick,->] (0,0) node[anchor= east] {0} -- (0,1.5) node[anchor=south ] {$x$};
\draw[dashed] (0,-1) -- (0,0); 

\draw[thick,] (5,0) -- (5,1.5) ;
\draw[dashed] (5,-1) -- (5,0); 

\draw [thick] (0 ,1)  node[anchor= east] {$r_1$} to [ curve through ={(2,1.2) .. (3,0.2) .. (4,0.5)} ] (5,1)  ; 
\filldraw[black]  (5,1) node[anchor= west] {$r_1$} ; 

\filldraw[black] (0,1) circle (2pt); 
\filldraw[black] (5,1) circle (2pt); 

\filldraw[black] (2.5,-0.4) node[anchor= north] {$p/\gamma$}; 
\filldraw[black] (-4.5,0) node[anchor= west] {\hspace{0.7cm} $\mathrm{Tr}(P_0 G_{p/\gamma}) =$}; 
\end{tikzpicture}
\end{center}

\item {\it Second cumulant}

From the formulae \eqref{third_order_expansion}, \eqref{AiP0Ai}
and \eqref{AiP02Ai}, using again the cyclicity of the trace together with
the reproducibility \eqref{repro} we obtain the second cumulant as
\begin{equation}
\begin{split}
\kappa_2(g)&=  \sum_{p_1,p_2 = 1}^{+\infty}
\frac{\sigma^{p_1+p_2}}{p_1! p_2!}g^{(p_1)}(0)g^{(p_2)}(0)  \left[  \mathrm{Tr}(P_0 G_{(p_1+p_2)/\gamma})-\mathrm{Tr}(P_0 G_{p_1/\gamma} P_0 G_{p_2/\gamma})
\right]  \\
&=  \sum_{p_1,p_2 = 1}^{+\infty}
\frac{\sigma^{p_1+p_2}}{p_1! p_2!}g^{(p_1)}(0)g^{(p_2)}(0)  \mathrm{Tr}(P_0 G_{p_1/\gamma} (1-P_0) G_{p_2/\gamma})
\label{secondc} 
\end{split}
\end{equation}
which has the following diagrammatic representation. Here the bridge must cross the zero line at least once.
\begin{center}
\begin{tikzpicture}
\draw[thick] (0,0) -- (5,0) ; 

\filldraw[black] (0,-0.7) node[anchor= east] {$r_2$};
\filldraw[black]  (5,1) node[anchor= west] {$r_1$} ; 

\draw [thick] (0 ,1)  node[anchor= east] {$r_1$} to [ curve through ={(0.5,0) ..(2,-0.7) .. (4,1)} ] (5,1) ; 

\draw[thick,->] (0,0)  node[anchor= east] {0}-- (0,1.5) node[anchor=south ] {$x$};
\draw[dashed] (0,-1.1) -- (0,0);  

\draw[thick,] (5,0) -- (5,1.5) ; 
\draw[dashed] (5,-1.1) -- (5,0); 

\draw[dashed] (2,0) -- (2,1.5) ; 
\draw[thick] (2,-1.1) -- (2,0) ;

\filldraw[black] (2,-0.7) circle (2pt);
\filldraw[black] (0,1) circle (2pt);
\filldraw[black] (5,1) circle (2pt); 

\filldraw[black] (1,-0.7) node[anchor= north] {$p_1/\gamma$};
\filldraw[black] (3.5,-0.7) node[anchor= north] {$p_2/\gamma$}; 

\filldraw[black] (-5.5,0) node[anchor= west] {\hspace{-1cm }$ \mathrm{Tr}(P_0 G_{p_1/\gamma} (1-P_0) G_{p_2/\gamma})=$};
\end{tikzpicture}
\end{center}

\item {\it Third cumulant}

From the second line in \eqref{third_order_expansion}, \eqref{AiP0Ai},
\eqref{AiP02Ai} and its generalization, using the cyclicity of the trace together with
the reproducibility \eqref{repro} we obtain the third cumulant as
\begin{equation}
\begin{split}
&\kappa_3(g) =\sum_{p_1,p_2,p_3 = 1}^{+\infty}
\frac{\sigma^{p_1+p_2+p_3}}{p_1! p_2! p_3!}g^{(p_1)}(0)g^{(p_2)}(0)g^{(p_3)}(0)\\
& \times \left[\mathrm{Tr}(P_0 G_{(p_1+p_2+p_3)/\gamma}) -3 \mathrm{Tr}(P_0 G_{(p_1+p_2)/\gamma} P_0 G_{p_3/\gamma})+2  \mathrm{Tr}(P_0 G_{p_1/\gamma} P_0 G_{p_2/\gamma}P_0 G_{p_3/\gamma})\right] \\
& = \sum_{p_1,p_2,p_3 = 1}^{+\infty}
\frac{\sigma^{p_1+p_2+p_3}}{p_1 !p_2 !p_3!}g^{(p_1)}(0)g^{(p_2)}(0)g^{(p_3)}(0)\\
& \times \left[ \mathrm{Tr}(P_0 G_{p_1/\gamma}(1-P_0)G_{p_2/\gamma} (1-P_0) G_{p_3/\gamma}) -\mathrm{Tr}((1-P_0) G_{p_1/\gamma} P_0 G_{p_2/\gamma}P_0 G_{p_3/\gamma}) \right]
\end{split}
\end{equation}
where we have used the $(p_1,p_2,p_3)$ permutation invariance. From the last line we see that the third cumulant is given by the antisymmetrization of the first trace, since in the second trace all projectors are complementary, so that
\begin{equation}
\label{thirdc} 
\begin{split}
\kappa_3(g)=\sum_{p_1,p_2,p_3 = 1}^{+\infty}&
\frac{\sigma^{p_1+p_2+p_3}}{p_1 !p_2 !p_3!}g^{(p_1)}(0)g^{(p_2)}(0)g^{(p_3)}(0)\\
&\times \mathcal{A}\left[\mathrm{Tr}(P_0 G_{p_1/\gamma}(1-P_0)G_{p_2/\gamma} (1-P_0) G_{p_3/\gamma})\right]
\end{split}
\end{equation}
where $\mathcal{A}(f)(r_1,r_2,r_3)=f(r_1,r_2,r_3)-f(-r_1,-r_2,-r_3)$.
It can be written diagrammatically as
\begin{center}
\begin{tikzpicture}
\draw[thick] (0,0) -- (5,0) ; 

\filldraw[black] (0,-0.6) node[anchor= east] {$r_2$};
\filldraw[black] (5,-1.05) node[anchor= west] {$r_3$};
\filldraw[black]  (5,1) node[anchor= west] {$r_1$} ; 

\draw [thick] (0 ,1)  node[anchor= east] {$r_1$} to [ curve through ={(0.4,0.15) ..(1.2,-0.6).. (1.8, -0.85) .. (3.2,-1.05) .. (4.5,0.2)} ] (5,1) ; 

\draw[thick,->] (0,0)  node[anchor= east] {0}-- (0,1.5) node[anchor=south ] {$x$};
\draw[dashed] (0,-1.4) -- (0,0);  

\draw[thick,] (5,0) -- (5,1.5) ; 
\draw[dashed] (5,-1.4) -- (5,0); 

\draw[dashed] (1.2,0) -- (1.2,1.5) ; 
\draw[thick] (1.2,-1.4) -- (1.2,0) ;

\draw[dashed] (3.2,0) -- (3.2,1.5) ; 
\draw[thick] (3.2,-1.4) -- (3.2,0) ;

\filldraw[black] (1.2,-0.6) circle (2pt);
\filldraw[black] (3.2,-1.05) circle (2pt);
\filldraw[black] (0,1) circle (2pt);
\filldraw[black] (5,1) circle (2pt); 

\filldraw[black] (0.6,-1) node[anchor= north] {$p_1/\gamma$};
\filldraw[black] (2.1,-1) node[anchor= north] {$p_2/\gamma$}; 
\filldraw[black] (4.1,-1) node[anchor= north] {$p_3/\gamma$}; 

\filldraw[black] (-7,0) node[anchor= west] {\hspace{-1.5cm }$\mathrm{Tr}(P_0 G_{p_1/\gamma}(1-P_0)G_{p_2/\gamma} (1-P_0) G_{p_3/\gamma})=$};

\filldraw[black] (6,0) node[anchor= east] {};
\end{tikzpicture}
\end{center}

\end{itemize}

The above algebra can be extended to any order, and is valid for both initial conditions,
upon replacing the propagator $G_p$ by their expressions 
due to the particular form of the two kernels \eqref{prop1} and \eqref{prop2}. We now start the explicit calculations for the droplet initial condition, the Brownian one is performed afterwards.

\subsection{Cumulants associated to the Airy kernel and application to 
KPZ with droplet initial condition}

\subsubsection{Evaluation of the first three cumulants} 

\begin{itemize}

\item {\it First cumulant evaluation} 

To calculate the first cumulant at any order, we start with
\be
\begin{split}
 \mathrm{Tr}(P_0G_{p/\gamma}) &= \int_0^{+\infty} \rmd r\,  G_{p/\gamma}(r,r) 
  =  \frac{\rme^{\frac{t }{12 }p^3 } }{\sqrt{4 \pi t} p^{3/2}} 
  \end{split}
\ee
After summation one obtains the first cumulant at any order in $t$.
\begin{equation}
\begin{split}
 \kappa_1(g)&  = \frac{1}{\sqrt{4 \pi t}}  
\sum_{p=1}^{+\infty} \frac{\sigma^{p}}{p!p^{3/2}} g^{(p)}(0) \rme^{\frac{t}{12 } p^3 }\\
&=  \frac{ e^{ \frac{t}{12} (\sigma \partial_\sigma)^3}  }{\pi \sqrt{t}}\int_{0}^{+\infty}\mathrm{d}x\sqrt{x} g(\sigma e^{-x}) \label{cum1} 
\end{split}
\end{equation}
\begin{appKPZ} Taking $g(x)=g_{\rm KPZ}(x)=\mathrm{Li}_1(x)$, the first cumulant for droplet initial condition reads
\begin{equation}
\label{KPZ_cum1}
\kappa_1(g)=\frac{ e^{ \frac{t}{12} (\sigma \partial_\sigma)^3}  }{ \sqrt{4\pi t}}\mathrm{Li}_{5/2}(\sigma)
\end{equation}
We provide two equivalent expressions for the first cumulant (see Appendix for the derivation)
\begin{equation}
\label{equiv_cum1}
\kappa_1(g)  = \frac{1}{\sqrt{4 \pi t}}  \int_\mathbb{R}\rmd u\, 
\mathrm{Li}_{5/2}(\sigma \rme^{t^{1/3} 2^{-2/3} u}) \Ai (u)
\end{equation}
and
\begin{equation}
\label{equiv_cum2}
\hspace*{-1.7cm} \kappa_1(g)=\frac{e^{ \frac{t}{12} (\sigma \partial_\sigma)^3} }{2\pi \sqrt{t}} \int_{\mathbb{R}}\rmd v\, \mathrm{Li}_{2}(\sigma \rme^{-v^2})
\end{equation}

\end{appKPZ}
\item {\it Second cumulant evaluation}

To obtain an exact expression for the second cumulant, we start by calculating the trace involved in \eqref{secondc}
\be
\begin{split} \label{start1} 
 \mathrm{Tr}&(P_0 G_{p_1/\gamma} (1-P_0) G_{p_2/\gamma}) =
\iint_{\mathbb{R}^+\times \mathbb{R}^-} \rmd r_1 \rmd r_2\,  G_{p_1/\gamma}(r_1,r_2)
G_{p_2/\gamma}(r_2,r_1) \\ 
& = 
\frac{\gamma \rme^{\frac{p_1^3 + p_2^3}{12 \gamma^3}}}{4 \pi \sqrt{p_1 p_2} } \iint_{\mathbb{R}^+\times \mathbb{R}^-}  \rmd r_1 \rmd r_2\, 
\rme^{- \frac{r_1+r_2}{2 \gamma} (p_1+p_2) - \gamma \frac{(r_1-r_2)^2}{4} (\frac{1}{p_1} + \frac{1}{p_2})} 
\end{split} 
\ee

This integral can be explicitly performed using the change of variable (note that there is a $1/2$ Jacobian)
\begin{equation} \label{changevariable} 
\lbrace r_1-r_2=u\in \mathbb{R}^+ \, ; \, 
r_1+r_2=v\in [-u,u] \rbrace 
\end{equation} 
which leads for the trace
\begin{equation} \label{erf} 
\begin{split}
\mathrm{Tr}(P_0 G_{p_1/\gamma}& (1-P_0) G_{p_2/\gamma})   \\
&= \frac{\gamma^{3/2}}{\sqrt{4 \pi }(p_1+p_2)^{3/2}} 
\rme^{\frac{(p_1+p_2)^3}{12 \gamma^3}} 
{\rm Erf}\left(\frac{ \sqrt{ p_1 p_2 (p_1+p_2)}}{2\gamma^{3/2} } \right) \\
&= \frac{\sqrt{p_1 p_2}}{2 \pi (p_1+p_2)} 
+ t \frac{ \sqrt{p_1 p_2}}{24 \pi }  (p_1^2 + p_1 p_2 + p_2^2) + \mathcal{O}(t^2)
\end{split}
\end{equation}
One can proceed to a general summation to obtain $\kappa_2(g)$ as a double series.
\begin{equation}
\begin{split}
\kappa_2(g)
&=\frac{e^{ \frac{t}{12} (\sigma \partial_\sigma)^3}  }{\sqrt{4\pi t}} \sum_{p_1,p_2 = 1} ^{+\infty}\frac{\sigma ^{p_1+p_2}}{p_1!p_2!(p_1+p_2)^{3/2}}g^{(p_1)}(0)g^{(p_2)}(0) {\rm Erf}\left(\frac{ \sqrt{t} \sqrt{p_1 p_2 (p_1+p_2)}}{2 } \right)  \\
\end{split}
\end{equation}
Keeping the two first terms of the expansion of the error function, we give the two first contributions to $\kappa_2(g)$ 
\begin{equation}
\begin{split}
\kappa_2(g)&= \frac{1}{2\pi}  \sum_{p_1,p_2 = 1}^{+\infty}
\frac{\sigma^{p_1+p_2}}{p_1! p_2!}g^{(p_1)}(0)g^{(p_2)}(0)  \left[ \frac{\sqrt{p_1 p_2}}{ (p_1+p_2)} 
+ \frac{t}{12} \sqrt{p_1 p_2}  (p_1^2 + p_1 p_2 + p_2^2) \right] + \mathcal{O}(t^2)\\
&=\frac{1}{2\pi^2} \int_0^\sigma \frac{\mathrm{d}u}{u}  \left[ u\partial_u \int_{-\infty}^{+\infty}\mathrm{d}p\, g(u e^{-p^2})\right]^2+\Big(\frac{\mathfrak{L}_{2}^{2}}{6}+\frac{\mathfrak{L}_{1}\mathfrak{L}_{3}}{3}\Big)t +\mathcal{O}(t^2)
\end{split}
\end{equation}
where we recall the notation $ \mathfrak{L}_{i}=(\sigma \partial_\sigma)^{i}
\int_{-\infty}^{+\infty}\frac{\mathrm{d}p}{2\pi}\, g(\sigma e^{-p^2}) $.
\begin{appKPZ}
Taking $g_{\rm KPZ}(x)=\mathrm{Li}_1(x)$, the second cumuland reads 
\begin{equation}
\kappa_2(g)=\frac{1}{2 \pi } \int_0^{\sigma} \frac{\mathrm{d}u}{u}  {\rm Li}_{1/2}(u)^2 
+\frac{t}{24\pi}\left[ 2 \, \Li_{-3/2}(\sigma)\Li_{1/2}(\sigma)+\mathrm{Li}_{-1/2}(\sigma)^2\right]+\mathcal{O}(t^2)
\end{equation}
\end{appKPZ}
We recall that $g_{KPZ}^{(p)}(0)=(p-1)!$.

\begin{remark}
It is possible to obtain an exact expression for $\kappa_2(g)$ to all orders. Starting instead from 
\eqref{secondc} and \eqref{start1} and inserting the form of the propagator given in the
last equation in \eqref{prop1}, we can perform the summation over $p_1$ and $p_2$, 
use the change of variable \eqref{changevariable} and a rescaling of $k$ to obtain
\begin{equation} \label{cum2exact} 
\begin{split}
& \kappa_2(g)= \frac{1}{2} \int_0^{+\infty} \mathrm{d}u\,  \int_{-u}^u \mathrm{d}v \, |F(u,v)|^2 \\
& F(u,v) = e^{\frac{t}{12} (\sigma \partial_\sigma)^3} \int_{-\infty}^{+\infty} \frac{\mathrm{d}k}{2 \pi} e^{-i k u} \, g(\sigma e^{- \frac{\sqrt{t}}{2} v - k^2}) 
\end{split}
\end{equation}

\end{remark}

\item {\it Third cumulant evaluation}

Let us first provide an exact formula for the summand in the third cumulant expression \eqref{thirdc}, in the form of an 
integral.
\begin{equation}
\begin{split}
 &\mathcal{A}\left(\mathrm{Tr}(P_0 G_{p_1/\gamma}(1-P_0) G_{p_2/\gamma} (1-P_0)G_{p_3/\gamma})\right)\\
 &=\iiint_{\mathbb{R}^+\times \mathbb{R}^- \times \mathbb{R}^-} \hspace{-1cm}\rmd r_1 \rmd r_2  \rmd r_3 \, \mathcal{A}\left(G_{p_1/\gamma}(r_1,r_2)
G_{p_2/\gamma}(r_2,r_3)G_{p_3/\gamma}(r_3,r_1)\right) \\
&=\frac{\rme^{t\frac{p_1^3+p_2^3+p_3^3}{12}}}{(4 \pi)^{3/2}t^{1/2}\sqrt{ p_1 p_2 p_3}}\\
& \times \iiint_{\mathbb{R}^+\times \mathbb{R}^- \times \mathbb{R}^-} \hspace{-1cm} \rmd r_1 \rmd r_2  \rmd r_3  \mathcal{A}\left(\rme^{ - \frac{r_1+r_2}{2\gamma } p_1- \frac{r_2+r_3}{2\gamma} p_2-\frac{r_3+r_1}{2\gamma} p_3 - \gamma \frac{(r_1-r_2)^2}{4 p_1} - \gamma \frac{(r_2-r_3)^2}{4 p_2} - \gamma \frac{(r_3-r_1)^2}{4 p_3}} \right)
 \end{split}
\end{equation}
We rescale the arguments so that the Gaussian exponential do not have any $\gamma$ dependence.
\begin{equation}
\begin{split}
 &\mathcal{A}\left(\mathrm{Tr}(P_0 G_{p_1/\gamma}(1-P_0) G_{p_2/\gamma} (1-P_0)G_{p_3/\gamma})\right)=-\frac{2 \rme^{t \frac{p_1^3+p_2^3+p_3^3}{12}}}{(4 \pi)^{3/2}\sqrt{ p_1 p_2 p_3}}\\
&\iiint_{\mathbb{R}^+\times \mathbb{R}^- \times \mathbb{R}^-}\hspace{-1cm} \rmd r_1 \rmd r_2  \rmd r_3 \sinh\left(\sqrt{t}\left[\frac{r_1+r_2}{2 } p_1+ \frac{r_2+r_3}{2} p_2+\frac{r_3+r_1}{2} p_3\right]\right)\\
& \times \rme^{ - \frac{(r_1-r_2)^2}{4 p_1} -  \frac{(r_2-r_3)^2}{4 p_2} -  \frac{(r_3-r_1)^2}{4 p_3}} 
 \end{split}
\end{equation}
Until now this expression is exact. We have not attempted to calculate explicitly this integral.
We now provide its small time expansion up to the second order, leading to 
a perturbative formula for the third cumulant. Expanding the hyperbolic sine up to the third order, we obtain
\begin{equation}
\mathcal{A}\left(\mathrm{Tr}(P_0 G_{p_1/\gamma}(1-P_0) G_{p_2/\gamma} (1-P_0)G_{p_3/\gamma})\right)=-\frac{2 \rme^{t \frac{p_1^3+p_2^3+p_3^3}{12}}}{(4 \pi)^{3/2}\sqrt{ p_1 p_2 p_3}} \left(\frac{\sqrt{t}}{2}\mathcal{I}_1+\frac{t^{3/2}}{48}\mathcal{I}_2 \right)
\end{equation}
where 
\begin{equation}
\begin{split}
\mathcal{I}_1&=\iiint_{\mathbb{R}^+\times \mathbb{R}^- \times \mathbb{R}^-} \hspace{-1cm}\rmd r_1 \rmd r_2  \rmd r_3   \left((r_1+r_2) p_1+ (r_2+r_3) p_2+(r_3+r_1) p_3\right)\\
&\hspace*{2cm} \times \rme^{ - \frac{(r_1-r_2)^2}{4 p_1} -  \frac{(r_2-r_3)^2}{4 p_2} -  \frac{(r_3-r_1)^2}{4 p_3}}\\
&= -4p_1 p_2 p_3
\end{split}
\end{equation}
and 
\begin{equation}
\begin{split}
\mathcal{I}_2&=\iiint_{\mathbb{R}^+\times \mathbb{R}^- \times \mathbb{R}^-} \hspace{-1cm}\rmd r_1 \rmd r_2  \rmd r_3   \left((r_1+r_2) p_1+ (r_2+r_3) p_2+(r_3+r_1) p_3\right)^3\\
&\hspace*{2cm} \times \rme^{ - \frac{(r_1-r_2)^2}{4 p_1} -  \frac{(r_2-r_3)^2}{4 p_2} -  \frac{(r_3-r_1)^2}{4 p_3}}\\
&=-16p_1 p_2 p_3(p_1 p_2 p_3+ p_1 p_2^2+ p_2 p_1^2+ p_1 p_3^2+ p_3 p_1^2+ p_3 p_2^2+ p_2 p_3^2)\\
&\overset{S_3}{=}-16 p_1 p_2 p_3(p_1 p_2 p_3+ 6p_1 p_2^2)
\end{split}
\end{equation}
The calculation of both integrals is non trivial but simply leads a homogeneous polynomial in the $p$ that is symmetric, hence easy to identify using Mathematica. The last equality has to be taken modulo the permutation of $p_1,p_2,p_3$ denoted by the symmetry group $S_3$.
We therefore evaluate the trace up to order $\mathcal{O}(t^{5/2})$ as 
\begin{equation}
\begin{split}
\mathcal{A}&\left(\mathrm{Tr}(P_0 G_{p_1/\gamma}(1-P_0) G_{p_2/\gamma} (1-P_0)G_{p_3/\gamma})\right)\\
&\overset{S_3}{=}\frac{ 1}{4\pi^{3/2}\sqrt{ p_1 p_2 p_3}} \left(2p_1 p_2 p_3 \sqrt{t} +p_1 p_2 p_3(\frac{1}{3} p_1 p_2 p_3+ 2p_1 p_2^2+\frac{1}{2}p_1^3)t^{3/2} +\mathcal{O}(t^{5/2})\right)
\end{split}
\end{equation}
 We summarize the two contributions we obtained for the third cumulant 
\begin{equation}
\begin{split}
\kappa_3(g)=&\sum_{p_1,p_2,p_3 = 1}^{+\infty}
\frac{\sigma^{p_1+p_2+p_3}\sqrt{ p_1 p_2 p_3}}{p_1!p_2!p_3!}g^{(p_1)}(0)g^{(p_2)}(0)g^{(p_3)}(0)\\
& \hspace*{1.3cm} \times \left[\frac{\sqrt{t}}{2 \pi^{3/2}}+\frac{p_1 p_2 p_3+ 6p_1 p_2^2+\frac{3}{2}p_1^3}{12 \pi^{3/2}}t^{3/2} \right]+\mathcal{O}(t^{5/2})\\
=&4\mathfrak{L}_{1}^{3}\sqrt{t}+\Big(\frac{2\mathfrak{L}_2^3}{3}+4\mathfrak{L}_{1}\mathfrak{L}_{3}\mathfrak{L}_2+\mathfrak{L}_{1}^{2}\mathfrak{L}_4\Big)t^{3/2}+\mathcal{O}(t^{5/2})\hspace*{2cm}
\end{split}
\end{equation} 
where we recall the notation  $\mathfrak{L}_{i}=(\sigma \partial_\sigma)^{i}
\int_{-\infty}^{+\infty}\frac{\mathrm{d}p}{2\pi}\, g(\sigma e^{-p^2})$.
\begin{appKPZ}
Taking $g_{\rm KPZ}(x)=\mathrm{Li}_1(x)$, the third cumuland reads
\begin{equation}
\begin{split}
\kappa_3(g)=\frac{\sqrt{t}}{2 \pi^{3/2}}&\mathrm{Li}_{1/2}^3(\sigma) + \frac{1}{8\pi^{3/2}}\Big( \frac{2}{3}\mathrm{Li}_{-1/2}^3(\sigma)+4\mathrm{Li}_{1/2}(\sigma) \mathrm{Li}_{-1/2}(\sigma) \mathrm{Li}_{-3/2}(\sigma) \\
&+\mathrm{Li}_{1/2}^2(\sigma) \mathrm{Li}_{-5/2}(\sigma)  \Big)   t^{3/2} +\mathcal{O}(t^{5/2})\\
\end{split}
\end{equation}
\end{appKPZ}

\begin{remark}
It is possible to obtain an exact (although a bit formal) expression for $\kappa_3(g)$ to all orders, see 
Appendix \eqref{sec:allorders}.
\end{remark}

\item{{\it Summary: first three terms in the cumulant expansion for droplet KPZ}}

Collecting the contributions above we obtain,
for the droplet initial condition, the following expansion
\bea
 &&\log  Q_t(\sigma) = \log \mathbb{E}_{\rm KPZ} \left[ \exp \left(\frac{\sigma}{\sqrt{4 \pi t} } \rme^{H(t)}  \right) \right]
=  \kappa_1(g) + \frac{1}{2} \kappa_2(g) +\frac{1}{6}\kappa_3(g)+\dots   \nonumber  \\
 && =   \frac{1}{\sqrt{4 \pi t}}   {\rm Li}_{5/2}(\sigma) 
+ \frac{1}{4 \pi } \int_0^{\sigma} \frac{\mathrm{d}u}{u}  {\rm Li}_{1/2}(u)^2 
+  \frac{\sqrt{t}}{24 \sqrt{ \pi}  } {\rm Li}_{-1/2}(\sigma)  +\frac{\sqrt{t}}{12 \pi^{3/2}}\mathrm{Li}_{1/2}^3(\sigma)  \nonumber \\
 &&\hspace*{1cm}+\frac{t}{48\pi}\left[ 2\Li_{-3/2}(\sigma)\Li_{1/2}(\sigma)+\mathrm{Li}_{-1/2}(\sigma)^2\right]+ \mathcal{O}(t)+ \mathcal{O}(t^{3/2})  \nonumber 
\eea 
This expression contains the first three orders in the small
time expansion. The leading order recovers the result of \cite{le2016exact}. 
It is interesting to note that the degree in terms of polylogarithm indicates from
which cumulant each term originates (i.e. a product of $n$ polylogarithms comes
from the $n$-th cumulant). The 
fourth order, i.e. the order $t$ is not complete as we miss the leading term from 
the fourth cumulant, not calculated here. By comparing with Eq. \eqref{G short time expansion sw t1/2}
we see that the two series agree to this order. Although it allows to obtain the few leading
orders painlessly, the systematics of this cumulant method remains to be improved
to match the method of Section \ref{section sw systematic}. We now turn to 
the cumulant calculation using the latter method.

\end{itemize} 

\subsubsection{Cumulants by the direct expansion method}

In Section \ref{section sw systematic} we introduced the parameter $\beta$ as a counting device. From the definition of the cumulants in Eq. \eqref{def_cum_general} it is easy to see that it exactly counts the order of the cumulant.
Hence we can now list the results for the cumulants from the direct method of Section \ref{section sw systematic}.
The first cumulants $\kappa_{n}(g)=(\partial_{\beta}^{n}q_{t,\beta}(\sigma))_{|\beta=0}$ are then equal to
		\begin{align}
			\kappa_{1}&=\frac{1}{\pi \sqrt{t}}\int_{0}^{+\infty}\mathrm{d}x\sqrt{x} g(\sigma e^{-x})+\frac{\mathfrak{L}_{2}}{12}\,\sqrt{t}+\frac{\mathfrak{L}_{5}}{288}\,t^{3/2}+\frac{\mathfrak{L}_8}{10368}\,t^{5/2}+\ldots\nonumber\\
			\kappa_{2}&=\frac{1}{2\pi^2} \int_0^\sigma \mathrm{d}u \, u \left[ \partial_u \int_{-\infty}^{+\infty}\mathrm{d}p\, g(u e^{-p^2})\right]^2+\Big(\frac{\mathfrak{L}_{2}^{2}}{6}+\frac{\mathfrak{L}_{1}\mathfrak{L}_{3}}{3}\Big)t+\Big(\frac{29}{360}\,\mathfrak{L}_{3}\mathfrak{L}_{4}+\frac{\mathfrak{L}_{2}\mathfrak{L}_{5}}{24}+\frac{\mathfrak{L}_{1}\mathfrak{L}_{6}}{72}\Big)t^{2}+\ldots\nonumber\\
			\kappa_{3}&=4\mathfrak{L}_{1}^{3}\sqrt{t}+\Big(\frac{2\mathfrak{L}_2^3}{3}+4\mathfrak{L}_{1}\mathfrak{L}_{3}\mathfrak{L}_2+\mathfrak{L}_{1}^{2}\mathfrak{L}_4\Big)t^{3/2}+\ldots\nonumber\\
			\kappa_{4}&=32\mathfrak{L}_{1}^{3}\mathfrak{L}_{2}t+\Big(4\mathfrak{L}_{2}^{4}+48\mathfrak{L}_{1}\mathfrak{L}_{3}\mathfrak{L}_{2}^{2}+24\mathfrak{L}_{1}^{2}\mathfrak{L}_{4}\mathfrak{L}_{2}+16\mathfrak{L}_{1}^{2}\mathfrak{L}_{3}^{2}+\frac{8}{3}\,\mathfrak{L}_{1}^{3}\mathfrak{L}_{5}\Big)t^{2}+\ldots \label{cum} \\
			\kappa_{5}&=(80\mathfrak{L}_{3}\mathfrak{L}_{1}^{4}+320\mathfrak{L}_{2}^{2}\mathfrak{L}_{1}^{3})t^{3/2}+\ldots\nonumber
		\end{align}
		where the $\mathfrak{L}_{j}$ are defined in 
		\eqref{Liexplicit} (setting $\beta=1$) for general $g(x)$  
		and in \eqref{Likpz} for KPZ with droplet initial condition. Note that an exact formula
		to all orders exists for the first cumulant $\kappa_1$, see Eq. \eqref{cum1}.

\subsubsection{Structure of the cumulant series in the large time limit}
\label{sec:structure} 
		
Since we are studying formally an expansion in $t$, but for arbitrary fixed $\sigma$,
it turns out that one can in fact obtain some results for arbitrary $t$ and even for 
large $t$. Obtaining such results assumes that the observed structure of the 
series in $t$ when $\sigma\to-\infty$ holds to arbitrary order (and in a non perturbative sense). Examination of our results \eqref{cum} for the cumulants, together with the conjecture 
formulated in Section \ref{sec:conjecture}
leads to the following guess for the structure of the cumulant expansion
\bea \label{guess1} 
&& \log Q_t(\sigma) = \sum_{n \geq 1} \frac{\kappa_n(\sigma,t)}{n!} \\
&& \kappa_n(\sigma,t)= t^{\frac{n}{2}-1} \kappa_n^0(\sigma) + \sum_{p \geq 1}
t^{\frac{n}{2}-1+p} \, \kappa_n^p(\sigma) \label{ser1} 
\eea
where the leading order is obtained from the conjecture $c_{r,0}(\mathbf{n})=(r-1)!$ of Section \ref{sec:conjecture}, as
\begin{equation} 
\label{cumconj} 
\kappa_n^0(\sigma)  = 
		2^{n-1} (\sigma\partial_{\sigma})^{n-3}\mathfrak{L}_{1}(\sigma)^{n}
\end{equation}
		for $n\geq 1$. Indeed it holds for $n=1$ since $\mathfrak{L}_{1}=
		(\sigma\partial_{\sigma})^{2} \mathfrak{L}_{-1}$ and $\kappa_1=t^{-1/2} 
		\mathfrak{L}_{-1}$ and for $n=2$, from \eqref{g_exp_lhs1} interpreting $(\sigma \partial_{\sigma})^{-1}=\int_0^{\sigma} \frac{du}{u}$.\\
		
		The Eq. \eqref{ser1} can now be studied at fixed $t$ but in the limit $\sigma \to - \infty$.
		In that limit the first term in \eqref{ser1}, $t^{\frac{n}{2}-1} \kappa_n^0(\sigma)$, is the dominant one, which gives the leading
		asymptotics of $Q_{t}(\sigma)$ for $\sigma\to-\infty$ for any $t$. Indeed we 
		see from \eqref{cum} and the above conjecture that the
	 term $t^{\frac{n}{2}-1+p} \, \kappa_n^p(\sigma)$ contains $n- 3 + 3 p$ derivatives $\sigma \partial_\sigma$. Each derivative makes the term more subdominant: more precisely, setting $\sigma = - e^{v}$, 
	   the condition on $g(x)$ for this to hold
		  is that $\lim_{v \to +\infty}
		 \frac{\log |g(- e^{v-x})|}{v} =0$, i.e. $g(- e^{v-x})$ is a sub-exponential function
		 of $v$. It is obeyed by the case $g(x)=-\log(1-x)$ relevant for the KPZ equation.
		 In that case $g(- e^{v-x}) \simeq_{v \to +\infty} - (v-x)_+$ and one obtains,
		 explicitly in the limit $v \to +\infty$. 
		\bea \label{res1} 
&& \mathfrak{L}_{-1}(- e^{v}) \simeq -\frac{1}{\pi} \int_0^{+\infty} \mathrm{d}x \,  \sqrt{x}(v-x)_+ = - \frac{4}{15 \pi} v^{5/2} \\
&& \mathfrak{L}_{1}(- e^{v}) = \partial_v^2 \mathfrak{L}_{-1}(- e^{v}) 
\simeq - \frac{1}{\pi} v^{1/2}
\eea

		 Remarkably, the structure of the series \eqref{ser1} is such that the 
		 {\it large time limit} is also controled by the term $t^{\frac{n}{2}-1} \kappa_n^0(\sigma)$. 
		 This can be seen as follows. Let us choose $\sigma = - e^{-z t}$, with fixed $z<0$.
		 From the conjecture above we can write in a symbolic form the general term
		 of \eqref{ser1} as
		 \bea \label{argument} 
		  t^{\frac{n}{2}-1+p} \, \kappa_n^p(\sigma=- e^{-z t}) &\sim&
		 t^{\frac{n}{2}-1+p} [(\sigma \partial_\sigma)^{n-3 + 3p}] [\mathfrak{L}_{1}(\sigma)^n]
		\\
		&  \sim& t^{\frac{n}{2}-1+p} [(- \frac{1}{t} \partial_z)^{n-3 + 3p}] [\mathfrak{L}_{1}(- e^{-z t})^n]
		\nonumber 
		 \eea
where here we just count the degree of homogeneity in derivatives and function $\mathfrak{L}_{1}$. Since the $p$ dependence in the power of $t$ behaves $t^{- 2p}$ we see that $p=0$ is the leading term for each fixed $n$ in the large $t$ limit. \\
		 
Until now our considerations where valid for general function $g(x)$. We now carry the explicit summation of all cumulants in the limit $t \to +\infty$ at fixed $z$, in the case of KPZ with droplet initial condition. 
		
		\begin{appKPZ}
		Using \eqref{cumconj} and \eqref{res1} we obtain for each cumulant
\bea
 \frac{\kappa_n}{n!} && \simeq \frac{t^{\frac{n}{2}-1}}{n!}  \kappa_n^0(-e^{-z t}) \\
 && \simeq
 (-1)^n \frac{2^{n-1}}{\Gamma(n+1) \pi^n} t^{\frac{n}{2}-1} (-\frac{1}{t} \partial_z)^{n-3} (-z t)^{n/2} 
 \\
&& = t^2 (-1)^n \frac{2^{n-1}}{\Gamma(n+1) \pi^n} 
\frac{\Gamma(\frac{n}{2}+1)}{\Gamma(4 - \frac{n}{2})} (-z)^{3 - \frac{n}{2}}
\eea 
and we note that all cumulants are proportional to $t^2$, which results from our
choice $\sigma=-e^{-z t}$, it is the only choice at large time which allows all cumulants to have the same homogeneity in the time variable. The summation over $n$ can then be explicitly performed, and we find
that at large time 
\be
\log Q_{t}(- e^{-z t}) \underset{t\gg 1}{=}  -t^2 \Phi_-(z) +\mathcal{O}(1)
\ee 
where $\Phi_-(z)$, $z<0$, is precisely the function obtained by completely different methods
in \cite{sasorov2017large} using a WKB analysis of a non-local Painleve equation
and in \cite{JointLetter} using a Coulomb gas method
\begin{equation}\label{eq:Phi}
\Phi_-(z)=\frac{4}{15\pi^6}(1-\pi^2 z)^{5/2}-\frac{4}{15\pi^6}+\frac{2}{3\pi^4}z -\frac{1}{2\pi^2}z^2
\end{equation}
This is a large deviation result for the left tail $H<0$ of the probability distribution of the KPZ height at large 
time \eqref{Plarget}, since for $t\gg 1$
\be 
\log Q_{t}(- e^{-z t}) = \mathbb{E}_{\rm KPZ}\left[ \exp\left(- e^{H-z t - \frac{1}{2} \log(4 \pi t)}\right) \right] \sim 
\mathbb{P}(\frac{H}{t} < z) \simeq  -t^2 \Phi_-(z)  
\ee
We recall that it predicts a crossover between the cubic tail of the Tracy Widom distribution
(matching the typical fluctuations regime) and 
$\Phi_{-}(z) \simeq_{z \to 0^-}\frac{1}{12} |z|^{3}$ and the $|H|^{5/2}$ outer left tail 
$\Phi_{-}(z) \simeq_{z \to -\infty} \frac{4}{15 \pi}  |z|^{5/2}$. Note that this
$|H|^{5/2}$ tail is identical (in power and prefactor) at short time, as seen
by the same cumulant method (see also 
\cite{Korshunov,MeersonParabola,le2016exact,KrajLedou2018}). 
\end{appKPZ}

We now turn to the Brownian initial condition, for which the cumulant
method is the only one presently to allow to obtain the next to leading 
term in the small time expansion. 

\subsection{Cumulants for KPZ with Brownian initial condition }
We proceed as in the previous section. For the Brownian initial condition
the propagator is given by Eq. \eqref{prop2}. We know, from the previous work on the 
Brownian initial condition \cite{krajenbrink2017exact}, 
where the leading order at small time was obtained, that one must scale
\be
w = \tilde w t^{-1/2} = \tilde w \gamma^{3/2}
\ee
to obtain a well defined short time limit with a continuously varying scaled drift
parameter $\tilde w$. We will study this limit here. Note also that it was found that the 
proper generating function is
\begin{equation}
Q_t(\sigma) = \mathbb{E}_{\, \mathrm{KPZ},\, B, \, \chi}\left[\exp\left(\frac{\sigma}{t} \rme^{H(t)+\chi}\right) \right] \end{equation}
which amounts to shift the height field by $\log t$. We now work with $g=g_{\rm KPZ}$.

\subsubsection{First cumulant evaluation}
We recall from section \ref{sec:expression} that the first cumulant is given by
\be
\label{1st_cum_brown}
 \kappa_1(g) =  \sum_{p=1}^{+\infty} \frac{(\sigma \gamma^3 )^{p}}{p} \,  \mathrm{Tr}(P_0 G_{p/\gamma}) 
\ee
The trace of the propagator \eqref{prop2} reads 
\begin{equation}
\begin{split}
\mathrm{Tr}(P_0 G_{p/\gamma})&=\int_{0}^{+\infty}\rmd r\, \int_{-\infty}^{+\infty}\frac{\rmd k}{2\pi}\, \frac{\Gamma(-i\gamma k+w-\frac{p}{2})\Gamma(i\gamma k+w-\frac{p}{2})}{\Gamma(-i\gamma k+w+\frac{p}{2})\Gamma(i\gamma k+w+\frac{p}{2})}\rme^{\frac{p^3}{12 \gamma^3}-\frac{p}{\gamma}(k^2+r)}\\
\end{split}
\end{equation}
One can simplify this trace by proceeding to the integration w.r.t $r$, further rescale $k$ by $\sqrt{\gamma}$ and introduce the rescaled drift $\tilde{w}$. This leads to an expression for the first cumulant \eqref{1st_cum_brown} valid at all orders in time.
\begin{equation}
\kappa_1(g)= \gamma^{3/2} e^{ \frac{(\sigma \partial_\sigma)^3}{12 \gamma^3} } \int_{-\infty}^{+\infty}\frac{\rmd k}{2\pi } \, \sum_{p=1}^{+\infty} \frac{(\sigma e^{-k^2} \gamma^3  )^{p}}{p^2} \frac{\Gamma(\gamma^{3/2} (\tilde w-ik)-\frac{p}{2})\Gamma(\gamma^{3/2}(\tilde w+i k)-\frac{p}{2})}{\Gamma(\gamma^{3/2} (\tilde w-ik)+\frac{p}{2})\Gamma(\gamma^{3/2} (\tilde w+ik)+\frac{p}{2})}
\end{equation}
The summation cannot be done per se, and we therefore employ Stirling's approximation to obtain an expansion of the four $\Gamma$ functions in powers of $\gamma^{-3/2}=\sqrt{t}$.
\begin{equation}
\frac{\Gamma(\gamma^{3/2} (\tilde w-ik)-\frac{p}{2})\Gamma(\gamma^{3/2}(\tilde w+i k)-\frac{p}{2})}{\Gamma(\gamma^{3/2} (\tilde w-ik)+\frac{p}{2})\Gamma(\gamma^{3/2} (\tilde w+ik)+\frac{p}{2})}
= \frac{\gamma^{-3p}}{(k^2 + \tilde w^2)^p} \left(1 + \frac{p \tilde w \gamma^{-3/2}}{(k^2 + \tilde w^2)} 
+ \mathcal{O}(\gamma^{-3}) \right)
\end{equation}
This allows to obtain the first two leading orders in the large $\gamma$ expansion of the
first cumulant
\begin{equation} \label{kappa1stat} 
\begin{split}
\kappa_1(g)&=\gamma^{3/2}  \int_{-\infty}^{+\infty}\frac{\rmd k}{2\pi} \, \sum_{p=1}^{+\infty} \left(\frac{\sigma e^{-k^2}  }{k^2 + \tilde w^2} \right)^{p} \frac{1}{p^2} \left(1 + \frac{p \tilde w \gamma^{-3/2}}{(k^2 + \tilde w^2)} + \mathcal{O}(\gamma^{-3}) \right)\\
& = \frac{1}{\sqrt{t}} \int_{-\infty}^{+\infty}\frac{\rmd k}{2\pi} \, 
{\rm Li}_2(\frac{\sigma \rme^{-k^2}}{ k^2 + \tilde w^2})  -  \int_{-\infty}^{+\infty}\frac{\rmd k}{2\pi} \, \frac{\tilde{w}}{k^2 + \tilde w^2} 
\log(1- \frac{\sigma \rme^{-k^2}}{ k^2 + \tilde w^2}) + \mathcal{O}(\sqrt{t}) \\
\end{split}
\end{equation}

\begin{remark}[Zero drift limit of the first cumulant]
\label{remark:zero_drift}
It was shown in \cite{krajenbrink2017exact} that the leading order of $\kappa_1$ with $\tilde{w}=0$ is well defined and can be written as a series of $\sqrt{-\sigma}$
\begin{equation}
\label{zero_drift_stat_short_time}
\int_{-\infty}^{\infty}\frac{\rmd k}{2\pi } \, \mathrm{Li}_2\Big(\frac{\sigma \rme^{-k^2}}{k^2}\Big)=\frac{1}{\sqrt{4\pi}}\sum_{n=1}^{+\infty}(-1)^n \frac{(-4\sigma)^{\frac{n}{2}}}{ \Gamma(n+1)}\Gamma\Big(\frac{n}{2}\Big)\Big(\frac{n}{2}\Big)^{\frac{n-3}{2}}
\end{equation}
From this the rate function $\Phi(H)$ was obtained for the stationary case $\tilde w=0$. 
To the next order, the limit $\tilde w=0$ is more tricky will be analyzed below. 
\end{remark}

\subsubsection{Second cumulant evaluation}
We now study the second cumulant, and we will restrict to its leading order in
the small time expansion. We recall the formula \eqref{secondc} 
\begin{equation} \label{22} 
\begin{split}
 \kappa_2(g)=  \sum_{p_1,p_2 = 1}^{+\infty}
\frac{(\sigma \gamma^3)^{p_1+p_2}}{p_1 p_2}  \mathrm{Tr}(P_0 G_{p_1/\gamma} (1-P_0) G_{p_2/\gamma})
\end{split}
\end{equation}
We only need here the leading order term of the propagator (see previous section)
\bea
G_{p/\gamma}(r_1,r_2) \simeq \rme^{-\frac{p}{\gamma}\frac{r_1+r_2}{2}}
\gamma^{\frac{1}{2}-3p}\int_{-\infty}^{+\infty}\frac{\rmd k}{2\pi}  \frac{\rme^{-pk^2-i\gamma^{1/2	}k(r_1-r_2)}}{(k^2 + \tilde w^2)^p} 
\eea
One has, after rescaling $(r_1,r_2) \to \gamma^{-1/2} (r_1,r_2)$,
\bea
&& \mathrm{Tr}(P_0 G_{p_1/\gamma} (1-P_0) G_{p_2/\gamma})  \\
&&\simeq  \frac{1}{4 \pi^2} \gamma^{-3(p_1+p_2)} \int_{\mathbb{R}^+} \mathrm{d}r_1 \int_{\mathbb{R}^-} \mathrm{d}r_2
\int_{\mathbb{R}} \mathrm{d}k_1 \int_{\mathbb{R}} \mathrm{d}k_2\, 
\rme^{-\frac{p_1+p_2}{\gamma^{3/2}}\frac{r_1+r_2}{2}}
 \frac{\rme^{-p_1 k_1^2 - p_2 k_2^2 -i
(k_1-k_2) (r_1-r_2)}}{(k_1^2 + \tilde w^2)^{p_1} (k_2^2 + \tilde w^2)^{p_2} } \nonumber \\
&& = \frac{1}{4 \pi^2} \gamma^{-3(p_1+p_2)}  \int_0^{+\infty}  \mathrm{d}u \, u 
\int_{\mathbb{R}} \mathrm{d}k_1 \int_{\mathbb{R}} \mathrm{d}k_2
 \frac{\rme^{-p_1 k_1^2 - p_2 k_2^2 -i
(k_1-k_2) u}}{(k_1^2 + \tilde w^2)^{p_1} (k_2^2 + \tilde w^2)^{p_2} } \label{calc1} 
\eea
where in the last line, to this order, we have discarded (i.e. set to unity) the term 
$\rme^{-\frac{p_1+p_2}{\gamma^{3/2}}\frac{r_1+r_2}{2}}$,
performed the
change of variable in \eqref{changevariable}, and integrated 
over $v$. We can insert in \eqref{22} and sum over $p_1,p_2$ to obtain
\begin{equation} \label{kappa2stat} 
\begin{split}
& \kappa_2(g) = \frac{1}{4 \pi^2}  \int_0^{+\infty} \mathrm{d}u \, u \, |\hat f(u)|^2 \\
& \hat f(u) =-  \int_{\mathbb{R}} \mathrm{d} k\,  \rme^{i k u} \, \log(1-\frac{\sigma \rme^{-k^2}}{k^2+ \tilde w^2 })\\
\end{split}
\end{equation}
\subsubsection{Summary}

We can now put together the previous calculations and obtain the two leading orders 
for the Brownian initial condition. Recalling that $\gamma=t^{-1/3}$ we obtain the following expansion
\bea
\label{G short time expansion stat t0}
 Q_t(\sigma) = \log \mathbb{E}_{\rm KPZ} \left[ \exp \left(\frac{\sigma}{t} \rme^{H(t)+\chi}  \right) \right]
&=&  \kappa_1(g) + \frac{1}{2} \kappa_2(g) +\dots  \\
&=& -  \frac{1}{\sqrt{ t}} \Psi(\sigma) +\Psi_0(\sigma) + \mathcal{O}(\sqrt{t}) \nonumber
\eea
with
\bea \label{PsiPsi0} 
 \Psi(\sigma) &=& -   \int_{-\infty}^{+\infty}\frac{\rmd k}{2\pi} \, 
{\rm Li}_2(\frac{\sigma \rme^{-k^2}}{k^2 + \tilde w^2}) \\
 \Psi_0(\sigma)  &=& -  \int_{-\infty}^{+\infty}\frac{\rmd k}{2\pi} \, \frac{\tilde{w}}{k^2 + \tilde w^2} 
\log(1- \frac{\sigma \rme^{-k^2}}{k^2 + \tilde w^2}) \nonumber\\
&& \hspace*{1cm}+ \frac{1}{8 \pi^2}  \int_0^{+\infty} \mathrm{d} u \, u \, \bigg|\int_{\mathbb{R}} \mathrm{d}k \,  \rme^{i k u} \, \log(1-\frac{\sigma \rme^{-k^2}}{k^2+ \tilde w^2}) \bigg|^2\nonumber
\eea 

The leading order recovers the result of \cite{le2016exact}
in an equivalent form, indeed one has
\be \label{id2}
\int_{-\infty}^{+\infty}\frac{\rmd k}{2\pi} \, 
{\rm Li}_2(\frac{\sigma \rme^{-k^2}}{k^2 + \tilde w^2}) 
= - \frac{1}{\pi} \int_0^{+\infty} \mathrm{d}y \,  (1 + \frac{1}{y+ \tilde w^2}) \sqrt{y} 
\log(1 - \frac{\sigma \rme^{-y}}{y+ \tilde w^2}) 
\ee
which is easily checked setting $y=k^2$ and performing
an integration by part. The r.h.s. of \eqref{id2} is precisely the
result  Eq.(16) of \cite{le2016exact}.  One can check that for $\tilde w \to +\infty$ one recovers the result for the
droplet initial condition. The limit $\tilde w \to 0$ is discussed in the next
subsection.

\subsubsection{Height distribution $P(H,t)$ at short time}
\label{subsubsec:Pstat} 

Let us proceed as in \cite{krajenbrink2017exact}, Supp. Mat. Section 4.1, and obtain the next order
in the short time expansion. We consider the leading behavior for fixed $\tilde w$, which implies that $w=\tilde w/\sqrt{t}$ 
is large. We perform the shift $\chi \to \chi + \ln (\sqrt{t})$, use Stirling's formula for $\Gamma(2w)$ factor 
in the PDF of $\chi$ given below Eq. \eqref{KAi}, and write, using the form
\eqref{expansionP} for $P(H,t)$
\begin{equation} \label{rep1} 
\begin{split}
&\mathbb{E}_{\mathrm{KPZ}}\left[  \exp \left(  \frac{\sigma e^H e^\chi}{t} \right) \right] =  {\cal C}(t) \int_{-\infty}^{+\infty} \mathrm{d} \chi \int_{-\infty}^{+\infty} \mathrm{d} H   \\
& \times  
\exp \left( - \frac{\Phi(H)+2 \tilde w \chi + e^{-\chi} -\sigma e^H e^{\chi}-2\tilde{w}+2\tilde{w}\ln(2\tilde{w})}{\sqrt{t}}+\Phi_0(H) -\frac{1}{2}\log(\frac{2\pi}{2w}) + \mathcal{O}(\frac{\sqrt{t}}{\tilde w}) \right)
\end{split}
\end{equation}
One first proceeds to the saddle point on $\chi$ at fixed $H$ and to that aim define 
\begin{equation}
\Xi(\chi)=2\tilde{w}\chi +e^{-\chi}-\sigma e^H e^\chi
\end{equation}
take $\chi_H$ to be the critical point, such that $\Xi'(\chi_H)=0$, then we have using the
saddle point method 
\begin{equation}
\begin{split}
&\mathbb{E}_{\mathrm{KPZ}}\left[  \exp \left(  \frac{\sigma e^H e^\chi}{t} \right) \right] = {\cal C}(t) \\
& \times \int_{-\infty}^{+\infty} \mathrm{d} H  
\exp \left( - \frac{\Phi(H)+\Xi(\chi_H)-2\tilde{w}+2\tilde{w}\ln(2\tilde{w})}{\sqrt{t}}+\Phi_0(H) +\frac{1}{2}\log(\frac{2\tilde{w}}{\Xi''(\chi_H)}) + \mathcal{O}(\sqrt{t}) \right)
\end{split}
\end{equation}
The critical point and the second derivative are given by
\begin{equation}
e^{\chi_H}=\frac{  \sqrt{ \tilde w^2 -\sigma e^H} - \tilde w}{-\sigma e^H}
\quad , \quad \Xi''(\chi_H)=2\sqrt{\tilde{w}^2-\sigma e^H}
\end{equation}
which leads to formula (79) in \cite{krajenbrink2017exact}, namely
\begin{equation}
\Phi(H) = \max_{\sigma \in ]-\infty,\tilde w^2]} \left( \Psi(\sigma) - 2 \sqrt{ \tilde w^2 -\sigma e^H}  + 2 \tilde w - 2 \tilde w \ln(2 \tilde w) 
+ 2 \tilde w \ln(\tilde w + \sqrt{ \tilde w^2 -\sigma e^H}) \right) \label{Phi200} 
\end{equation}
and comparing with \eqref{G short time expansion stat t0} we
obtain at fixed $\sigma$
\begin{equation}
\begin{split}
\Phi_0(H)+\frac{1}{2}\log(\frac{\tilde{w}}{\sqrt{\tilde{w}^2-\sigma e^H}}) -\frac{1}{2}\log(\frac{\mathrm{d}^2}{\mathrm{d}H^2} \Xi(\chi_H)+\Phi''(H))
= \Psi_0(\sigma)
\end{split}
\end{equation}
with ${\cal C}(t)=1/\sqrt{2 \pi \sqrt{t}}$. Here 
$\Psi$ and $\Psi_0$ are given explicitly in \eqref{PsiPsi0}.
The evaluation of the second derivative gives 
$\frac{d^2}{dH^2} \Xi(\chi_H)
=-  \frac{\sigma e^H}{2 \sqrt{\tilde w^2- \sigma e^H}}$.
This gives
\be \label{Phi0stat} 
\Phi_0(H) = \Psi_0(\sigma) - \frac{1}{2}\log \tilde{w}+ \frac{1}{2}\log(-  \frac{\sigma e^H}{2}+\Phi''(H) \sqrt{\tilde w^2- \sigma e^H})
\ee
which must be complemented with the relation between $\sigma$, $H$, $\Phi(H)$ in the parametric representation \cite{krajenbrink2017exact} 
\begin{equation}
\label{parametric}
                 e^{H}= -\sigma \Psi'(\sigma )^2-2\tilde{w}\Psi'(\sigma) \quad , \quad 
                \Phi'(H)=-\sigma\Psi'(\sigma)
\end{equation}

Note that the above results are valid for $H < H_c(\tilde w)$ 
with
\be \label{Hcw} 
H_c(\tilde w) = \log \left(  - \tilde w^2 \Psi'(\tilde w^2 )^2-2\tilde{w}\Psi'(\tilde w^2) \right)
\ee
and we will not attempt here to extend the analysis of \cite{krajenbrink2017exact}
for $\Phi(H)$ (which involves two successive analytical continuations
one exhibiting a phase transition) 
to obtain the expression for $H > H_c(\tilde w)$ for $\Phi_0(H)$. It 
would be interesting to see the signature of the phase transition on
the subleading rate function $\Phi_0(H)$.

\medskip

{\bf Stationary limit}

Let us now discuss the stationary KPZ limit $\tilde w \to 0$. The leading term $\Phi(H)$ is well defined 
in this limit as was discussed in \cite{krajenbrink2017exact}. Indeed, $\Psi(\sigma)$ has the
following limit
\bea \label{Psistat} 
 \Psi(\sigma) &=& -   \int_{-\infty}^{+\infty}\frac{\rmd k}{2\pi} \, 
{\rm Li}_2(\frac{\sigma \rme^{-k^2}}{k^2}) 
\eea
see remark \ref{remark:zero_drift}, and $\Phi(H)$ is thus determined by
the parametric equation
\begin{equation}
\label{parametric2}
                 e^{H}= -\sigma \Psi'(\sigma )^2 \quad , \quad 
                \Phi'(H)=-\sigma\Psi'(\sigma)
\end{equation}
for $H<H_c$. The continuations for $H>H_c$ are obtained in \cite{krajenbrink2017exact}. We now examine the limit of the subleading rate function $\Phi_0(H)$. It is given formally by 
\be 
\Phi_0(H) = \tilde \Psi_0(\sigma) + \frac{1}{2}\log(-  \frac{\sigma e^H}{2}+\Phi''(H) \sqrt{- \sigma e^H})
\ee
where
\be \label{limPhi0} 
 \tilde \Psi_0(\sigma) = \lim_{\tilde w \to 0} \Psi_0(\sigma) - \frac{1}{2}\log \tilde{w}
\ee
with the same relation \eqref{parametric2} between $\sigma$ and $H$. We now examine $\Psi_0(\sigma)$ as given by \eqref{PsiPsi0}
and show that the limit \eqref{limPhi0} exists, which is non-trivial.
The first term in \eqref{PsiPsi0} behaves when $\tilde w \to 0$ as
\begin{equation}
-  \int_{-\infty}^{+\infty}\frac{\rmd k}{2\pi} \, \frac{\tilde{w}}{k^2 + \tilde w^2} 
\log(1- \frac{\sigma \rme^{-k^2}}{ k^2 + \tilde w^2}) = \frac{1}{2}\log ( -\frac{4\tilde{w}^2}{\sigma} )+\mathcal{O}(\tilde{w})
\end{equation}
where the $\mathcal{O}(\tilde w)$ term is given by
\begin{equation}
\begin{split}
 \int_{-\infty}^{+\infty}\frac{\rmd k}{2\pi} \, \frac{1}{k^2 + 1} 
\log(  \rme^{-\tilde{w}^2 k^2} -\frac{(k^2+1)\tilde{w}^2}{\sigma} ) &= \int_{-\infty}^{+\infty}\frac{\rmd k}{2\pi} \, \frac{\tilde{w}}{k^2 + \tilde{w}} 
\log(  \rme^{- k^2} -\frac{(k^2+\tilde{w}^2)}{\sigma} ) \\
&= \tilde{w} \int_{-\infty}^{+\infty}\frac{\rmd k}{2\pi} \, \frac{1}{k^2} 
\log(  \rme^{- k^2} -\frac{k^2}{\sigma} ) +\mathcal{O}(\tilde{w}^2)
\end{split}
\end{equation}
We thus see that there is a logarithmic divergence as $\tilde w \to 0$
in the first term. It does not exactly cancel the $-\frac{1}{2} \ln(\tilde w)$ in
the definition of $\tilde \Phi_0$, hence we hope that the second term in
\eqref{PsiPsi0} will cancel the remaining divergence. We now show that the cancellation does occur, i.e. the second term
is $\simeq - \frac{1}{2} \ln(\tilde w)$. For this let us call $F(\sigma,\tilde w)$ 
this second term and apply
$\tilde w \partial_{\tilde w}$ on the second term. After one
integration by part we obtain
\bea
\tilde w \partial_{\tilde w} F(\sigma,\tilde w) && = - \frac{1}{\pi^2} \int_0^{+\infty} \mathrm{d}u \big( \int_{-\infty}^{+\infty} \frac{\mathrm{d}k}{k^2 + \tilde w^2} e^{i k u} \frac{\sigma}{(k^2 + \tilde w^2)e^{k^2}-\sigma} \big) \\
&& \times \big( \int_{-\infty}^{+\infty} \mathrm{d}p \, p \sin(p u) 
 \frac{\sigma (1 + \frac{1}{p^2 + \tilde w^2})  }{(p^2 + \tilde w^2)e^{p^2}-\sigma} \big) 
\eea
We are interested in the limit $\tilde w \to 0$. For this we use the following rescaling
$k \to \tilde w k$, $p \to \tilde w p$, $u \to u/\tilde w$.
In the limit $\tilde w \to 0$ with fixed $\sigma$ we obtain
\bea
\tilde w \partial_{\tilde w} F(\sigma,\tilde w) = - \frac{1}{\pi^2} \int_0^{+\infty} \mathrm{d}u \big( \int_{-\infty}^{+\infty} \frac{\mathrm{d}k}{k^2 + 1} e^{i k u} \big) \big( \int_{-\infty}^{+\infty} \mathrm{d}p \,  \sin(p u) 
 \frac{p}{p^2 + 1} \big) = - \frac{1}{2}
\eea
which shows the cancellation. 

Calculating the derivative $\partial_\sigma F(\sigma,\tilde w) $ 
we obtain after integration by part and taking the limit $\tilde w=0$
we obtain the derivative of the limit
\be
\tilde \Psi'_0(\sigma) \simeq  \frac{1}{2 \pi^2} 
\int_0^{+\infty} \mathrm{d}u 
\big( \int_{-\infty}^{+\infty} \mathrm{d}k e^{i k u} \frac{1}{k^2 e^{k^2} - \sigma} \big) 
\big( \int_{-\infty}^{+\infty} \mathrm{d}p  \sin(p u) 
 \frac{ \sigma p (1+ \frac{1}{p^2}) }{p^2 e^{p^2} - \sigma} \big) 
\ee
from which one can obtain $\Phi'_0(H)$ using the above equations. 

\subsubsection{Application to the limit $t \to +\infty$}
\label{subsec:statlargetime}

Let us now evaluate the first two cumulants in the large $t$ limit 
setting $\sigma = -e^{-z t}$ as in Section \ref{sec:structure}.
From Eq. \eqref{kappa1stat} we find
\be
\begin{split}
\kappa_1(g) &= t^2 \frac{4}{15 \pi} (-z)^{5/2}  -  \int_{\mathbb{R}} \frac{\mathrm{d} k}{2\pi} \frac{\tilde w}{k^2 + \tilde w^2}
(-z-k^2)_+ \\
&=t^2 \frac{4}{15 \pi} (-z)^{5/2} -\frac{\left(\tilde{w} ^2-z\right) \arctan \left(\frac{\sqrt{-z}}{\tilde{w}}\right)-\tilde{w}    \sqrt{-z}}{\pi }\\
   \end{split}
\ee 
The first term $\mathcal{O}(t^2 |z|^{5/2})$ is the leading behavior of the rate function $\Phi^{\rm Brownian}_-(z)$,
which thus has the same leading behavior (for any $\tilde w$) that $\Phi^{\rm droplet}_-(z)=\Phi_-(z)$.
The second term is a $\mathcal{O}(1)$ correction at large $t$, which furthermore has the finite
limit $z/2$ as $\tilde w \to 0$. 

For the second cumulant \eqref{kappa2stat}, we find,
replacing at large time $ \log(1-\frac{ \rme^{-k^2-z t}}{k^2+ \tilde w^2 }) \simeq (- z t - k^2)_+$
and rescaling $k$ and $u$
\be
\kappa_2(g) = \frac{t^2 z^2}{4 \pi^2} \int_0^{+\infty} \mathrm{d} u \, u |\int_{-1}^{1} \mathrm{d} k \, e^{i k u} (k^2-1) |^2 = \frac{t^2 z^2}{\pi^2} 
\ee
which recovers exactly the subleading behavior at large $|z|$ of $\Phi_-(z)$,
hence is in agreement with the conjecture $\Phi^{\rm stat}(z)=\Phi^{\rm drop}(z)$
based on Coulomb gas arguments \cite{inprep2}.

\section{Comparison with stationary large deviations of TASEP} 

In this section, we point out unexpected similarities between short time expansions for the KPZ equation defined on an infinite or semi-infinite line and stationary large deviations of TASEP in a periodic or open domain.

\subsection{Leading terms at short time for the KPZ equation}
We consider the solution of the KPZ equation on the semi-infinite line $\mathbb{R}^{+}$ with sharp wedge initial condition at $0$ and prescribed boundary density $\partial_x h(0,t)=\rho$. In both cases, the generating function $Q_{t}(\sigma)$ of \eqref{G sw} (with some specific definition of the shifted height $H(t)$, see \cite{krajenbrink2018large} ) has the short time behaviour $\log Q_{t}(\sigma)\simeq-\Psi(\sigma)/\sqrt{t}$. Exact formulas for $\rho=-1/2,0,+\infty$ \cite{krajenbrink2018large} seem to indicate that the function $\Psi$ only depends on whether the boundary density $\rho$ is finite or not. For finite density (reflective wall), one has \cite{krajenbrink2018large}
\begin{equation}
\label{Psi reflective wall}
\Psi(\sigma)=-\frac{1}{4\sqrt{\pi}}\sum_{n=1}^{+\infty}\frac{\sigma^n}{n^{\frac{5}{2}}}=-\frac{\mathrm{Li}_{5/2}(\sigma)}{4\sqrt{\pi}}=-\int_{-\infty}^{\infty}\frac{\rmd k}{4\pi} \, \mathrm{Li}_2(\sigma \rme^{-k^2})\;,
\end{equation}
which is the same as for the droplet on the full line up to a factor $2$, while when $\rho\to+\infty$ (hard wall), one finds instead \cite{krajenbrink2018large}
\begin{equation}
\label{Psi hard wall}
\Psi(\sigma)=-\frac{1}{4\sqrt{\pi}}\sum_{n=1}^{+\infty} \frac{(2n)!}{n!n^{n+\frac{5}{2}}}\Big(\frac{\sigma}{4}\Big)^n=-\int_{-\infty}^{\infty}\frac{\rmd k}{4\pi} \, \mathrm{Li}_2(\sigma k^2 \rme^{-k^2})\;.
\end{equation}

\subsection{Stationary large deviations for periodic and open TASEP}

The totally asymmetric simple exclusion process (TASEP) is a microscopic model of driven hard core particles which belongs to KPZ universality. We summarize here some known results for the stationary large deviations of the corresponding height field.

The first result for stationary large deviations of TASEP was obtained by Derrida and Lebowitz \cite{DerridaLebowitz} in the case of a system with periodic boundaries. In terms of the KPZ fixed point with periodic boundaries, which describes a random field $h(x,t)$, $x\equiv x+1$ obtained as the limit of strong non-linearity of the solution of the KPZ equation, the result can be stated in a parametric form as (see \textit{e.g.} \cite{mallick2018brownian})
\begin{equation}
	\label{LDF TASEP periodic leading}
	\log\mathbb{E}_{\, \mathrm{KPZ}}[\rme^{sh(x,t)}]\underset{t\to\infty}{\simeq}-t\,\frac{\Li_{5/2}(\sigma)}{\sqrt{2\pi}}
\end{equation}
with
\begin{equation}
	\label{s[sigma] TASEP periodic}
	s=-\frac{\Li_{3/2}(\sigma)}{\sqrt{2\pi}}\;.
\end{equation}
The result is the same for any initial condition. It is also independent of $x$ by translation invariance of the stationary state.

Subleading corrections to \eqref{LDF TASEP periodic leading} depend on the initial condition. They have been studied in \cite{P2016} for special cases, see also \cite{mallick2018brownian}. For periodic sharp wedge initial condition, one has in particular
\begin{equation}
\label{LDF TASEP periodic subleading}
\log\mathbb{E}_{\, \mathrm{KPZ}}[\rme^{sh(x,t)}]\underset{t\to\infty}{\simeq}-t\,\frac{\Li_{5/2}(\sigma)}{\sqrt{2\pi}}+\int_{0}^{\sigma}\frac{\rmd u}{u}\,\Li_{1/2}(u)^{2}+\log\frac{\Li_{3/2}(\sigma)}{\Li_{1/2}(\sigma)}\;.
\end{equation}
up to exponentially small corrections in time, with $\sigma$ still defined as in \eqref{s[sigma] TASEP periodic}.

\begin{figure}[h!]
	\begin{picture}(60,60)
	\put(6,4){\vector(1,0){60}}
	\put(6,4){\vector(0,1){60}}
	\put(36,3){\line(0,1){2}}
	\put(5,34){\line(1,0){2}}
	\put(33.6,5.5){$0.5$}
	\put(-1,32.5){$0.5$}
	\put(63,6){$\rho_{a}$}
	\put(-6,60){$1-\rho_{b}$}
	\put(6,4){\thicklines\line(1,1){30}}
	\put(36,34){\thicklines\line(1,0){30}}
	\put(36,34){\thicklines\line(0,1){30}}
	\put(13,45){
		\begin{tabular}{c}
		Low\\[-3mm]
		density
		\end{tabular}
	}
	\put(35,20){
		\begin{tabular}{c}
		High\\[-3mm]
		density
		\end{tabular}
	}
	\put(42,49){
		\begin{tabular}{c}
		Maximal\\[-3mm]
		current
		\end{tabular}
	}
	\end{picture}
	\caption{Phase diagram of TASEP with open boundaries connected to reservoirs of particles $A$ and $B$ at respective densities $\rho_{a}$, $\rho_{b}$. The particles move in the direction from $A$ to $B$. The average density of particles $\rho$ in the stationary state is equal to $\rho_{a}$ in the low density phase, $\rho_{b}$ in the high density phase and $1/2$ in the maximal current phase. The corresponding average current is $\rho(1-\rho)$. The transitions from the maximal current phase to either the low or high density phases are second order, while the transition between the low and high density phases (shock line) is first order.}
	\label{fig phase diagram open TASEP}
\end{figure}
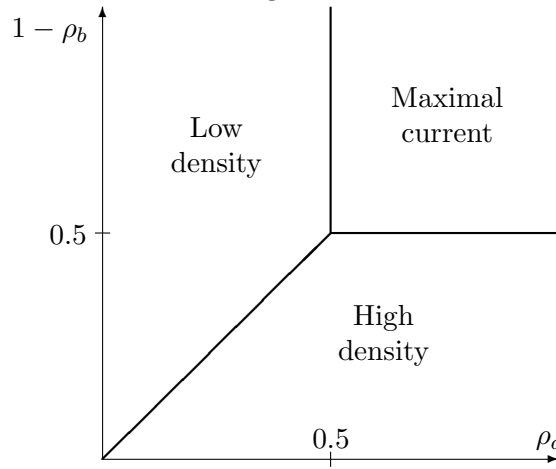

Stationary large deviations for open TASEP connected to reservoirs of particles at both ends have been obtained by Lazarescu and Mallick in \cite{LazarescuMallick}, see also \cite{LazarescuMallick2,Lazarescu,CrampeNepomechie}. The results depend on the density of particles $\rho_{a}$, $\rho_{b}$ in the left and right reservoir (with particles hopping from left to right) according to the phase diagram of the model drawn in figure \ref{fig phase diagram open TASEP}. Only the leading term in time of the large deviations is known so far.

The results below are given in terms of the random field $h(x,t)$ characterizing the KPZ fixed point normalized in the same way as in the periodic case, and with $\partial_{x}h(0,t)$ and $\partial_{x}h(1,t)$ prescribed by the boundary densities. Two cases are of particular interest for comparison with large deviations at short time for the KPZ equation on $\mathbb{R}^{+}$: the edge and the bulk of the maximal current phase.

At the transition between the low density and the maximal current phase, corresponding formally to $\partial_{x}h(0,t)=-1/2$ and $\partial_{x}h(1,t)=-\infty$ \cite{CorwinShen,Parekh}, the generating function is essentially the same as for the periodic system,
\begin{equation}
\label{LDF TASEP open MC/LD}
\log\mathbb{E}_{\, \mathrm{KPZ}}[\rme^{sh(x,t)}]\underset{t\to\infty}{\simeq}-t\,\frac{\Li_{5/2}(\sigma)}{4\sqrt{4\pi}}
\end{equation}
with
\begin{equation}
	\label{s[sigma] TASEP open MC/LD}
	s=-\frac{\Li_{3/2}(\sigma)}{2\sqrt{4\pi}}\;.
\end{equation}
The result is the same at the transition between the high density and the maximal current phase, where $\partial_{x}h(0,t)=+\infty$ and $\partial_{x}h(1,t)=1/2$.

In the maximal current phase $0\leq\rho_{b}<\frac{1}{2}<\rho_{a}\leq1$, corresponding formally to $\partial_{x}h(0,t)=+\infty$ and $\partial_{x}h(1,t)=-\infty$, the generating function has the parametric expression
\begin{equation}
\label{LDF TASEP open MC}
\log\mathbb{E}_{\, \mathrm{KPZ}}[\rme^{sh(x,t)}]\underset{t\to\infty}{\simeq}-\frac{t}{8\sqrt{\pi}}\,\sum_{n=1}^{\infty}\frac{(2n)!}{n!n^{n+5/2}}\,\sigma^{n}
\end{equation}
with 
\begin{equation}
\label{s[sigma] TASEP open MC}
s=-\frac{1}{8\sqrt{\pi}}\sum_{n=1}^{\infty}\frac{(2n)!}{n!n^{n+3/2}}\,\sigma^{n}\;.
\end{equation}

\subsection{Hints of a duality for the KPZ equation in finite volume}
We discuss now the similarities between large deviations at short time for the KPZ equation and at long time for TASEP. We consider separately the case of systems with and without boundaries.

\subsubsection{No boundary}
The similarity between the expression \eqref{G short time expansion sw t1/2}, \eqref{F short time expansion sw t0}, \eqref{s[sigma]} for the KPZ equation on $\mathbb{R}$ with droplet initial condition at short time and the expressions \eqref{s[sigma] TASEP periodic}, \eqref{LDF TASEP periodic subleading} for periodic TASEP with sharp wedge initial condition is striking. This was already noted in \cite{le2016exact} (see Supp. Mat. Section 5) for the leading order, but here we see that the similarity persists to the next order. This similarity suggests the existence of a duality for the KPZ equation with periodic boundaries between large deviations around the Edwards-Wilkinson and the KPZ fixed point. Indeed, the KPZ equation on $\mathbb{R}$ with sharp wedge initial condition is expected to have the same short time behaviour as the KPZ equation with periodic boundaries and sharp wedge initial condition since the correlation length is much smaller than the system size at short time. Thus, \eqref{G short time expansion sw t1/2}, \eqref{F short time expansion sw t0} should also hold for the KPZ equation with periodic boundaries. On the other hand, as mentioned above, \eqref{LDF TASEP periodic subleading} characterizes large deviations at long time for the KPZ fixed point with periodic boundaries. The similarity between \eqref{G short time expansion sw t1/2}, \eqref{F short time expansion sw t0} and \eqref{LDF TASEP periodic subleading} thus hints at the existence of a duality for the KPZ equation with periodic boundaries, at least at the level of large deviations.

\subsubsection{Boundary with fixed density}
Similar observations hold for the KPZ equation on the interval $[0,1]$ with prescribed density $\partial_{x}h(0,t)=\rho$ at the left boundary and infinite density $\partial_{x}h(0,t)=-\infty$ at the right boundary. The striking resemblance between \eqref{LDF TASEP open MC/LD} and \eqref{Psi reflective wall} for finite $\rho$ and between \eqref{LDF TASEP open MC} and \eqref{Psi hard wall} for $\rho\to\infty$ again suggest the existence of a duality for the large deviations around the Edwards-Wilkinson and the KPZ fixed point of the KPZ equation on an open interval.

\section{Application to the linear statistics of the Airy point process (edge of the GUE)}
\label{sec:appGUE} 

In this Section we apply the general expansion formula obtained in Sections \ref{section sw systematic} and 
\ref{sec:cumulants} to obtain 
information about (i) typical fluctuations, from the cumulant expansion and (ii) large deviations
of the linear statistics of the Airy point process, i.e. of the GUE eigenvalues at the edge as
explained in the Introduction.

\subsection{Linear statistics at the edge of the GUE: cumulant expansion at the level of the Gaussian free field and beyond}

Consider the following linear statistics problem for the Airy point process, i.e. we are interested
in the average 
\be \label{linstat1} 
Q_t(\sigma) = \mathbb{E}_{\Ai}\left[ \exp\left( \sum_{i=1}^\infty f(t^{1/3} \mathbf{a}_i)\right) \right]
=  \mathbb{E}_{\Ai} \left[\exp\left(\int_{\mathbb{R}} \mathrm{d}a \, g(\sigma e^{t^{1/3}a})\mu(a) \right)\right]
\ee
over the Airy point process, for a large class of functions $f$. To relate to our previous results 
we write $f(a)=g(\sigma e^{a})$ ($\sigma$ being here just a convenient parameter). We have defined
\begin{equation} \label{defH} 
\mu(a)=\sum_{i=1}^{+\infty} \delta(a-\mathbf{a}_i) \qquad 
\mathsf{H}(a)=-\int_a^{+\infty}\mu(a')\mathrm{d}a'
\end{equation}
respectively the empirical measure $\mu$, and the height field $\mathsf{H}$ of the Airy point process.
The height field counts the number of points above a certain level with $\mathsf{H}'(a)=\mu(a)$. Eq. \eqref{linstat1} can be written in the form of an expansion in the cumulants of the empirical
measure as 
\bea \label{eq2}
&& \mathbb{E}\left[\exp\left(\int_\mathbb{R} \mathrm{d}a \, g(\sigma e^{t^{1/3}a})\mu(a) \right)\right]=\exp\left(\sum_{n=1}^{+\infty} \frac{1}{n!} \int_\mathbb{R} \mathrm{d}a_1 \dots \mathrm{d}a_n
\prod_{j=1}^n g(\sigma e^{t^{1/3}a_j})
\overline{\mu(a_1) \dots \mu(a_n)}^c  \right) \nonumber \\
&& =\exp\left(\sum_{n=1}^{+\infty} \frac{1}{n!} \int_\mathbb{R} \mathrm{d}a_1 \dots \mathrm{d}a_n
\prod_{j=1}^n g(\sigma e^{a_j}) t^{-n/3}
\overline{\mu(t^{-1/3} a_1) \dots \mu(t^{-1/3} a_n)}^c  \right)
\eea
where here the $a_i$ are just integration variables, not the points of the Airy point process. On the other hand
\eqref{linstat1}-\eqref{eq2} are nothing but our generating function $Q_t(\sigma)$, and from
\eqref{G sw2} and \eqref{deter_process} we see that the cumulants of the empirical measure
can be extracted from the cumulants defined in Section \ref{sec:cumpres}, since
\bea \label{mucum} 
\int_\mathbb{R} \mathrm{d}a_1\dots \mathrm{d}a_n
\prod_{j=1}^n g(\sigma e^{a_j}) t^{-n/3}
\overline{\mu(t^{-1/3} a_1) \dots \mu(t^{-1/3} a_n)}^c = \kappa_n(g)
\eea
The short time series expansion for the cumulants $\kappa_n(g)$ were
obtained in \eqref{cum}, where we recall that the $\mathfrak{L}_{i}$ have the 
explicit expression
\begin{equation} \label{Linew} 
\mathfrak{L}_{i}=(\sigma \partial_\sigma)^{i}
\int_{-\infty}^{+\infty}\frac{\mathrm{d}p}{2\pi}\, g(\sigma e^{-p^2}) = \frac{1}{2 \pi} 
\int_{-\infty}^{+\infty} \mathrm{d}a \frac{\theta(-a)}{\sqrt{-a}} f^{(i)}(a)
\end{equation}
Explicit comparison of \eqref{cum} and \eqref{mucum} for a generic fonction $g$ (i.e. for a generic $f$)
allows, by identification, to read off the expansion of the scaled cumulant 
$t^{-n/3} \overline{\mu(t^{-1/3} a_1) \dots \mu(t^{-1/3} a_n)}^c$ in powers of $t^{1/2}$ at small $t$.
This expansion clearly corresponds to studying each cumulant in the large negative $a$ region of the Airy point process,
i.e. starting from the matching region with the bulk of the GUE, and expanding towards the edge.
This was the purpose of inserting a power $t^{1/3}$ in the argument of $f$ in \eqref{linstat1}.

Note that from our guess \eqref{guess1} for the general structure of the cumulants
we can conjecture that the proper form of the series expansion of the scaled cumulants 
is, schematically
\be
t^{1 - \frac{5 n}{6}} \overline{\mu(t^{-1/3} a_1) \dots \mu(t^{-1/3} a_n)}^c = \text{series ~ of~ the~ form} ~ \sum_{p\geq 0} \delta_p(a_1,.. a_n) t^p
\ee
where the $\delta_p$ are distributions, some of them discussed below. 

\begin{remark}
We note that the leading term for the first and second cumulants, i.e. $\kappa_1$ 
at order $t^{-1/2}$ and $\kappa_2$ at order $t^0$ in \eqref{cum}, 
have previously been obtained by Basor and Widom \cite{basor1999determinants}
for arbitrary $g(x)$, by quite different methods. Here we obtain the terms in the expansion to a much higher order. While the $\mathcal{O}(t^{-1/2})$ term for $\kappa_1$ readily coincide, we obtain the
$\mathcal{O}(t^0)$ term of $\kappa_2$ in \eqref{cum}, i.e. $\kappa_2^0$, in the form
\begin{equation}
\label{g_exp_lhs}
 \kappa_2^0 = 2 \int_0^\sigma \frac{\mathrm{d}u}{u} \mathfrak{L}_1^2=\frac{1}{2\pi^2} \int_0^\sigma \mathrm{d}u \, u \left[ \partial_u \int_{-\infty}^{+\infty}\mathrm{d}p\, g(u e^{-p^2})\right]^2 
\end{equation}
However, it can be rewritten equivalently in the form
obtained in \cite{basor1999determinants} as
\begin{equation}
\label{g_exp_rhs}
\kappa_2^0 =  2 \int_0^\sigma \frac{\mathrm{d}u}{u} \mathfrak{L}_1^2 = \frac{1}{4\pi^2}\int_0^\infty \mathrm{d}u \, u \abs{\int_{-\infty}^{+\infty}\mathrm{d}p \, e^{iup }g(\sigma e^{-p^2}) }^2
\end{equation}				
The equality is checked by Taylor expanding the function $g$ in both r.h.s of \eqref{g_exp_lhs} and \eqref{g_exp_rhs} and checking that the series match. We can also observe that 
\eqref{g_exp_rhs} is the leading order in $t$ of the exact result given in \eqref{cum2exact}.
We will use the form 
\eqref{g_exp_rhs} below, as it is more convenient for our present purpose.
\end{remark}

\subsubsection{First cumulant: mean density at the edge seen from the bulk}

Let us write the identification formula \eqref{mucum} to the first order, 
using our result \eqref{cum1} for the first cumulant to all orders
\be \label{cum1new} 
\int  \mathrm{d}a \,g(\sigma e^{a}) t^{-1/3} \overline{\mu(a t^{-1/3})} =
\kappa_1(g) =
\frac{ e^{ \frac{t}{12} (\sigma \partial_\sigma)^3}  }{\pi \sqrt{t}}\int_{0}^{+\infty}\mathrm{d}x\sqrt{x} g(\sigma e^{-x}) 
\ee 
It is straightforward to identify the leading term  
\begin{equation}
t^{1/6}\overline{\mu(at^{-1/3})}=\frac{\sqrt{-a}}{\pi}\theta(-a) + \mathcal{O}(t) 
\end{equation}
a well known result, since the mean density of the Airy point process must match the 
semi-circle density of the GUE for $a \to -\infty$. Identifying the next orders in the small $t$ expansion is more delicate, 
since they only admit a distributional limit.
Consider a function $g(\sigma e^{-x})$ which vanishes fast, as well as all its derivatives,
near $x=0^+$. Then we can rewrite
\be
\kappa_1(g)=  
\frac{1}{\pi \sqrt{t}}\int_{0}^{+\infty}\mathrm{d}x\sqrt{x} e^{ - \frac{t}{12} (\partial_x)^3}  g(\sigma e^{-x})
\ee 
and integrate by parts. This leads to
\bea \label{meanall} 
 t^{-1/6} \overline{\mu(a t^{-1/3})}&& = \frac{\theta(-a)}{\pi} e^{-\frac{t}{12} (\partial_a)^3} \sqrt{-a}  \\
&& = \frac{\theta(-a)}{\sqrt{4\pi}}  \sum_{p \geq 0} \frac{t^p }{12^p p! \Gamma(\frac{3}{2}-3p)} (-a)^{\frac{1}{2}- 3 p} \\
&&
= \frac{\theta(-a)}{\pi}  \Big( \sqrt{-a} + \frac{t}{32}  (-a)^{-5/2} -\frac{105t^2 }{2048} (-a)^{-\frac{11}{2}} + \dots \Big) 
\eea
where we have used the formula $\partial_x^a x^k=\frac{\Gamma(k+1)}{\Gamma(k-a+1)} x^{k-a}$. On the other hand, we know the exact result for the mean density of the Airy point process
$\overline{\mu(a)}=K_{\Ai}(a,a)= \Ai'(a)^2 - a \Ai(a)^2$. Asymptotics of the
Airy functions allow indeed to recover \eqref{meanall}, but only upon discarding terms which are 
fast oscillating for large negative $a$. 
Hence it is valid in the weak sense and only for smooth test functions, the general formula
being \eqref{cum1new}.

\subsubsection{Second cumulant to leading order and the Gaussian free field}

We now write the identification formula \eqref{mucum} to the second 
cumulant level, at leading order in $t$, that is $\kappa_2(g) \simeq \kappa_2^0$
at order $t^0$, using \eqref{g_exp_rhs}. Let us define $h(a)=g(\sigma e^{-a^2})$, $\hat{h}(u)=\frac{1}{2\pi}\int_\mathbb{R} \mathrm{d}a\, e^{iua}h(a)$ and recall that $f(a)=g(\sigma e^{a})$. We rewrite (recalling that $h$ is even) 
\begin{equation}
\begin{split}
\kappa_2^0&= \int_0^{+\infty} \mathrm{d}u \, u \, \hat{h}(u)\hat{h}(-u) =\frac{1}{8\pi^2}\iint_{\mathbb{R}^2}\mathrm{d}a_1 \mathrm{d}a_2 \, \left[\frac{h(a_1)-h(a_2)}{a_1-a_2}\right]^2\\
&=\frac{1}{4\pi^2}\iint_{\mathbb{R}^+\times \mathbb{R}^+}\mathrm{d}a_1 \mathrm{d}a_2 \, (h(a_1)-h(a_2))^2 \left[\frac{1}{(a_1-a_2)^2}+\frac{1}{(a_1+a_2)^2}\right]\\
&=-\frac{1}{2\pi^2}\iint_{\mathbb{R}^+\times \mathbb{R}^+}\mathrm{d}a_1 \mathrm{d}a_2 \, h'(a_1)h'(a_2) \log\abs{\frac{a_1-a_2}{a_1+a_2}}\\
& =  -\frac{1}{2\pi^2}\iint_{\mathbb{R}^+\times \mathbb{R}^+}\mathrm{d}a_1 \mathrm{d}a_2 \, f'(-a_1)f'(-a_2) \log\abs{\frac{\sqrt{a_1}-\sqrt{a_2}}{\sqrt{a_1}+\sqrt{a_2}}}
\end{split}
\end{equation}
From \eqref{mucum} we want to identify this expression, to leading order at small $t$ with
\begin{equation}
\begin{split}
\kappa_2^0 =  \iint_{\mathbb{R}\times \mathbb{R}} &\mathrm{d}a_1
\mathrm{d}a_2 \,f(a_1)f(a_2) t^{-2/3}\overline{\mu(a_1t^{-1/3})\mu(a_2t^{-1/3})}^c\\
&= \iint_{\mathbb{R}\times \mathbb{R}} \mathrm{d}a_1\mathrm{d}a_2 \,f(a_1)f(a_2) t^{-2/3}\overline{\mathsf{H}'(a_1t^{-1/3})\mathsf{H}'(a_2t^{-1/3})}^c\\
&= \iint_{\mathbb{R}\times \mathbb{R}} \mathrm{d}a_1\mathrm{d}a_2 \,f'(a_1)f'(a_2) \overline{\mathsf{H}(a_1t^{-1/3})\mathsf{H}(a_2 t^{-1/3})}^c\\
\end{split}
\end{equation}
Hence we obtain, by identification, for the second cumulant of the height field
\bea \label{resGFF} 
\overline{\mathsf{H}(a_1 t^{-1/3})\mathsf{H}(a_2 t^{-1/3})}^c = 
-\frac{1}{2\pi^2} \log\abs{\frac{\sqrt{-a_1}-\sqrt{-a_2}}{\sqrt{-a_1}+\sqrt{-a_2}}} \theta(-a_1)
\theta(-a_2) + \mathcal{O}(t)
\eea
This logarithmic correlator is a particular case of the correlator of the Gaussian free field, as we now discuss. In \cite{BorodinFerrari,borodinclt} it is shown that the height field associated to the GUE, $H_{\rm GUE}$ defined
likewise by counting the number of eigenvalues above a certain level, is
described upon rescaling by a Gaussian free field with a specific correlator
(see also \cite{BWZ}). Taking $y=1$ in the formula in pages 2, 3 and 4 of \cite{borodinclt}, the interval $[-2,2]$ on the real line (which is
the support of the GUE measure) is mapped to a semi-circle of radius $1$ in the upper half-plane $\mathbb{H}$ by the transformation
\begin{equation}
\Omega : x \to \frac{x}{2}+i \sqrt{1-(\frac{x}{2})^2}
\end{equation}
Its inverse has the form $\Omega^{-1}(z)=2\Re(z)$. The theorem of 
Borodin \cite{borodinclt} states that $H_{\rm GUE}(x,y)$ with $\lambda_{\rm GUE}=x$ and $y=1$
satisfies 
\be
\overline{ H_{\rm GUE}(x,1) H_{\rm GUE}(x',1) }^c = C(\Omega(x),\Omega(x'))
\ee
with the Gaussian free field correlator on the upper half plane with Dirichlet boundary condition
		\begin{equation}
		C(z,w)=-\frac{1}{2\pi}\log \abs{\frac{z-w}{z-\bar{w}}}
		\end{equation}
		
		We now take the limit of Borodin's formula near the edge $x=2+\frac{a}{N^{2/3}}$.
		The matching from the bulk going to the edge, with the
		edge going to the bulk, allows to identify $H_{\rm GUE}(x,1) \sim \sqrt{\pi} {\sf H}(a)$ for 
		large negative $a$. Hence
		it predicts to the first non zero order for large negative $a$
				\begin{equation}
		\overline{ \mathsf{H}(a_1) \mathsf{H}(a_2) }^c =-\frac{1}{2\pi}\log \abs{\frac{\sqrt{-a_1}-\sqrt{-a_2}}{\sqrt{-a_1}+\sqrt{-a_2}}} \theta(-a_1)
\theta(-a_2) 
		\end{equation}
		which is perfectly consistent with our result \eqref{resGFF}.

\subsubsection{Deeper towards the edge: beyond the Gaussian free field}
\label{subsub:deeperGFF}
As we go deeper towards the edge there are two types of corrections to the Gaussian free field. The first one are still Gaussian with corrections to the logarithmic variance of the Gaussian free field. The second arise from higher cumulants. We use repeatedly Eq. \eqref{cum} and Eq. \eqref{Linew}.

\begin{itemize} 

\item

Let us first study the former, that is higher corrections in $t$ to $\kappa_2$. Using \eqref{cum} it is straighforward to obtain the distribution result, for appropriate test functions $f$
\begin{equation}
\begin{split}
&  \iint_{\mathbb{R}\times \mathbb{R}} \mathrm{d}a_1
\mathrm{d}a_2 \,f(a_1)f(a_2) t^{-2/3}\overline{\mu(a_1t^{-1/3})\mu(a_2t^{-1/3})}^c|_{\mathcal{O}(t,t^2)} =
\\
&  \iint_{\mathbb{R}\times \mathbb{R}} \mathrm{d}a_1
\mathrm{d}a_2 \,f'(a_1)f'(a_2) \overline{{\sf H}(a_1t^{-1/3}) {\sf H}(a_2t^{-1/3})}^c|_{\mathcal{O}(t,t^2)}
\\
& = \frac{1}{4 \pi^2} \iint_{\mathbb{R}\times \mathbb{R}} \mathrm{d}a_1
\mathrm{d}a_2 \,
 \frac{\theta(-a_1) \theta(-a_2)}{\sqrt{-a_1} \sqrt{-a_2} } \bigg[  t \left( \frac{1}{6} f''(a_1) f''(a_2) + \frac{1}{3}  f'(a_1) f^{(3)}(a_2) \right)  \\
& + t^2 \left( \frac{29}{360} f^{(3)}(a_1) f^{(4)}(a_2) + \frac{1}{24} f''(a_1) f^{(5)}(a_2) 
+ \frac{1}{72}  f'(a_1) f^{(6)}(a_2) \right) + \mathcal{O}(t^3) \bigg] \label{secondcum2} 
\end{split}
\end{equation}
One sees that the next contributions in the second cumulant are sum of separable contributions in the variables $a_1$ and $a_2$, and therefore the fully entangled part comes the zero-th order given by the Gaussian free field.\\

\item

Let us now study the contribution from the next cumulant, i.e. $\kappa_3$. At leading order, 
from \eqref{cum}, we obtain the following correction from the third cumulant of the height field of the Airy process, which reads
\begin{equation}
\begin{split}
&  \iiint_{\mathbb{R}^3} \mathrm{d}a_1
\mathrm{d}a_2 \mathrm{d}a_3 \,f(a_1)f(a_2)f(a_3) t^{-1}\overline{\mu(a_1t^{-1/3})\mu(a_2t^{-1/3})\mu(a_3t^{-1/3})}^c
\\
& = \frac{1}{2 \pi^3} \iiint_{\mathbb{R}^3} \mathrm{d}a_1
\mathrm{d}a_2\mathrm{d}a_3 \,
 \frac{\theta(-a_1) \theta(-a_2)\theta(-a_3)}{\sqrt{-a_1} \sqrt{-a_2}\sqrt{-a_3} } \bigg[  \sqrt{t} f'(a_1) f'(a_2)f'(a_3) + \mathcal{O}(t^{3/2}) \bigg]
\end{split}
\end{equation}
leading to the following expression for the third cumulant of the Airy point process height field 
\be
\overline{{\sf H}(a_1 t^{-1/3}) {\sf H}(a_2 t^{-1/3})  {\sf H}(a_3 t^{-1/3}) }^c
= -  \frac{1}{2 \pi^3}  \frac{\theta(-a_1) \theta(-a_2)\theta(-a_3)}{\sqrt{-a_1} \sqrt{-a_2}\sqrt{-a_3} } \sqrt{t}
+ \mathcal{O}(t^{3/2}) 
\ee 
which, remarkably is quite simple to this order, and factorized. Higher cumulants and
corrections in small $t$ are evaluated similarly. The $n$-th cumulant can be read from $\kappa_n$ in
\eqref{cum} quite simply: according to \eqref{Linew},
each factor $\mathfrak{L}_{i}$ there leads to a factor $f^{(i)}$
in the $n$-uple integral generalising \eqref{secondcum2}.

\end{itemize} 

\subsection{Large deviations in the linear statistics of the Airy point process (edge of GUE)} 
\label{sec:largedevAPP} 
				
Here we generalize the large time calculation of Section \ref{sec:structure}, performed there in
the context of droplet KPZ, to a more general large deviation linear statistics problem 
for the Airy point process. The general aim is to study, in the large time limit
\be \label{linlarge}
\mathbb{E}_{\Ai} [ e^{- t \sum_i \phi(t^{-2/3} a_i) } ] \sim e^{- t^2 {\cal F}(\phi) }
\ee 
and calculate the rate functional ${\cal F}(\phi)$, for a given set of functions $\phi$.  We start with the following choice of functions parameterized by $z<0$ and $\gamma$
\be
\phi(x) = \phi_z(x)=  (-z+x)^\gamma_+
\ee
which for $\gamma=1$ correspond to the calculation performed in Section \ref{sec:structure}.
We will adopt the same notation and denote the rate function ${\cal F}(\phi) = \Phi_-(z)$. To carry out the calculation we first find a suitable function $g$ so we can apply the considerations
of Section \ref{sec:structure}. Consider thus the following choice (with $\sigma = - e^{- z t}$)
\begin{equation}
\begin{split}
 \beta g(- e^{-z t + t^{1/3} a}) & =  t^{1-\gamma} \Gamma(\gamma+1) {\rm Li}_\gamma(-e^{-z t + t^{1/3} a}) \\
\; &  \underset{t \gg 1}{\simeq}
- t^{1-\gamma} (-z t + t^{1/3} a)_+^\gamma
= - t \phi_z(t^{-2/3} a) 
\end{split}
\end{equation}
Note that there is a time dependence in $g$, equivalently a factor $t^{1-\gamma}$ in the parameter $\beta$.  We aim to evaluate \eqref{linlarge} by applying the following result from Section \ref{sec:structure}, i.e. 
\begin{equation}
\begin{split}
\log Q_t(\sigma) &= \log \mathbb{E}_{\Ai} [ e^{\beta g(- e^{-z t + t^{1/3} a_i})  } ]\\
&\underset{t\gg1}{\simeq}\sum_{n \geq 1} \frac{1}{n!} t^{\frac{n}{2}-1+ n(1-\gamma)}  2^{n-1} (\sigma\partial_{\sigma})^{n-3}\mathfrak{L}_{1}(\sigma)^{n} 
+ \dots
\end{split}
\end{equation}
with 
\be
\mathfrak{L}_{1}(\sigma) = (\sigma \partial_\sigma)^2  \mathfrak{L}_{-1}(\sigma)
\quad , \quad \mathfrak{L}_{-1}(\sigma) = \frac{\Gamma(\gamma+1)}{\sqrt{4 \pi}} {\rm Li}_{\gamma+ \frac{3}{2}}(\sigma) 
\ee
where we have used again only the leading term in each cumulant (the $\dots$ represents
the higher order terms, subdominant at large $t$). This can be again justified
by the same argument as in Section \ref{sec:structure}, simply 
multiplying the whole equation \eqref{argument} by $t^{n(1-\gamma)}$, 
which does not change the fact that the higher
values of $p$ are subdominant.\\

Writing $\sigma = -e^{-z t} = - e^v$, the asymptotics for large positive $v$ are (up to exponentially small corrections in $v$) and assuming $\gamma>-1$
\bea
&& \mathfrak{L}_{-1}(- e^{v}) \simeq -\frac{1}{\pi} \int_0^{+\infty} \mathrm{d}x \,  \sqrt{x}(v-x)_+^\gamma  = - \frac{\Gamma(\gamma+1)}{\sqrt{4\pi} \Gamma(\frac{5}{2}+\gamma)} v^{3/2+\gamma} \\
&& \mathfrak{L}_{1}(- e^{v}) = \partial_v^2 \mathfrak{L}_{-1}(- e^{v}) 
\simeq - \frac{\Gamma(\gamma+1)}{\sqrt{4 \pi}\Gamma(\frac{1}{2}+\gamma)} v^{\gamma-1/2}
\eea
The choice $v=-zt$ is made such that the main contribution of all cumulants is now homogeneous to $t^2$ and can be therefore summed together. We can thus identify the 
function $\Phi_-(z)$ as the following series, summed over all cumulants 
\begin{equation}
\begin{split}
 t^2 \Phi_-(z) &= -t^2 \sum_{n \geq 1} 
 (-1)^n \frac{ \Gamma(\gamma+1)^n}{2\Gamma(n+1)\Gamma(\frac{1}{2}+\gamma)^n \pi^{n/2}}  \partial_u^{n-3} u^{n(\gamma-\frac{1}{2})}|_{u=-z}  \\
& = -t^2 \sum_{n \geq 1}  (-1)^n \frac{  \Gamma(\gamma+1)^n}{2\Gamma(n+1)\Gamma(\frac{1}{2}+\gamma)^n \pi^{n/2}}\frac{\Gamma(n(\gamma-\frac{1}{2})+1)}{\Gamma(4-n(\frac{3}{2}-\gamma))} 
(-z)^{3-n(\frac{3}{2}-\gamma)} 
\end{split}
\end{equation} 
Defining
\begin{equation}
\Omega= \frac{\Gamma(\gamma+1)}{ \sqrt{\pi}\Gamma(\frac{1}{2}+\gamma)}
\end{equation}
we obtain the large deviation rate function
\begin{equation}
\Phi_-(z) =-\frac{1}{2} \sum_{n \geq 1}  \frac{(-\Omega)^n }{\Gamma(n+1)}\frac{\Gamma(n(\gamma-\frac{1}{2})+1)}{\Gamma(4-n(\frac{3}{2}-\gamma))} (-z)^{3-n(\frac{3}{2}-\gamma)} 
\end{equation}

We give in Table \ref{Table1_large_rate} a few examples (in increasing order in $\gamma$) for which the summation
of the series is easy. The resulting function $\Phi_-(z)$ is found to be positive as required.\\ 
\begin{table}[h!]
\makebox[\textwidth][c]{
\begin{tabular}{|c||c||c|}
\hline 
\hspace*{0.2cm}$\gamma$\hspace*{0.2cm} & $\Phi_-(z)$ & Expansion around $z=0^-$ \\ 
\hline 
\hline 
&&\\[-3ex]
$\dfrac{1}{2} $ & $\dfrac{z^2}{8}+\dfrac{z}{16}+\dfrac{1}{96}$ & $\dfrac{z^2}{8}+\dfrac{z}{16}+\dfrac{1}{96}$  \\ [1ex]
\hline 
&&\\[-3ex]
$1$ & $\dfrac{4}{15\pi^6}(1-\pi^2 z)^{5/2}-\dfrac{4}{15\pi^6}+\dfrac{2}{3\pi^4}z -\dfrac{1}{2\pi^2}z^2$  &  $-\dfrac{z^3}{12}-\dfrac{\pi^2 z^4}{96} -\dfrac{\pi^4 z^5}{320}  + \mathcal{O}(z^6) $\\ [1ex]
\hline 
&&\\[-3ex]
$\dfrac{3}{2} $ & $-\dfrac{z^3}{28}$ & $-\dfrac{z^3}{28}$   \\  [1ex]
\hline 
&&\\[-3ex]
$2$ & \makecell{\hspace*{-2cm}$\dfrac{16 (-z)^{7/2} }{105 \pi } \, _2F_1\left(\dfrac{5}{6},\dfrac{7}{6};\dfrac{9}{2};-\dfrac{48 z}{\pi ^2}\right)$\\ \vspace*{-0.5cm}\\\hspace*{0.3cm}
$+\dfrac{81 \pi ^6 }{29360128} \,  _2F_1\left(-\dfrac{8}{3},-\dfrac{7}{3};-\dfrac{5}{2};-\dfrac{48 z}{\pi ^2}\right)$\\\vspace*{-0.5cm}\\  \hspace*{1cm}$-\dfrac{z^3}{12}-\dfrac{3 \pi ^2
   z^2}{256}-\dfrac{27 \pi ^4 z}{81920}-\dfrac{81 \pi ^6}{29360128}$} & $\dfrac{16(-z)^{7/2}}{105\pi}-\dfrac{4 z^4}{9\pi^2}+\dfrac{128 (-z)^{9/2}   }{81\pi^3} +\mathcal{O}(z^5) $ \\ [1ex]
   &&\\[-3ex]
\hline 
&&\\[-3ex]
$\dfrac{5}{2}$ & $-\dfrac{z^3}{12}+\dfrac{2 z^2}{15}-\dfrac{32 z}{675}-\dfrac{8 (4-15 z)^{5/2}}{50625}+\dfrac{256}{50625}$ & $\dfrac{5z^4 }{128} + \dfrac{45 z^5}{1024}+\dfrac{1125 z^6}{16384} +\mathcal{O}(z^7)	 $ \\  [1ex]
\hline
&&\\[-3ex]
$\dfrac{7}{2} $ &$-\dfrac{8z}{105}  \, _2F_1\left(-\dfrac{2}{3},-\dfrac{1}{3};\dfrac{3}{2};-\dfrac{945 z^2}{128}\right)-\dfrac{z^3}{12}+\dfrac{8 z}{105}$  & $ -\dfrac{7z^5}{256}+\dfrac{175z^7}{4096}-\dfrac{42875z^9}{393216}  +\mathcal{O}(z^{11})$\\ [1ex]
\hline 
\end{tabular} 
} \caption{Variety of rate functions for different values of $\gamma$ for $z<0$ with the three first orders of their expansion around $z=0^-$. }
\label{Table1_large_rate}
\end{table}

Remarkably, these results exactly agree with calculations using a completely different method related to the Coulomb gas \cite{inprep2}. This comes in strong support of the validity of the "conjectures" discussed above for the present method. We can generalize to a larger set of functions 
functions $\phi(x)$ by choosing, with $\sigma = - e^{-z t}$ 
\bea
&& \beta g(\sigma e^{t^{1/3} a} ) = \sum_i c_i t^{1-\gamma_i} 
{\rm Li}_{\gamma_i}(\sigma e^{-\delta_i  t + t^{1/3} a} ) \simeq_{t \to +\infty}
 - t \phi_z(t^{-2/3} a)  \\
&&  \phi_z(x)= \tilde \phi(-z+x)= \sum_i c_i  (-z -\delta_i +x)^{\gamma_i}_+
\eea
with $\delta_i>0$ and $c_i>0$. For this class of functions we find
\bea
&& {\cal F}(\phi) = - \sum_{n \geq 1} \frac{2^{n-1}}{n! \pi^n}  \partial_{z}^{n-3}
		(\partial_z^2 \int_0^{+\infty} dx \sqrt{x} \phi_z(-x) )^n
	\eea
	
It remains an open problem whether this formula can be extended to
a larger class of function $\phi(x)$. 


\section{Application to the counting statistics of trapped fermions at finite temperature}
\label{fermion_large_dev}

Here we apply the connection between non-interacting fermions near the edge of a smooth trap at finite temperature and the solution of the KPZ equation with droplet initial condition for arbitrary time \cite{dean2015finite,dean2016noninteracting}. The short time expansion for KPZ corresponds to the 
{\it high temperature expansion} for the fermions. The leading order at high temperature was studied in Ref. \cite{le2016exact} and here we obtain many higher orders. In addition we also study some properties in the {\it low temperature} which corresponds to the large time limit. 

\subsection{Edge fermions} 

Let us summarize the problem, for more details see \cite{dean2015finite,dean2016noninteracting}
(see also \cite{KrajLedou2018} section 9). 
Consider the quantum system of $N$ non-interacting spinless fermions of mass $m$ in an harmonic trap at finite temperature $T$ described by the Hamiltonian $H=\sum_{i=1}^N\frac{p_i^2}{2m}+\frac{1}{2}m\omega^2 x_i^2$. We use $x^*=\sqrt{\hbar /m\omega}$ and $T^*=\hbar \omega$ as units of length and energy. At $T=0$, in the ground state and for large $N$, the average fermion density is given by the semi-circle law with an edge at $x_{\mathrm{edge}}=\sqrt{2N}$. At finite temperature, the behavior of the physical quantities in the bulk changes on a temperature scale $T\sim N$ (bulk scaling), while near the edge it varies on a scale $T=N^{1/3}/b$, where $b$ is the {\it inverse reduced temperature}, a parameter of order one. Here we study counting statistics properties at the edge,
as a function of $b$. \\

We consider the grand canonical ensemble, where the mean total number of fermions $N$, is large. However from
the equivalence between ensembles for local observables (as considered here) we expect the conclusions to hold also for the canonical ensemble at fixed $N$ \cite{dean2016noninteracting}
(see Ref. \cite{grabsch2018fluctuations} for cases where deviations occur).
It was shown in \cite{dean2015finite,dean2016noninteracting} that if one defines the reduced fermion
coordinates
\be
\tilde \xi_i = \frac{x_i-x_{\mathrm{edge}}}{w_N}
\ee 
where $w_N=N^{-1/6} /\sqrt{2}$, then the set of $\tilde \xi_i$ form a determinantal point process with an associated
kernel which,
in the limit $N, T \to +\infty$ with $b$ fixed, takes the form
\be \label{Kt} 
 K_{b}(r,r') := \int_{-\infty}^{+\infty} \mathrm{d}u \, \Ai(r+u) \Ai(r'+u) \frac{1}{1 +  e^{- b u } } 
\ee

To study the high temperature expansion, it is more convenient to define as in \cite{le2016exact} the reduced fermion coordinates $\xi_i = b \,\tilde \xi_i$, \cite{footnote10}, i.e. 
\be
\xi_i = \frac{x_i-x_{\mathrm{edge}}}{w_N/b} \quad , \quad w_N/b = T N^{-1/3} w_N = \frac{T}{\sqrt{2 N}} 
\ee 

\subsection{Counting statistics} 

We study here $N(s) \equiv N(\xi > s)$, the total number of fermions in the
interval $[s,+\infty[$ (i.e. with $\xi_i \geq s$). From standard properties of determinantal
processes \cite{borodin2009determinantal}, the Laplace transform of the PDF of $N(s)$ can be written in the form
of the generating function \eqref{G sw2} studied in this paper as
\bea \label{LTN} 
&& \langle e^{- \lambda N(s) } \rangle = {\rm Det}[I - (1- e^{- \lambda}) P_{s/b} K_{b} ] 
= \mathrm{Det}\left[1- (1- e^{- \lambda})  (1- \rme^{\beta \hat g_{t,\sigma}}) K_\Ai \right] \\
&& = \mathrm{Det}\left[1- (1- \rme^{\beta \hat g^\lambda_{t,\sigma}}) K_\Ai \right] 
= Q_{t}(\sigma)
\eea 
where 
\bea
&& t = b^3 \quad , \quad \sigma =  - e^{- s} \\
&& \hat g_{t,\sigma}(a)=g_{\rm KPZ}(\sigma e^{t^{1/3} a}) \quad , \quad g_{\rm KPZ}(x) =- \log(1-x) \\
&& \hat g^\lambda_{t,\sigma}(a)=g^\lambda(\sigma e^{t^{1/3} a}) 
\quad , \quad g^\lambda(x)= g_{\rm KPZ}(x) - g_{\rm KPZ}(x e^{-\lambda}) = \log \frac{1- x e^{\lambda}}{1-x} 
\eea
Hence we can use our results for $Q_{t}(\sigma)$ with the choice $g(x) \to g^\lambda(x)$ and obtain immediately the 
Laplace transform \eqref{LTN}. From \eqref{G short time expansion}
we obtain the high temperature expansion up to $\mathcal{O}(b^{6})$ as
\begin{align}
			\label{Nexp1}
			\log \langle e^{- \lambda N(s) } \rangle & = \frac{1}{b^{3/2}} 
			 \frac{1}{\sqrt{4\pi}}\,( \Li_{5/2}(- e^{- s})- \Li_{5/2}(- e^{- \lambda - s})) \\
			&
			+\frac{1}{4\pi}\int_{0}^{1}\!\frac{\rmd v}{v}\,( \Li_{1/2}(- v e^{-s}) -  \Li_{1/2}(- v e^{-s -\lambda}))^{2}
			+\Big(\frac{2(\mathfrak{L}^\lambda_{1})^{3}}{3}+\frac{\mathfrak{L}^\lambda_{2}}{12}\Big) b^{3/2}
			+ \dots \nonumber
\end{align}
where the $\dots$ are all terms beyond $\mathcal{O}(b^{3/2})$ in \eqref{G short time expansion}
obtained using the replacements
		\bea \label{Lifermions} 
&& \mathfrak{L}_{i}(\sigma) \to 	\mathfrak{L}_{i}^\lambda(\sigma,\Lambda)=\frac{1}{\sqrt{4\pi}}\,( \mathfrak{L}_{i}(\sigma) - \mathfrak{L}_{i}(\Lambda) ) \\
&& \mathfrak{L}_{i}(\sigma) = \frac{1}{\sqrt{4\pi}}\,  \Li_{\frac{3}{2}-i}(\sigma) 
			  ~~ , ~~  \sigma=-e^{-s} 
			~~ , ~~ \Lambda=-e^{-\lambda-s} ~~ , ~~ t = b^3 \nonumber 
		\eea
		and setting $\beta=1$ there. The leading term $\mathcal{O}(1/b^{3/2})$ was
		obtained in \cite{le2016exact}, and we display here considerably many more orders. 
Several type of information can be now extracted 
from this formula (i) about the cumulants of $N(s)$ and its PDF and (ii) about the position of the
right-most fermion. Indeed, setting $\lambda \to +\infty$ one obtains the CDF of the (reduced) position of 
the rightmost fermion, $\xi_{\max}= \max_i \xi_i$. Indeed
\be
\lim_{\lambda \to +\infty} \log \langle e^{- \lambda N(s) } \rangle = 
\log \mathbb{P}(\xi_{\max} < s) 
\ee
In \eqref{Nexp1} all terms containing $e^{-\lambda}$ then vanish. We now study these observables for different regions as the interval $[s,+\infty[$ is varied.
If $s$ is chosen near the typical location of the rightmost fermion then $N(s) = \mathcal{O}(1)$.

\subsection{High temperature: near the typical rightmost fermion, $s \sim \xi_{\rm typ}$, $N(s) \sim \mathcal{O}(1)$} 

As discussed in \cite{le2016exact} 
the leading term for small $b$ (high temperature) 
shows that the CDF of $\xi_{\rm max}$ is peaked around the typical value $\xi_{\rm typ}= \log( \frac{1}{b^{3/2} \sqrt{4 \pi}})$ 
and has the form of a Gumbel distribution. Indeed keeping only the leading term
\be
\log \mathbb{P}(\xi_{\max} - \xi_{\rm typ} < \hat s) \simeq - e^{- \hat s}
\ee
using that $\Li_{a}(y) \sim y$ at small $y$. Now we can use \eqref{Nexp1} and 
obtain the systematic corrections away from the Gumbel distribution. One finds
\be \label{pmax}
\log \mathbb{P}(\xi_{\max} - \xi_{\rm typ} < \hat s) = - e^{- \hat s} 
+ \frac{1}{2} \sqrt{\frac{\pi}{2}}  e^{-2 \hat s} b^{3/2} 
+  (- \frac{4 \pi}{9 \sqrt{3}} e^{-3 \hat s} + \frac{1}{2} e^{-2 \hat s} - \frac{1}{12} e^{- \hat s} ) b^3
+ \mathcal{O}(b^{9/2}) 
\ee
Note that to each order in $b$ there is are contributions from the various cumulants. Note also that the expansion \eqref{pmax} works at fixed $\hat s$
but fails when $\hat s \sim - \xi_{\rm typ} \sim \frac{3}{2} \ln b$, i.e. when $\xi_{\max}$ is of order one or negative.
This is the region $s=\mathcal{O}(1)$ which is the region of large deviations for the PDF of $\xi_{\max}$ as
discussed below. 

Similarly one can study the PDF of $\hat N(\hat s) = N(s= \xi_{\rm typ} + \hat s)$, i.e
the number fluctuations around the typical position of the rightmost fermion. One finds
\bea
\label{Ntyp}
&& \log \langle e^{- \lambda \hat N(\hat s) } \rangle = - e^{- \hat s}  (1- e^{-\lambda}) 
+ \frac{1}{2} \sqrt{\frac{\pi}{2}}  e^{-2 \hat s}  (1- e^{-2 \lambda})  b^{3/2} \\
&& + \bigg( \frac{1}{2} (1-e^{-\lambda })^2 e^{ -2 \hat s}- \frac{1}{12} (1-e^{-\lambda
   }) e^{- \hat s}-\frac{4 \pi (1-e^{-3 \lambda
   }) e^{ -3 \hat s}}{9 \sqrt{3}} \bigg) b^3 + \mathcal{O}(b^{9/2}) \nonumber 
\eea
which recovers \eqref{pmax} for $\lambda=+\infty$. The leading term at high temperature $b \to 0$ corresponds to the Poisson distribution $p_{\bar n}(n)=\frac{\bar n^n}{n!} e^{-\bar n}$ which
leads to $\log \langle e^{- \lambda n} \rangle = - \bar n (1- e^{-\lambda})$ with
a mean $\bar n=e^{-\hat s}$. This reflects the fact that the
positions of the fermions become independent random variables at high temperature. The next orders
quantify the deviations from the Poisson distribution (it can formally be written as
a convolution of distributions $N=q n$ where $n$ is Poisson and $q=1,2,3..$
with however non-positive or random parameters $\bar n$).
The first cumulants (mean and variance) are
\bea
&& \langle \hat N(\hat s) \rangle= 
e^{-\hat s}-\sqrt{\frac{\pi }{2}} b^{3/2} e^{-2
   \hat s}+\frac{1}{36} b^3 e^{-\hat s} \left(16 \sqrt{3} \pi 
   e^{-2 \hat s}+3\right)+ \mathcal{O}(b^{9/2}) \\
   && \langle \hat N(\hat s)^2 \rangle^c= 
e^{-\hat s}-\sqrt{2 \pi } b^{3/2} e^{-2 \hat s}+b^3 \left(e^{-2
   \hat s}+\frac{e^{- \hat s}}{12}+\frac{4 \pi  e^{-3
   \hat s}}{\sqrt{3}}\right)+\mathcal{O}\left(b^{7/2}\right)
\eea
and are of order unity in the region $\hat s=\mathcal{O}(1)$ (i.e. near the typical location
of the rightmost fermion). 

\subsection{High temperature: cumulants of $N(s)$ in the region $s \sim \mathcal{O}(1)$} 

As we show below, and obtained in \cite{le2016exact}, in the region $s \sim \mathcal{O}(1)$ the mean number of fermions in $[s,+\infty[$ is large, and behaves at high temperature as $\mathcal{O}(b^{-3/2})$, more precisely
\be \label{meanN} 
\langle N(s) \rangle \simeq - \frac{1}{b^{3/2}} 
  \frac{1}{\sqrt{4\pi}} \Li_{\frac{3}{2}}(- e^{ - s}) \simeq_{s \to - \infty}  \frac{2}{3 \pi}  (-\frac{s}{b})^{3/2}
\ee
which is only the leading term. Here we obtain the systematic small $b$ expansion of the
cumulants in that region $s \sim \mathcal{O}(1)$. Since $N(s)$ is typically large, the
probability of $N(s)=0$ i.e. of $\xi_{\max}<s$ is small, and accordingly, \eqref{Nexp1} also contains the information about the large deviations of the CDF of $\xi_{\max}$ at high temperature. The leading term exhibits a
$|s|^{5/2}$ tail which was discussed in \cite{le2016exact}. The formula \eqref{Nexp1} 
with $e^{-\lambda} \to 0$ thus contains all the systematic corrections in the small $b$ expansion to the
rate function of these large deviations (which we will not further discuss). 

The cumulants of $N(s)$ in an expansion at small $b$ and fixed $s$ can be obtained from \eqref{Nexp1}
as, for $p \geq 1$ 
\begin{equation}
\begin{split}
 \langle  N(s) ^p  \rangle^c & = (-\partial_{\lambda})^p \log \langle 
 e^{- \lambda N(s) } \rangle |_{\lambda=0 } \\
 &= (\Lambda \partial_{\Lambda})^p \log \langle 
 e^{- \lambda N(s) } \rangle |_{\Lambda=\sigma=-e^{-s} } \\
 \end{split}
\end{equation}
where we recall that $\Lambda=-e^{-\lambda-s}$. This leads to
\bea \label{expcumN} 
 && \langle  N(s) ^p  \rangle^c =  - \frac{1}{b^{3/2}} 
  \frac{1}{\sqrt{4\pi}} \Li_{\frac{5}{2}-p}(- e^{ - s}) 
 \\
 && + \frac{1}{4\pi} \sum_{p_1,p_2 \geq 1,p_1+p_2=p} \frac{p!}{p_1! p_2!} 
  \int_{0}^{1}\!\frac{\rmd v}{v}\,  \Li_{\frac{1}{2}-p_1}(- v e^{ - s})
 \Li_{\frac{1}{2}-p_2}(- v e^{ - s}) \nonumber \\
 && - b^{3/2} \bigg(
 \frac{2}{3} 
\frac{1}{(4\pi)^{3/2}} \sum_{p_1,p_2,p_3 \geq 1,p_1+p_2+p_3=p} \frac{p!}{p_1! p_2! p_3!}
  \Li_{\frac{1}{2}-p_1}(- e^{ - s})  \Li_{\frac{1}{2}-p_2}(- e^{ - s})  \Li_{\frac{1}{2}-p_3}(- e^{ - s}) \nonumber \\
&& + \frac{1}{12} 
\frac{1}{(4\pi)^{1/2}} \Li_{-\frac{1}{2}-p}(- e^{ - s})   \bigg) + \dots \nonumber
\eea 
This formula recovers, as in \cite{le2016exact} to the leading order in $b$, that 
the typical $N(s)$ at fixed $s$ is large with $N(s)_{\rm typ} \sim 1/b^{3/2}$.
All cumulants are also of the order $1/b^{3/2}$ hence the relative fluctuations
are small (but not the absolute ones). As $s \to - \infty$, the distribution becomes
peaked around  $\langle N(s) \rangle  \simeq \sqrt{\frac{1}{4 \pi b^3}} \frac{4 (-s)^{3/2}}{3 \sqrt{\pi }} =
 \frac{2}{3 \pi}  (-s/b)^{3/2}$ 
(which is also the asymptotic integrated density of the Airy process) with Gaussian fluctuations with variance
$\langle N(s)^2 \rangle^c  \simeq 
\frac{ \sqrt{-s}}{\pi b^{3/2}} $,
as the higher cumulants tend to zero (using ${\rm Li}_a(z)  \simeq - \frac{(\ln(-z))^{a}}{\Gamma(a+1)}$
for $z \to -\infty$). The formula \eqref{expcumN} thus provides the correction terms in the 
small $b$ expansion of all cumulants. Note that one can equivalently write
\bea
 \langle  N(s) ^p  \rangle^c = (\Lambda \partial_{\Lambda})^p 
 \sum_{n \geq 1} \frac{\kappa_n}{n!} |_{\Lambda=\sigma=-e^{-s} } 
\eea
where the cumulants $\kappa_n$ are given in \eqref{cum}, with the same substitution 
\eqref{Lifermions}. One observes that the $p$-th cumulant of $N(s)$ is only determined by the 
cumulants $\kappa_n$ with $1 \leq n \leq p$. Since we also have an exact
expansion to all orders for $\kappa_1$ we have, using \eqref{cum1}
and \eqref{KPZ_cum1}, or alternatively \eqref{equiv_cum1}, to all orders in $b$,
\bea \label{cum1exact1} 
\langle N(s) \rangle = - \frac{ e^{ - \frac{b^3}{12} \partial_s^3}  }{b^{3/2} \sqrt{4\pi}}
\mathrm{Li}_{3/2}(- e^{-s}) 
\eea 
where we have used the identity \eqref{Li3/2identity}. Let us also give the variance of $N(s)$ to $\mathcal{O}(b^6)$, either from \eqref{expcumN} 
or from \eqref{cum}
\bea
&& \langle N(s)^2 \rangle^c = - \frac{ e^{ - \frac{b^3}{12} \partial_s^3}  }{b^{3/2} \sqrt{4\pi}}
\mathrm{Li}_{1/2}(- e^{-s}) 
+ \frac{1}{2 \pi}  \int_{0}^{1}\!\frac{\rmd v}{v}\,  [ \Li_{-\frac{1}{2}}(- v e^{ - s}) ]^2 \\
&&  
+\Big(\frac{\mathfrak{L}_{3}^{2}}{6}+\frac{\mathfrak{L}_{2}\mathfrak{L}_{4}}{3}\Big) b^3+\Big(\frac{29}{360}\,\mathfrak{L}_{4}\mathfrak{L}_{5}+\frac{\mathfrak{L}_{3}\mathfrak{L}_{6}}{24}+\frac{\mathfrak{L}_{2}\mathfrak{L}_{7}}{72}\Big) b^6 +\ldots\nonumber
\eea
where here $\mathfrak{L}_{k}= \frac{1}{\sqrt{4 \pi}} \Li_{\frac{3}{2}- k }(-e^{ - s})$. We note that
the terms are obtained by simply shifting $\mathfrak{L}_{k} \to - \mathfrak{L}_{k+1}$ in 
\eqref{cum} (we recall that $\sigma \partial_\sigma \mathfrak{L}_{k}(\sigma) = 
\mathfrak{L}_{k+1}(\sigma)$). Similar manipulations lead to explicit expressions for 
the expansion of the higher cumulants $p \geq 3$.  Finally, there is an exact formula for the variance to all orders in $b$, from \eqref{cum2exact} we find
\bea \label{cum2Nexact}
&& \langle N(s)^2 \rangle^c = 
- \frac{ e^{ - \frac{b^3}{12} \partial_s^3}  }{b^{3/2} \sqrt{4\pi}}
\mathrm{Li}_{1/2}(- e^{-s})
+ \frac{1}{4} \int_0^{+\infty} \mathrm{d}u\,  \int_{-u}^u \mathrm{d}v \,
|F(u,v)|^2 \\
&&  F(u,v) = e^{-\frac{b^3}{12} \partial_s^3} \int_{-\infty}^{+\infty} \frac{\mathrm{d}k}{2 \pi} e^{-i k u} \, \mathrm{Li}_{0}(- e^{-s- \frac{b^{3/2}}{2} v - k^2}) 
\eea

\subsection{Low temperature: large deviation of the PDF of $N(s)$ in the 
region $s \sim b^3$}

We now study the low temperature limit of large $b$. At strictly zero temperature $b=+\infty$ 
the typical fluctuations of the position of the rightmost fermion are well known to be
given by the Tracy-Widom distribution \cite{TW} and some results for the counting statistics 
near this position have also been obtained 
\cite{dean2015finite,dean2016noninteracting,eisler_prl,mehta,marino_prl,castillo,CDM14,marino_pre}.
There are however interesting and non-trivial large deviations for large but not strictly infinite $b$. First of all it is known that (see \cite{KrajLedou2018} section 9) for any fixed $z<0$
\begin{equation}
- \lim_{b \to +\infty}\frac{1}{b^6}  \log \mathbb{P}\left(\frac{\xi_{\rm max}}{b^3} = \frac{x_{\mathrm{max}}-x_{\mathrm{edge}}}{b^2 w_N} < z \right) =  \Phi_-(z) \label{sb2} 
\end{equation}
where $\Phi_-(z)$ is the large deviation rate function \eqref{eq:Phi} for KPZ. 
Hence there is a large deviation scale $\sim b^2 w_N$ \cite{footnote10} of fluctuations for $x_{\max}$ 
at low temperature. These large deviations are thus observed for $s \sim - b^3$ very negative
(i.e. towards the bulk). 

We will thus study now the number of fermions, denoted ${\cal N}$, in an interval $[s,+\infty[$ with $s$ within that regime 
\be
{\cal N}=N(s=z b^3)=N(\xi > z b^3)    
\ee
in the limit of large $b$ with fixed $z<0$. 
The typical value of ${\cal N}$ is given by its average, from \eqref{meanN},
$\langle {\cal N} \rangle \simeq \frac{2}{3 \pi} (-z)^{3/2} b^3 \gg 1$, i.e. it is very large. However the
number of fermions can fluctuate on the same scale $b^3$. As we show below the PDF of
${\cal N}$ takes the following large deviation form at large $b$
\bea \label{rateN} 
P({\cal N}) \sim \exp \left( - b^6 F(\nu) \right) \quad , \quad \nu = {\cal N}/b^3
\eea 
This implies that we expect that the Laplace transform \eqref{Nexp1} of
the PDF of ${\cal N}$ should take the form, denoting $\lambda = b^3 \tilde \lambda$
\bea
\langle e^{- b^3 \tilde \lambda {\cal N} } \rangle = \int d{\cal N} P({\cal N}) e^{- b^3 \tilde \lambda {\cal N} } 
\sim \exp \left( - b^6 \Phi(z,\tilde \lambda) \right)
\eea
where the large deviation rate function $\Phi(z,\tilde \lambda)$ is related to
$F(\nu)$ by the Legendre transform
\bea \label{legendre1} 
 \Phi(z,\tilde \lambda) = \min_{\nu} \left[ F(\nu) + \tilde \lambda \nu \right]   \quad , \quad \nu = {\cal N}/b^3
\eea 

We now show that there is indeed such a rate function $\Phi(z,\tilde \lambda)$ and we
calculate it. As for the calculation of $\Phi_-(z)$ leading \eqref{eq:Phi} we can retain only
the leading order of the cumulants $\kappa_n^0(\sigma)$ given in \eqref{cumconj}. In fact the
calculation is very similar to the one of Section \ref{sec:structure}. We must make the
replacement \eqref{Lifermions} and use again the asymptotics \eqref{res1}. We obtain, replacing $t=b^3$,
\bea
&& \log \langle e^{- b^3 \tilde \lambda N(s=z b^3) } \rangle = 
\sum_{n \geq 1} \frac{\kappa_n}{n!}  \simeq \sum_{n \geq 1}  
\frac{b^{3(\frac{n}{2}-1)}}{n!}  \kappa_n^0(\sigma=-e^{-z b^3}) \\
&&  = b^6 \sum_{n \geq 1}  (-1)^n \frac{2^{n-1}}{\Gamma(n+1) \pi^n}  (- \partial_z)^{n-3} ( (-z)^{1/2} - (-z-\tilde \lambda)_+^{1/2})^n
\eea 
where we performed the replacement \eqref{Lifermions} inside the formula for the leading 
order of the cumulants \eqref{cumconj} and used that 
$\mathfrak{L}_{1}(- e^{- z t}) \simeq_{t \to +\infty} - \frac{1}{\pi} (-z t)_+^{1/2}$. Clearly we find that
\bea
\Phi(z,\tilde \lambda) = \Phi_-(z) \quad , \quad \tilde \lambda > -z 
\eea 
The interpretation of that fact is that when $\tilde \lambda > -z$ the minimum in 
the Legendre transform \eqref{legendre1} must be attained for $\nu=0$, a fact
that we will check later. We now obtain the rate function $\Phi(z,\tilde \lambda)$ for $\tilde \lambda < -z$
as follows. It is convenient
to define $u_1=-z>0$ and $u_2=-z-\tilde \lambda>0$. It is written in the 
form of a series in an expansion in $z,\tilde \lambda$ simultaneously large 
(equivalently $u_1,u_2$ large) as
\bea \label{Philambda} 
&& \Phi(z,\tilde \lambda)= 
 \frac{4}{15 \pi}  (u_1^{5/2} - u_2^{5/2})  \\
&& 
- \frac{1}{2 \pi^2} \left( u_1^2 + u_2^2  - \sqrt{u_1 u_2} (u_1+u_2) +  (u_1-u_2)^2  \log \frac{\sqrt{u_1} + \sqrt{u_2}}{\sqrt{u_1-u_2}} \right) \nonumber 
\\
&& - 
\sum_{n \geq 3}  (-1)^n \frac{2^{n-1}}{\Gamma(n+1) \pi^n}  (\partial_{u_1}+ \partial_{u_2} )^{n-3} ( u_1^{1/2} - 
u_2^{1/2})^n \bigg|_{u_1=-z, u_2=-z - \tilde \lambda} \nonumber
\eea
where we have separated the terms $n=1,2$. The only delicate term is $n=2$, let us 
indicate its calculation. It reads (for $\tilde \lambda < -z$ and $\sigma=-e^{- b^3 z}$)
\bea
&& - \frac{1}{2b^6}  \kappa_2^0(\sigma) = - \frac{1}{b^6} (\sigma \partial_\sigma)^{-1} 
(\mathfrak{L}_{1}(\sigma) - \mathfrak{L}_{1}(\sigma
e^{-\lambda}))^2  = - \frac{1}{b^6} \int_0^{-e^{-z b^3}} \frac{\mathrm{d}u}{u}(\mathfrak{L}_{1}(u) - \mathfrak{L}_{1}(u
e^{-b^3 \tilde \lambda}))^2 \nonumber \\
&& = - \frac{1}{b^6} \int_{-\infty}^{-z b^3} \mathrm{d}v (\mathfrak{L}_{1}(-e^{v}) - \mathfrak{L}_{1}(- 
e^{v-b^3 \tilde \lambda}))^2 
\simeq - \frac{1}{\pi^2} \int_{0}^{-z} \mathrm{d}v (v_+^{1/2} - (v- \tilde \lambda)_+^{1/2})^2 
\eea 
Calculation of this integral gives the result for $n=2$ in \eqref{Philambda}.
Note that \eqref{Philambda} is only a series expansion (to all orders) and 
we have not obtained $\Phi(z,\tilde \lambda)$ in a form as explicit at $\Phi_-(z)$.\\

Let us now extract some information about the PDF $P({\cal N})$ in \eqref{rateN}.
The associated rate function is given by the Legendre transform
\bea \label{legendre2} 
F(\nu)  = \max_{\tilde \lambda} \left[  \Phi(z,\tilde \lambda) - \tilde \lambda \nu \right]   \quad , \quad \nu = {\cal N}/b^3
\eea 
Although we do not have the full explicit form we can use the expansion \eqref{Philambda} to
obtain $F(\nu)$ in an expansion in the limit of large $|z|$. In that limit $\nu$ is 
large with $\nu \sim (-z)^{3/2}$, hence we define $\tilde \nu$ through
\be
\nu = \frac{3 \pi}{2} (-z)^{3/2}  \, \tilde \nu
\ee
so that for $z \to - \infty$ the typical value is $\tilde \nu_{\rm typ}=1$. Then one has the expansion
at large negative $z$ and fixed $\tilde \nu$
\bea \label{expandF} 
F(\nu) = \frac{4}{15 \pi} (-z)^{5/2} F_0(\tilde \nu) - \frac{z^2}{2 \pi^2} F_1(\tilde \nu) 
+ \dots 
\eea 
where each term corresponds to a term in the expansion of $\Phi_-(z)$ at large $|z|$
and $F_n(0)=1$. Keeping only the leading term in 
\eqref{Philambda} we obtain to leading order 
\bea \label{maximize} 
 F(\nu) && \simeq \max_{\tilde \lambda} \left[ \frac{4}{15 \pi} ( (-z)^{5/2} - (-z-\tilde \lambda)_+^{5/2} ) - \tilde \lambda \nu \right] \\
&& = \frac{4}{15 \pi} (-z)^{5/2} F_0(\tilde \nu) \quad , \quad F_0(\tilde \nu)=1 - \frac{5}{2} \tilde \nu + \frac{3}{2} \tilde \nu^{5/3} 
\eea 
We find that $F_0(\tilde \nu)$ is convex, with the minimum attained at the typical (i.e. mean) value $\tilde \nu_{\rm typ}=1$, such that $F_0(\tilde \nu_{\rm typ})=0$. The 
maximum in \eqref{maximize} is achieved at $\tilde \lambda=\tilde \lambda^*=-z (1 - \tilde \nu^{2/3})$.
Note that the part $\tilde \nu < 1$ corresponds to $\tilde \lambda >0$, and 
$\tilde \nu > 1$ corresponds to $\tilde \lambda < 0$. This second part seems 
also correct: while our calculation adressed $\lambda >0$ its analytical continuation
to $\lambda >0$ seems to hold.\\

We can easily obtain the next correction in \eqref{expandF} from performing the
Legendre transform \eqref{legendre2} in presence of the $n=2$ term in
\eqref{Philambda} and we obtain
\bea
&& F_1(\tilde \nu) = 
1 + \tilde \nu^{4/3}  - \tilde \nu^{1/3} (1+\tilde \nu^{2/3}) 
+  (1-\tilde \nu^{2/3})^2  \log \frac{1 + \tilde \nu^{1/3} }{\sqrt{1-\tilde \nu^{2/3}}} 
\eea
The calculation of the next orders $F_n(\tilde \nu)$, $n \geq 2$ is possible along the same
line but is more tedious.

\section{Conclusions}

We have studied in this paper short time expansions for large deviations of the height in the KPZ equation on $\mathbb{R}$ with either droplet (sharp wedge) or stationary initial condition. In the droplet case, we obtained a systematic algorithm giving the short time expansion. For the Brownian initial condition, our results give the first two orders in time. It would be interesting to see whether the weak noise theory could be extended beyond the leading order, and compare with our results. \\

Our results for the droplet case extend straightforwardly to more general linear statistics of the Airy point process describing the edge of the spectrum in the Gaussian unitary ensemble. It allows to test the matching to the Gaussian free field in the bulk regime, and to describe corrections due to the edge of the spectrum. Applications to the counting statistics of fermions at the edge of a trap are also discussed, e.g. deviations from the high temperature Gumbel and Poisson distributions are systematically obtained.\\

Finally, we have pointed out several unexpected similarities hinting at the possible existence of a duality between large deviations around the (perturbative) Edwards-Wilkinson fixed point and the (non-perturbative) KPZ fixed point in a finite volume. A more precise formulation of this hypothetical duality would be extremely nice. It might be helpful for this to understand precisely how the results of \cite{derrida2009current} for the symmetric simple exclusion process, which have somewhat similar expressions involving half-integer polylogarithms, may fit within the duality.

\bigskip

{\bf Acknowledgments}

We thank A. Borodin, V. Gorin, S. N. Majumdar and G. Schehr for interesting discussions. 
We are grateful to S. N. Majumdar for pointing Ref. \cite{basor1999determinants} to us.
PLD and AK acknowledge support from ANR grant ANR-17-CE30-0027-01 RaMaTraF.

\newpage
\appendix
\begin{section}{First coefficients of the short time expansion for droplet initial condition}
	\label{Appendix coeffs crq}
	In this Appendix, we give in table \ref{table crq} the first coefficients $c_{r,q}(\mathbf{n})$ of \eqref{conjecture G} corresponding to \eqref{G short time expansion}. 
	
	\begin{table}[h!]
		\renewcommand{\arraystretch}{0.5}
		\begin{tabular}{lllll}
			\begin{tabular}{llcr}
				$r$ & $q$ & $\mathbf{n}$ & $c_{r,q}(\mathbf{n})$\\
				1 & 0 & (0,0,0) & 1 \\
				1 & 1 & (1) & 1\\\\
				2 & 0 & (1,0,0,0) & 1 \\
				2 & 1 & (1,1) & 1 \\
				2 & 1 & (2,0) & 2\\\\
				3 & 0 & (1,1,0,0,0) & 2 \\
				3 & 0 & (2,0,0,0,0) & 2 \\
				3 & 1 & (1,1,1) & 2 \\
				3 & 1 & (2,1,0) & 4 \\
				3 & 1 & (3,0,0) & 6 \\
				3 & 2 & (4) & 40 \\
				\vspace{81mm}
			\end{tabular}
			&\hspace*{7mm}&
			\begin{tabular}{llcr}
				$r$ & $q$ & $\mathbf{n}$ & $c_{r,q}(\mathbf{n})$\\
				4 & 0 & (1,1,1,0,0,0) & 6 \\
				4 & 0 & (2,1,0,0,0,0) & 6 \\
				4 & 0 & (3,0,0,0,0,0) & 6 \\
				4 & 1 & (1,1,1,1) & 6 \\
				4 & 1 & (2,1,1,0) & 12 \\
				4 & 1 & (2,2,0,0) & 16 \\
				4 & 1 & (3,1,0,0) & 18 \\
				4 & 1 & (4,0,0,0) & 24 \\
				4 & 2 & (3,2) & 116 \\
				4 & 2 & (4,1) & 120 \\
				4 & 2 & (5,0) & 200\\\\
				5 & 0 & (1,1,1,1,0,0,0) & 24 \\
				5 & 0 & (2,1,1,0,0,0,0) & 24 \\
				5 & 0 & (2,2,0,0,0,0,0) & 24 \\
				5 & 0 & (3,1,0,0,0,0,0) & 24 \\
				5 & 0 & (4,0,0,0,0,0,0) & 24 \\
				5 & 1 & (1,1,1,1,1) & 24 \\
				5 & 1 & (2,1,1,1,0) & 48 \\
				5 & 1 & (2,2,1,0,0) & 64 \\
				5 & 1 & (3,1,1,0,0) & 72 \\
				5 & 1 & (3,2,0,0,0) & 84 \\
				5 & 1 & (4,1,0,0,0) & 96 \\
				5 & 1 & (5,0,0,0,0) & 120 \\
				5 & 2 & (2,2,2) & 448 \\
				5 & 2 & (3,2,1) & 464 \\
				5 & 2 & (3,3,0) & 696 \\
				5 & 2 & (4,1,1) & 480 \\
				5 & 2 & (4,2,0) & 704 \\
				5 & 2 & (5,1,0) & 800 \\
				5 & 2 & (6,0,0) & 1200 \\
				5 & 3 & (7) & 19600 \\
				\vspace{5mm}
			\end{tabular}
			&\hspace*{7mm}&
			\begin{tabular}{llcr}
				$r$ & $q$ & $\mathbf{n}$ & $c_{r,q}(\mathbf{n})$\\
				6 & 0 & (1,1,1,1,1,0,0,0) & 120 \\
				6 & 0 & (2,1,1,1,0,0,0,0) & 120 \\
				6 & 0 & (2,2,1,0,0,0,0,0) & 120 \\
				6 & 0 & (3,1,1,0,0,0,0,0) & 120 \\
				6 & 0 & (3,2,0,0,0,0,0,0) & 120 \\
				6 & 0 & (4,1,0,0,0,0,0,0) & 120 \\
				6 & 0 & (5,0,0,0,0,0,0,0) & 120 \\
				6 & 1 & (1,1,1,1,1,1) & 120 \\
				6 & 1 & (2,1,1,1,1,0) & 240 \\
				6 & 1 & (2,2,1,1,0,0) & 320 \\
				6 & 1 & (2,2,2,0,0,0) & 384 \\
				6 & 1 & (3,1,1,1,0,0) & 360 \\
				6 & 1 & (3,2,1,0,0,0) & 420 \\
				6 & 1 & (3,3,0,0,0,0) & 504 \\
				6 & 1 & (4,1,1,0,0,0) & 480 \\
				6 & 1 & (4,2,0,0,0,0) & 528 \\
				6 & 1 & (5,1,0,0,0,0) & 600 \\
				6 & 1 & (6,0,0,0,0,0) & 720 \\
				6 & 2 & (2,2,2,1) & 2240 \\
				6 & 2 & (3,2,1,1) & 2320 \\
				6 & 2 & (3,2,2,0) & 3200 \\
				6 & 2 & (3,3,1,0) & 3480 \\
				6 & 2 & (4,1,1,1) & 2400 \\
				6 & 2 & (4,2,1,0) & 3520 \\
				6 & 2 & (4,3,0,0) & 4896 \\
				6 & 2 & (5,1,1,0) & 4000 \\
				6 & 2 & (5,2,0,0) & 5120 \\
				6 & 2 & (6,1,0,0) & 6000 \\
				6 & 2 & (7,0,0,0) & 8400 \\
				6 & 3 & (4,4) & 77696 \\
				6 & 3 & (5,3) & 80480 \\
				6 & 3 & (6,2) & 86240 \\
				6 & 3 & (7,1) & 98000 \\
				6 & 3 & (8,0) & 156800
			\end{tabular}
		\end{tabular}
		\caption{First coefficients $c_{r,q}(\mathbf{n})$ of \eqref{conjecture G} obtained from the short time expansion \eqref{G short time expansion}.}
		\label{table crq}
	\end{table}
\end{section}

\section{General identities, theorems and proofs of various results}
\subsection{General identities}
\subsubsection{Polylogarithm}
We introduce the polylogarithm ${\rm Li}_{s}(z)= \sum\limits_{p=1}^{+\infty} \dfrac{z^k}{p^s}$ and note the following identity
\begin{equation}
\mathrm{Li}_{s+\frac{1}{k}}(\sigma)=\frac{1}{\Gamma(1+\frac{1}{k})}\int_{0}^{\infty}\rmd x \, \mathrm{Li}_s(\sigma \rme^{-x^k})
\end{equation}
This is seen by expanding the polylogarithm as a series, integrating each term using the definition of the $\Gamma$ function and summing back the series. Two important cases of this identity are
\begin{equation}
\label{int_to_half_inf}
\mathrm{Li}_{s+\frac{1}{2}}(\sigma)= \frac{1}{\sqrt{\pi}}\int_{-\infty}^{\infty}\rmd x \, \mathrm{Li}_s(\sigma\rme^{-x^2})
\end{equation}
and
\bea
\label{Li3/2identity}
\frac{1}{\sqrt{4 \pi}} \mathrm{Li}_{s+3/2}(\sigma)=\frac{1}{\pi} \int_0^{+\infty} \mathrm{d} x \sqrt{x} \mathrm{Li}_{s}(\sigma e^{-x})
\eea 
We also recall the asymptotics for $z \to -\infty$
\bea \label{largezLi} 
{\rm Li}_s(z) = \sum_{k=0}^{+\infty} (-1)^k (1-2^{1-2 k}) (2 \pi)^{2k} \frac{B_{2k}}{(2 k)!} \frac{[\log(-z)]^{s-2 k}}{\Gamma(s+1-2 k)}
\eea 
where $B_{2k}$ are the Bernoulli numbers.\\

 Finally we need the following expansion around $x=0$.
\bea \label{Liexp} 
{\rm Li}_s(e^{-x}) = \Gamma(1-s) x^{s-1} + \sum_{k=0}^{+\infty} \frac{(-1)^k}{k!} \zeta(s-k) x^k
\eea 

\subsubsection{Airy function}
We recall the definition of the Airy function
\begin{equation} \label{ai_function} 
\Ai(v) = \int\limits_{-\infty+i \epsilon}^{+\infty+i \epsilon}\frac{\rmd\eta}{2\pi} \; \mathrm{exp}\Big(i \frac{\eta^3}{3}+ i v\eta\Big)
\end{equation}
where $\epsilon=0^+$. The Airy function is normalized to 1, i.e. $\int_{-\infty}^{\infty}\rmd v\,\Ai(v)=1$,
and enjoys a duplication identity
\begin{equation}
\label{ai_duplication}
\Ai(u)^2=\frac{1}{2^{2/3}\pi}\int_{-\infty}^{\infty}\rmd v \, \Ai(v^2+ 2^{2/3}u)
\end{equation}

\subsection{Sparre Andersen theorem}
\label{sparre_andersen_theorem}
The Sparre-Andersen theorem lies in the framework of random partial  sums $S_i=X_1+\dots+X_i$ of a sequence of random variables $\lbrace X_i\rbrace $. Here we make the hypothesis that our process forms a bridge, i.e. $S_0=S_{n+1}=0$ and we are interested in $N_n^*$ be the number of points $(j,S_j)$, $j=1,\dots,n$ which lie above the straight line from $(0,0)$ to $(n+1,S_{n+1})$.
\begin{theorem}[Sparre Andersen Corollary 1, Ref. \cite{andersen1954fluctuations}]
If the random variables $X_1,\dots,X_{n+1}$ are independent and each has a continuous distribution, or if the random variables are symmetrically dependent and the joint distribution function is absolutely continuous, then for any $C$ which is symmetric with respect to $X_1, \dots, X_{n+1}$ and has $\mathbb{P}(C)>0$, we have
\begin{equation}
\forall m\in [0,n], \; \mathbb{P}(N_n^*=m \mid C)=\frac{1}{n+1}
\end{equation}
\end{theorem}
For our case of interest, the event $C$ will be our hypothesis that the process is a bridge.
\begin{equation}
 C=\lbrace S_0=0 \rbrace \cap \lbrace S_{n+1}=0 \rbrace
 \end{equation} 
\subsection{Proof of the identities for the functions $L_{\beta}$}
\label{Appendix identities L}
In this section, we prove the identities \eqref{L[Li]}-\eqref{sdsdaL} for the functions $L_{\beta}(\sigma,a,b)$.

\subsubsection{Proof of the identity \eqref{L[Li]}}
		From the definition \eqref{L[int]}, one has
		\begin{equation}
			L_{\beta}(\sigma,0,0)=\sum_{k=1}^{\infty}(-1)^{k-1}\sum_{n_{1},\ldots,n_{k}=1}^{\infty}\bigg(\prod_{\ell=1}^{k}a_{n_l,\beta}\,\sigma^{n_{\ell}}\bigg) \int_{0}^{\infty}\rmd z_{1}\ldots\rmd z_{k-1}\,\bigg(\prod_{\ell=1}^{k}K_{n_{j}}(z_{\ell}-z_{\ell-1})\bigg)_{\substack{z_{0}=0\\z_{k}=0}} 
		\end{equation}
	where $K_{n}(z)=\frac{\rme^{-\frac{z^{2}}{4n}}}{\sqrt{4\pi n}}$ is the Gaussian kernel. The integral in the expression above can be interpreted as the probability that a random walker, starting initially at position $0$, ends at position $0$ after $k-1$ steps after staying only on positive positions for all intermediate steps. The transition probabilities depend on the variables $n_{j}$, but in a symmetric way since each $n_{j}$ has the same weight. Thus, it is possible to use the Sparre Andersen theorem (see Appendix \ref{sparre_andersen_theorem}) to replace the integral by $\frac{1}{k}\,K_{n_{1}+\ldots+n_{k}}(0)$. This leads to

		\begin{equation}
		\begin{split}
			L_{\beta}(\sigma,0,0)&=\frac{1}{\sqrt{4\pi}}\sum_{k=1}^{\infty}\frac{(-1)^{k-1}}{k}\sum_{n_{1},\ldots,n_{k}=1}^{\infty}\frac{\prod_{\ell=1}^{k}a_{n_l,\beta}\,\sigma^{n_{\ell}}}{\sqrt{n_{1}+\ldots+n_{k}}}\\
			&=\frac{1}{2\pi}\int_{0}^{+\infty}\frac{\mathrm{d}x}{ \sqrt{x} } \sum_{k=1}^{\infty}\frac{(-1)^{k-1}}{k}\sum_{n_{1},\ldots,n_{k}=1}^{\infty}\prod_{\ell=1}^{k}a_{n_\ell,\beta}\,(\sigma e^{-x})^{n_{\ell}}\\
&=\frac{1}{2\pi }\int_{0}^{+\infty}\frac{\mathrm{d}x}{ \sqrt{x}} \sum_{k=1}^{\infty}\frac{(-1)^{k-1}}{k}(\exp(\beta g(\sigma e^{-x}))-1)^{k}\\
			&=\frac{\beta}{2\pi }\int_{0}^{+\infty}\frac{\mathrm{d}x}{ \sqrt{x}} g(\sigma e^{-x}) =\frac{\beta}{2\pi }\int_{-\infty}^{+\infty}\mathrm{d}p\, g(\sigma e^{-p^2})\\
	&		=\frac{\beta }{\pi }(\sigma \partial_\sigma) \int_{0}^{+\infty}\mathrm{d}x\sqrt{x} g(\sigma e^{-x})
			\end{split}
		\end{equation}
With no boundary term using that $g(0)=0$ as long as $\sqrt{x}g(\sigma e^{-x})\to 0$ for $x\to +\infty$. 
We precisely obtain \eqref{L[Li]}.

\subsubsection{Proof of the identity \eqref{L sym}}
	The identity \eqref{L sym} simply follows from renaming $z_{j}\to z_{k-j}$ in the definition \eqref{L[int]}.

	\begin{subsubsection}{Proof of the identity \eqref{dadbL}}
		The identity \eqref{dadbL} is a consequence of \eqref{L sym} and \eqref{sdsdaL}. However, since our derivation of \eqref{sdsdaL} below is slightly involved, we prove here \eqref{dadbL} directly from the definition \eqref{L[int]} of $L_{\beta}(\sigma,a,b)$. For any function $f$, the identity $\sum_{j=0}^{k}\partial_{z_{j}}\prod_{\ell=1}^{k}f(z_{\ell}-z_{\ell-1})=0$ holds. After insertion into \eqref{L[int]}, this implies
		\begin{align}
			(\partial_{a}+\partial_{b})&L_{\beta}(\sigma,a,b)\\
			&+\sum_{k=1}^{\infty}(-1)^{k-1}\int_{0}^{\infty}\rmd z_{1}\ldots\rmd z_{k-1}\,\sum_{j=1}^{k-1}\partial_{z_{j}}\prod_{\ell=1}^{k}\Big(\sum_{n=1}^{\infty}\frac{a_{n,\beta}\,\sigma^{n}}{\sqrt{4\pi n}}\,\rme^{-\frac{(z_{\ell}-z_{\ell-1})^{2}}{4n}}\Big)_{\substack{z_{0}=a\\z_{k}=b}}=0 \nonumber
		\end{align}
		Performing the integration with respect to $z_{j}$, exchanging summations as $\sum_{k=1}^{\infty}\sum_{j=1}^{k-1}=\sum_{j=1}^{\infty}\sum_{k=j+1}^{\infty}$ and making the change of variable $k\to k+j$ leads to \eqref{dadbL}.

	\end{subsubsection}

	\begin{subsubsection}{Proof of the identity \eqref{sdsdaL}}
		From the definition \eqref{L[int]} of $L_{\beta}(\sigma,a,b)$, one has
		\begin{equation}
			\label{sdsL[int]}
			\sigma\partial_{\sigma}L_{\beta}(\sigma,a,b)=\sum_{k=1}^{\infty}\sum_{j=1}^{k}(-1)^{k-1}\int_{0}^{\infty}\rmd z_{1}\ldots\rmd z_{k-1}\,\prod_{\ell=1}^{k}\Big(\sum_{n=1}^{\infty}\frac{n^{\delta_{j,\ell}}a_{n,\beta}\,\sigma^{n}}{\sqrt{4\pi n}}\,\rme^{-\frac{(z_{\ell}-z_{\ell-1})^{2}}{4n}}\Big)_{\substack{z_{0}=a\\z_{k}=b}}\;,
		\end{equation}
		with $\delta_{j,\ell}$ the Kronecker delta function. We then apply $\partial_{a}=\partial_{z_{0}}=(\sum_{r=0}^{j-1}\partial_{z_{r}})-(\sum_{r=1}^{j-1}\partial_{z_{r}})$ on the previous expression. The contributions of $\sum_{r=0}^{j-1}\partial_{z_{r}}$ and $\sum_{r=1}^{j-1}\partial_{z_{r}}$ will eventually correspond to the two terms in the right hand side of \eqref{sdsdaL}. For the term with $\sum_{r=0}^{j-1}\partial_{z_{r}}$, we use the identity
		\begin{equation}
			\sum_{r=0}^{j-1}\partial_{z_{r}}\prod_{\ell=1}^{k}f((z_{\ell}-z_{\ell-1})^{2})=-2(z_{j}-z_{j-1})\,\frac{f'((z_{j}-z_{j-1})^{2})}{f((z_{j}-z_{j-1})^{2})}\,\prod_{\ell=1}^{k}f((z_{\ell}-z_{\ell-1})^{2})\;,
		\end{equation}
		valid for any function $f$, while for the term with $\sum_{r=1}^{j-1}\partial_{z_{r}}$, we compute the integral with respect to $z_{r}$. We obtain
		\begin{align}
			&\sigma\partial_{\sigma}\partial_{a}L_{\beta}(\sigma,a,b)\\
			&=\sum_{k=1}^{\infty}\sum_{j=1}^{k}(-1)^{k-1}\int_{0}^{\infty}\rmd z_{1}\ldots\rmd z_{k-1}\,\frac{z_{j}-z_{j-1}}{2}\,\prod_{\ell=1}^{k}\Big(\sum_{n=1}^{\infty}\frac{a_{n,\beta}\,\sigma^{n}}{\sqrt{4\pi n}}\,\rme^{-\frac{(z_{\ell}-z_{\ell-1})^{2}}{4n}}\Big)_{\substack{z_{0}=a\\z_{k}=b}}\nonumber\\
			&+\sum_{k=1}^{\infty}\sum_{j=1}^{k}\sum_{r=1}^{j-1}(-1)^{k-1}\bigg(\int_{0}^{\infty}\rmd z_{1}\ldots\rmd z_{r-1}\,\prod_{\ell=1}^{r}\Big(\sum_{n=1}^{\infty}\frac{n^{\delta_{j,\ell}}a_{n,\beta}\,\sigma^{n}}{\sqrt{4\pi n}}\,\rme^{-\frac{(z_{\ell}-z_{\ell-1})^{2}}{4n}}\Big)_{\substack{z_{0}=a\\z_{r}=0}}\bigg)\nonumber\\
			&\hspace{35mm}\times\bigg(\int_{0}^{\infty}\rmd z_{r+1}\ldots\rmd z_{k-1}\,\prod_{\ell=r+1}^{k}\Big(\sum_{n=1}^{\infty}\frac{n^{\delta_{j,\ell}}a_{n,\beta}\,\sigma^{n}}{\sqrt{4\pi n}}\,\rme^{-\frac{(z_{\ell}-z_{\ell-1})^{2}}{4n}}\Big)_{\substack{z_{r}=0\\z_{k}=b}}\bigg)\;,\nonumber
		\end{align}
		Using $\sum_{j=1}^{k}\frac{z_{j}-z_{j-1}}{2}=\frac{z_{k}-z_{0}}{2}=\frac{b-a}{2}$, the first term becomes equal to $\frac{b-a}{2}\,L_{\beta}(\sigma,a,b)$. In the second term, we exchange the summations as $\sum_{k=1}^{\infty}\sum_{j=1}^{k}\sum_{r=1}^{j-1}=\sum_{r=1}^{\infty}\sum_{k=r+1}^{\infty}\sum_{j=r+1}^{k}$ and make the change of variable $k\to k+r$. Using \eqref{sdsL[int]}, we finally obtain the identity \eqref{sdsdaL}.
	\end{subsubsection}

\subsection{Proof of the identity \eqref{prop2} for propagator of the Brownian initial condition }

Starting from the definition in Eq. \eqref{prop2}, one wants to calculate the deformed Airy propagator
\begin{equation}
G_{p/\gamma}(r_1,r_2) =\int_{-\infty}^{+\infty} \rmd v \, \rme^{p v/\gamma} \Ai_\Gamma^\Gamma(v+r_1,\gamma,w,w) \Ai_\Gamma^\Gamma(v+r_2,\gamma,w,w) 
\end{equation}
To achieve this, one calls Lemma 6 and Eqs. (4.15), and (4.17) of \cite{SasamotoStationary2} which reads (the function $\Ai_{\Gamma \Gamma}^{\Gamma \Gamma}$ is defined in Eq. (4.16) of  \cite{SasamotoStationary2}, as it only serves as an intermediate notation we do not recall its definition)
\begin{equation}
\begin{split}
&\Ai_\Gamma^\Gamma(v+r_1,\gamma,w,w) \Ai_\Gamma^\Gamma(v+r_2,\gamma,w,w) =\\
&\frac{1}{2^{1/3}\pi}\int_{-\infty}^{+\infty}\rmd k \, \Ai_{\Gamma \Gamma}^{\Gamma \Gamma}(2^{2/3}k^2+2^{2/3}v +2^{-1/3}(r_1+r_2),2^{-1/3}\gamma,-i\gamma k+w,i\gamma k+w)\rme^{ik(r_1-r_2)}
\end{split}
\end{equation}
and
\begin{equation}
\begin{split}
&2^{2/3}\int_{-\infty}^{+\infty}\rmd v\, \Ai_{\Gamma \Gamma}^{\Gamma \Gamma}(2^{2/3}k^2+2^{2/3}v+ 2^{-1/3}(r_1+r_2),2^{-1/3}\gamma,-i\gamma k+w,i\gamma k+w)\\
& \times \rme^{2^{2/3}p(k^2+v+\frac{r_1+r_2}{2})/(2^{2/3}\gamma)} \\
&=\int_{-\infty}^{+\infty}\rmd v\, \Ai_{\Gamma \Gamma}^{\Gamma \Gamma}(v,2^{-1/3}\gamma,-i\gamma k+w,i\gamma k+w)\rme^{2^{1/3}pv/(2\gamma)}\\
&=\frac{\Gamma(-i\gamma k+w-\frac{p}{2})\Gamma(i\gamma k+w-\frac{p}{2})}{\Gamma(-i\gamma k+w+\frac{p}{2})\Gamma(i\gamma k+w+\frac{p}{2})}\rme^{\frac{p^3}{12 \gamma^3}}\\
\end{split}
\end{equation}
Combining both expressions leads to the result expressed in \eqref{prop2}.
\begin{equation}
G_{p/\gamma}(r_1,r_2) =\rme^{\frac{p^3}{12 \gamma^3}-\frac{p}{\gamma}\frac{r_1+r_2}{2}}\int_{-\infty}^{+\infty}\frac{\rmd k}{2\pi}\frac{\Gamma(-i\gamma k+w-\frac{p}{2})\Gamma(i\gamma k+w-\frac{p}{2})}{\Gamma(-i\gamma k+w+\frac{p}{2})\Gamma(i\gamma k+w+\frac{p}{2})}\rme^{-\frac{p}{\gamma}k^2-ik(r_1-r_2)}
\end{equation}
\section{Additional identities for the cumulants}
\subsection{Additional identities \eqref{equiv_cum1} and \eqref{equiv_cum2} for the first cumulant of the droplet initial condition}
To obtain Eq. \eqref{equiv_cum1} one starts from the definition \eqref{third_order_expansion}. With this definition the first cumulant for $g=g_{\rm KPZ}$ reads 
	\begin{equation}
	\begin{split}
	\kappa_1&=\int_{\mathbb{R}}\rmd u \, \mathrm{Li}_1(\sigma \rme^{u/\gamma})K_{\rm Ai}(u,u)\\
		\end{split}
	\end{equation}
	Integrating this expression by part using that $\partial_u K_{\Ai}(u,u)=-\Ai(u)^2$ and then using the duplication identity \eqref{ai_duplication} leads to 
		\begin{equation}
	\begin{split}
		\kappa_1&=\gamma \int_{\mathbb{R}}\rmd u \, \mathrm{Li}_2(\sigma \rme^{u/\gamma})\Ai^2(u)\\
	&=\frac{\gamma}{2^{2/3}\pi} \iint_{\mathbb{R}^2}\rmd u \rmd v\mathrm{Li}_2(\sigma \rme^{u/\gamma})\Ai(v^2+ 2^{2/3}u)\\
	&=\frac{\gamma^{3/2}}{2\pi} \iint_{\mathbb{R}^2}\rmd u \rmd v\mathrm{Li}_2(\sigma \rme^{2^{-2/3}u/\gamma-v^2})\Ai(u)\\
		\end{split}
	\end{equation}
Recalling that $\gamma=t^{-1/3}$, the identity \eqref{int_to_half_inf} for polylogarithms allows to obtain Eq. \eqref{equiv_cum1} as
	\begin{equation}
	\kappa_1= \frac{\gamma^{3/2}}{\sqrt{4 \pi}}  \int_\mathbb{R}\rmd u\, 
	\mathrm{Li}_{5/2}(\sigma \rme^{2^{-2/3} u/\gamma}) \Ai(u)
	\end{equation}
We observe from this expression that the dominant order is obtained by setting the exponential in the argument of the polylogarithm to 1 and using the normalization of the Airy function.\\

To obtain Eq. \eqref{equiv_cum2}, one starts from Eq. \eqref{KPZ_cum1} and uses the identity \eqref{int_to_half_inf} for polylogarithms.

\subsection{All order expressions}
\label{sec:allorders} 

It is possible to obtain an exact expression for $\kappa_3(g)$ to all orders. Starting instead from 
\eqref{thirdc} and inserting the form of the propagator given in the
last equation in \eqref{prop1}, we can perform the summation over $p_1$, $p_2$, $p_3$
and a rescaling of $k_1,k_2,k_3$ to obtain 
\bea
&& \kappa_3(g) = \\
&&   \left[ \int_{\mathbb{R}^+ \times \mathbb{R}^- \times\mathbb{R}^-} -  \int_{\mathbb{R}^- \times \mathbb{R}^+ \times\mathbb{R}^+} \right] \mathrm{d}r_1 \mathrm{d}r_2 \mathrm{d}r_3 
\int_\mathbb{R} \frac{\mathrm{d}k_1}{2 \pi} \int_\mathbb{R} \frac{\mathrm{d}k_2}{2 \pi} \int_\mathbb{R} \frac{\mathrm{d}k_3}{2 \pi}
e^{i k_1 (r_1-r_2) + i k_2 (r_2-r_3) + i k_3 (r_3-r_1)} \nonumber \\
&& \times 
[ e^{\frac{t}{12} (\sigma \partial_\sigma)^3} 
g(\sigma e^{- \frac{\sqrt{t}}{2} (r_1+r_2) - k_1^2}) ] 
[ e^{\frac{t}{12} (\sigma \partial_\sigma)^3} 
g(\sigma e^{- \frac{\sqrt{t}}{2} (r_2+r_3) - k_2^2}) ] 
[ e^{\frac{t}{12} (\sigma \partial_\sigma)^3} 
g(\sigma e^{- \frac{\sqrt{t}}{2} (r_3+r_1) - k_3^2}) ] \nonumber 
\eea 

We can now give the general cumulant

We first note that using \eqref{prop1}
\bea
&& [ {\sf Ai} \, \hat g_{t,\sigma}^m \, {\sf Ai}](r,r')= \sum_{p_1,.. p_m=1}^{+\infty}  
\frac{\sigma^{p_1+.. +p_m}}{p_1! \dots p_m!}g^{(p_1)}(0) \dots g^{(p_m)}(0)
G_{(p_1+\dots +p_m)/\gamma}(r,r') \\
&& = e^{\frac{t}{12} (\sigma \partial_\sigma)^3} \int_\mathbb{R} \frac{\mathrm{d}k}{2 \pi}  e^{i k (r-r')}
[ g(\sigma e^{- \frac{r+r'}{2 \gamma} - \frac{k^2}{\gamma}}) ]^m 
\eea
Hence we have
\bea
\kappa_n(g)=\sum_{\ell=1}^n \frac{(-1)^{\ell+1}}{\ell}\sum_{\substack{m_1,\dots,m_\ell \geq 1 \\ m_1+\dots+m_\ell=n}}\frac{n!}{m_1!\dots m_\ell !}
\int_{\mathbb{R}^+} \mathrm{d}r_1 \dots \int_{\mathbb{R}^+} \mathrm{d}r_\ell
\int_\mathbb{R} \frac{\mathrm{d}k_1}{2 \pi} \dots \int_\mathbb{R}  \frac{\mathrm{d}k_\ell}{2 \pi} \\
\times e^{i k_1(r_1-r_2)+ik_2(r_2-r_3)+ \dots + ik_\ell (r_\ell-r_1)}
\prod_{j=1}^\ell e^{\frac{t}{12} (\sigma \partial_\sigma)^3} 
[ g(\sigma e^{- \frac{\sqrt{t} (r_j+r_{j+1})}{2}  - k_j^2}) ]^{m_j}
\eea

\section{Analytical continuation} 
\label{sec:analytical} 

For $0<\sigma<1$, let us denote $\sigma=e^{-y^2}$ with $y>0$.
Using \eqref{Liexp} and parametrizing $\sigma =e^{- y^2}<1$ let us first note that 
both sides of \eqref{change} are simple power series in $y$ near $y=0$ 
related by the change $y \to -y$, which will be useful below.\\

Denoting $h=H-H_c$, the first equation in \eqref{param_phi_0_1} which describes the branch $h<0$, can be written as
\bea
h=\log \frac{\mathrm{Li}_{3/2}(e^{-y^2})}{\zeta(3/2) e^{-y^2}} = f(y) = \sum_{n \geq 1} f_n y^n 
\eea
where $f(y)$ has a simple power series expansion near $y=0$, with $f_1=- \frac{2 \sqrt{\pi}}{\zeta(3/2)}$ etc$\dots$ 
The first equation in \eqref{param_phi_0_2} which describes the branch $h>0$, can be written as
\bea
h=\log \frac{\mathrm{Li}_{3/2}(e^{-y^2})+\sqrt{16 \pi} y }{\zeta(3/2) e^{-y^2}} = f(- y) 
= \sum_{n \geq 1} (-1)^n f_n y^n 
\eea
Hence the two branches are simply $h=f(\pm y)$. The series $f(y)$ can be inverted leading
to $y= \pm g(h)$ where $g(h)=\sum_{n \geq 1} g_n h^n$ is again simple power series in $h$. Similarly one can check that the formulae for $\Phi(H)$ in \eqref{param_phi_0_1} and in \eqref{param_phi_0_2} respectively, are also identical up to the change $y \to -y$, i.e. $\Phi(H)= \phi(y)$ for $h<0$ and
$\Phi(H)= \phi(-y)$ for $h>0$, where $\phi(y)$ has a simple power series expansion near $y=0$.
Hence $\Phi(H)=\phi(\pm y)$ and inserting $y= \pm g(h)$ we find that 
$\Phi(H)=\phi(g(h))$ which is analytic near $h=0$. The analytical continuation is thus correct.\\

The analytical continuation for the the function $\Phi_0(H)$ is more delicate. 
Let us first rewrite, for $0<\sigma<1$, the third equation in \eqref{param_phi_0_1}, by adding and substracting
$\frac{1}{2} \log \sqrt{- \log \sigma}$ and splitting some integrals, as
\be
  \Phi_0(H)= K + \frac{1}{4\pi}\int_{1}^{\sigma}\!\frac{\rmd u}{u}\, \Big[ \Li_{1/2}(u)^{2} - \frac{\pi}{- \log u}\Big]
 +\frac{1}{2}\log \Big( \frac{\mathrm{Li}_{3/2}(\sigma)^2}{(\mathrm{Li}_{1/2}(\sigma)-\mathrm{Li}_{3/2}(\sigma))
\sqrt{- \log \sigma}} \Big)
\ee
where $K$ is the constant given in \eqref{constanteK}.
Let us denote again $\sigma=e^{-y^2}$ with $y>0$. We have
\bea
&& K = \frac{1}{2\pi}\int_{1}^{+\infty} \mathrm{d}y \, y \,\Li_{1/2}(e^{-y^2})^{2} + 
\frac{1}{2\pi}\int_{0}^{1}  \mathrm{d}y \, y \, \Big[ \Li_{1/2}(e^{-y^2})^{2} - \frac{\pi}{y^2}\Big] - \log( 2 \pi^{3/2}) \\
&& = 0.00814542 -0.638714 - \log( 2 \pi^{3/2})
\eea
and, for $h<0$
\be
  \Phi_0(H)= K - \frac{1}{2\pi}\int_{0}^{y}  \mathrm{d}y' \, y' \, \Big[ \Li_{1/2}(e^{-(y')^2})^{2} - \frac{\pi}{(y')^2}\Big] 
+\frac{1}{2}\log \Big( \frac{\mathrm{Li}_{3/2}(e^{-y^2})^2}{(\mathrm{Li}_{1/2}(e^{-y^2})-\mathrm{Li}_{3/2}(e^{-y^2})) y } \Big)
\ee
The analytic continuation for $h>0$ is then 
\bea \label{cont3} 
&&  \Phi_0(H)= K  - \frac{1}{2\pi}\int_{0}^{y}  \mathrm{d}y' \, y' \, \Big[ \Big(\Li_{1/2}(e^{-(y')^2}) - \frac{\sqrt{4\pi}}{y'}\Big)^{2} - \frac{\pi}{(y')^2}\Big] \\
&& + \frac{1}{2}\log \Bigg( \frac{\Big(\mathrm{Li}_{3/2}(e^{-y^2})+y \sqrt{16 \pi} \Big)^2}{\Big(-\mathrm{Li}_{1/2}(e^{-y^2})+\mathrm{Li}_{3/2}(e^{-y^2})+y \sqrt{16 \pi}+\frac{\sqrt{4\pi}}{y}\Big)y} \Bigg) \nonumber 
\eea
One can indeed check that the logarithmic term has the form $\tilde \phi_0(\pm y)$ in the two branches, 
where $\tilde \phi_0(y)$ is a simple series in $y$ near $y=0$. Next we find that 
\bea
&& y \Big[ \Li_{1/2}(e^{-y^2})^{2} - \frac{\pi}{y^2} \Big] = \psi(y) = \sum_{n \geq 0} \psi_n y^n  \\
&& y \Big[\Big( \Li_{1/2}(e^{-y^2}) - \frac{\sqrt{4\pi}}{y}\Big)^{2} - \frac{\pi}{y^2} \Big] = - \psi(- y) = \sum_{n \geq 0} (-1)^{n+1} \psi_n y^n
\eea 
Upon integration of these series $\int_0^y \mathrm{d}y'$ we see the property that 
$\Phi_0(H)=\phi_0(\pm y)$, where $\phi_0(y)$ is a simple series in $y$ near $y=0$.
Hence $\Phi_0(H)=\phi_0(g(h))$ is indeed analytic near $h=0$.
Restoring the variable $\sigma=e^{-y^2}$, one sees that Eq. \eqref{cont3} leads to the expression for
$\Phi_0(H)$ in Eq. \eqref{param_phi_0_2}.

\end{document}